\documentclass[aps,prc,amsfonts,showpacs,superscriptaddress,groupedaddress,nofootinbib,floatfix``]{revtex4}  
\pdfoutput=1
\usepackage{graphics}  
\graphicspath{{.}{./figures/}{./iowa/}}

\usepackage{dcolumn}   
\usepackage{bm}        
\usepackage{amssymb}   
\usepackage{color}

\usepackage{psfrag}
\usepackage{epsfig}
\usepackage{epsf}
\usepackage{float}
\usepackage{slashed}

\newcommand{\beq}{\begin{equation}}
\newcommand{\eeq}{\end{equation}}
\newcommand{\beqa}{\begin{eqnarray}}
\newcommand{\eeqa}{\end{eqnarray}}

\newcommand{\totdiff}[2]{\frac{\mathrm{d}#1}{\mathrm{d}#2}}

\newcommand{\comm}[2]{ \left[ #1 \, \text{,} \, #2 \right] }  
\newcommand{\opme}[3]{{#1}^{#2}_{#3}}
\newcommand{\tpme}[5]{{#1}^{#2#3}_{#4#5}}
\newcommand{\ichange}[2]{\left[ #1 \leftrightarrow #2 \right]}
\newcommand{\ket}[1]{| #1 \rangle}

\begin{document}

\title{Few- and many-nucleon systems with semilocal coordinate-space regularized\\
  chiral nucleon-nucleon forces}


\author{S.~Binder}
\affiliation{Department of Physics and Astronomy, University of Tennessee, Knoxville, TN 37996, USA}
\affiliation{Physics Division, Oak Ridge National Laboratory, Oak Ridge, TN 37831, USA}

\author{A.~Calci}
\affiliation{TRIUMF, 4004 Wesbrook Mall, Vancouver, British Columbia, V6T 2A3 Canada}

\author{E.~Epelbaum}
\affiliation{Institut f\"ur Theoretische Physik II, Ruhr-Universit\"at
  Bochum, D-44780 Bochum, Germany}

\author{R.J.~Furnstahl}
\affiliation{Department of Physics, The Ohio State University, 
Columbus, Ohio 43210, USA}

\author{J.~Golak}
\affiliation{M. Smoluchowski Institute of Physics, Jagiellonian
University,  PL-30348 Krak\'ow, Poland}

\author{K.~Hebeler}
\affiliation{Institut f\"ur Kernphysik, Technische Universit\"at 
Darmstadt, 64289 Darmstadt, Germany}

\author{T.~H\"uther}
\affiliation{Institut f\"ur Kernphysik, Technische Universit\"at 
Darmstadt, 64289 Darmstadt, Germany}

\author{H.~Kamada}
\affiliation{Department of Physics, Faculty of Engineering,
Kyushu Institute of Technology, Kitakyushu 804-8550, Japan}

\author{H.~Krebs}
\affiliation{Institut f\"ur Theoretische Physik II, Ruhr-Universit\"at
  Bochum, D-44780 Bochum, Germany}


\author{P.~Maris}
\affiliation{Department of Physics and Astronomy, Iowa State
  University, Ames, Iowa 50011, USA}

\author{Ulf-G.~Mei{\ss}ner}
\affiliation{Helmholtz-Institut~f\"{u}r~Strahlen-~und~Kernphysik~and~Bethe~Center~for~Theoretical~Physics,
~Universit\"{a}t~Bonn,~D-53115~Bonn,~Germany}
\affiliation{Institut f\"ur Kernphysik, Institute for Advanced Simulation 
and J\"ulich Center for Hadron Physics, Forschungszentrum J\"ulich, 
D-52425 J\"ulich, Germany}
\affiliation{JARA~-~High~Performance~Computing,~Forschungszentrum~J\"{u}lich,~D-52425~J\"{u}lich,~Germany}


\author{A.~Nogga}
\affiliation{Institut f\"ur Kernphysik, Institute for Advanced Simulation 
and J\"ulich Center for Hadron Physics, Forschungszentrum J\"ulich, 
D-52425 J\"ulich, Germany}


\author{R.~Roth}
\affiliation{Institut f\"ur Kernphysik, Technische Universit\"at 
Darmstadt, 64289 Darmstadt, Germany}

\author{R.~Skibi\'nski}
\affiliation{M. Smoluchowski Institute of Physics, Jagiellonian
University,  PL-30348 Krak\'ow, Poland}

\author{K.~Topolnicki}
\affiliation{M. Smoluchowski Institute of Physics, Jagiellonian
University,  PL-30348 Krak\'ow, Poland}

\author{J.P.~Vary}
\affiliation{Department of Physics and Astronomy, Iowa State
  University, Ames, Iowa 50011, USA}

\author{K.~Vobig}
\affiliation{Institut f\"ur Kernphysik, Technische Universit\"at 
Darmstadt, 64289 Darmstadt, Germany}

\author{H.~Wita{\l}a}
\affiliation{M. Smoluchowski Institute of Physics, Jagiellonian
University,  PL-30348 Krak\'ow, Poland}

\collaboration{LENPIC Collaboration}

\date{\today}

\begin{abstract}
We employ a variety of \emph{ab initio} methods including Faddeev-Yakubovsky
equations, No-Core Configuration Interaction Approach, Coupled-Cluster
Theory and In-Medium Similarity Renormalization Group  to perform a
comprehensive analysis of  the nucleon-deuteron
elastic and breakup reactions and selected properties of
light and medium-mass nuclei up to $^{48}$Ca using the recently
constructed semilocal coordinate-space regularized chiral
nucleon-nucleon potentials. We compare the results with those based on
selected phenomenological and chiral EFT two-nucleon potentials,
discuss the convergence pattern of the
chiral expansion and estimate the achievable theoretical accuracy at
various chiral orders using the novel approach to quantify truncation
errors of the chiral expansion without relying on cutoff
variation. We also address the robustness of this method and explore  
alternative ways to estimate the theoretical uncertainty
from the truncation of the chiral expansion. 
\end{abstract}

\maketitle

\vspace{-0.2cm}

\section{Introduction}
\def\theequation{\arabic{section}.\arabic{equation}}
\label{sec:intro}

Nuclear forces have been extensively studied within the framework
of chiral effective field theory (EFT) over the past two decades; see
Refs.~\cite{Epelbaum:2008ga,Machleidt:2011zz} for review articles. In this approach, two-,
three- and more-nucleon forces are calculated from the most general
effective Lagrangian order by order in the
chiral expansion, i.e., a perturbative expansion in powers of $Q \in
\{p/\Lambda_b, \; M_\pi / \Lambda_b \}$ with $p$, $M_\pi$ and $\Lambda_b$  referring
to the magnitude of the typical nucleon three-momenta, the pion
mass and the breakdown scale, respectively. 

Most of the calculations available so far utilize the heavy-baryon
formulation of chiral EFT with pions and nucleons being the only
active degrees of freedom and make use
of Weinberg's power counting for contact interactions based on naive
dimensional analysis \cite{Weinberg:1990rz,Weinberg:1991um}; see, however, 
Refs.~\cite{Nogga:2005hy,Birse:2005um,Valderrama:2009ei,Epelbaum:2012ua,Gasparyan:2012km,Oller:2014uxa}
and 
references therein for alternative formulations. Much progress has
been made within this framework in the past two years to improve the description of the
nucleon-nucleon (NN) force.  First, the order-$Q^5$
(i.e., N$^4$LO) \cite{Entem:2014msa} and even most of the 
order-$Q^6$ (N$^5$LO) contributions to the
NN force have been worked out \cite{Entem:2015xwa}. Second, a new
generation of NN potentials up to N$^4$LO has been developed 
using semilocal \cite{Epelbaum:2014efa,Epelbaum:2014sza} and
nonlocal \cite{Entem:2017gor} regularization schemes; see also
Refs.~\cite{Piarulli:2014bda,Carlsson:2015vda,Piarulli:2016vel,Ekstrom:2017koy}
for related studies along similar lines.  
In contrast to the previous
order-$Q^4$  (N$^3$LO)  chiral NN potentials of
Refs.~\cite{Entem:2003ft,Epelbaum:2004fk}, the long-range part of 
the interaction introduced in
Refs.~\cite{Epelbaum:2014efa,Epelbaum:2014sza}  is regularized in
coordinate space by multiplying 
with the function 
\begin{equation}
\label{DefReg}
f \bigg( \frac{r}{R} \bigg) = \bigg[ 1 - \exp \bigg( -\frac{r^2}{R^2}
\bigg) \bigg]^n\,, \quad \quad n=6\,,
\end{equation}
while the contact interactions are regularized in momentum space using a non-local
Gaussian regulator with the cutoff $\Lambda = 2 R^{-1}$. 
(See Refs.~\cite{Gezerlis:2013ipa,Gezerlis:2014zia,Piarulli:2014bda,Piarulli:2016vel}
for recently constructed chiral potentials with
locally regularized long-range interactions.)  
The resulting semilocal coordinate-space regularized (SCS) chiral
potentials are available for $R=0.8$, $0.9$, $1.0$,  $1.1$,  and
$1.2$~fm. The use of a local regulator for the short-range part of
the interaction allows one to reduce the amount of finite-cutoff
artifacts; see, however, Ref.~\cite{Dyhdalo:2016ygz} for a discussion of regulator
artifacts in uniform matter. Furthermore, in contrast to the
first generation of the chiral N$^3$LO potentials, all pion-nucleon ($\pi$N)
low-energy constants (LECs) were determined from the $\pi$N system
without any fine tuning. Consequently, the long-range part of the
NN force is predicted in a parameter-free way. In fact, clear
evidence of the resulting (parameter-free) contributions to the two-pion
exchange at orders $Q^3$ (i.e., N$^2$LO) and $Q^5$ was found in NN
phase shifts \cite{Epelbaum:2014efa,Epelbaum:2014sza}. We further
emphasize that the approximate independence of the results for phase
shifts on the functional form of the coordinate-space regulator in Eq.~(\ref{DefReg})
was demonstrated in Ref.~\cite{Epelbaum:2014efa} at N$^3$LO by employing different
exponents $n=5$ and $n=7$ and introducing an additional spectral
function regularization with the momentum cutoff in the range of
$\Lambda = 1$ to $2$~GeV.

Very recently, a new family of semilocal
momentum-space regularized (SMS) chiral 
NN potentials was introduced \cite{Reinert:2017usi}. In addition to
employing a momentum-space version of a local regulator for the
long-range interactions and using the $\pi N$ LECs from matching 
pion-nucleon Roy-Steiner equations to chiral perturbation theory
\cite{Hoferichter:2015tha} (see also 
Refs.~\cite{Krebs:2012yv,Alarcon:2012kn,Chen:2012nx,Wendt:2014lja,Yao:2016vbz,Siemens:2016hdi,Siemens:2016jwj,Siemens:2017opr,Hoferichter:2015hva,Hoferichter:2015tha} 
for related work on
the determination of the $\pi N$ LECs),
the SMS potentials of Ref.~\cite{Reinert:2017usi} differ from the SCS
ones of Ref.~\cite{Epelbaum:2014efa,Epelbaum:2014sza} in the  
determination of the NN contact interactions.
That is, the SMS potentials of Ref.~\cite{Reinert:2017usi} were fitted
directly to NN scattering data rather than to the Nijmegen partial wave
analysis \cite{NIJMI}. Another important difference concerns the
implementation of the contact interactions. In particular, 
the SMS potentials of Ref.~\cite{Reinert:2017usi} utilize a
specific choice for $3$ redundant N$^3$LO contact operators out of
$15$, 
which parametrize the unitary ambiguity in the
short-range part of the 
nuclear force at this chiral order. This is in contrast to the 
 potentials of
Refs.~\cite{Entem:2003ft,Epelbaum:2004fk,Epelbaum:2014efa,Epelbaum:2014sza,Entem:2017gor},
where all $15$
order-$Q^4$ contact interactions were fitted to Nijmegen PWA and/or NN
scattering data.    

Another important recent development is the establishment of a simple
algorithm for estimating the theoretical uncertainty from the
truncation of the chiral expansion \cite{Epelbaum:2014efa}. The new
method uses the available information on the chiral expansion of a
given observable to estimate the magnitude of neglected higher
order terms. To be specific, consider some few-nucleon observable $X(p)$ with $p$
being the corresponding momentum scale. The chiral expansion of $X$ up
to order $Q^n$ can be written in the form 
\begin{equation}  
X^{(n)} = X^{(0)} + \Delta X^{(2)} + \ldots + \Delta X^{(n)} \,,
\end{equation}
where we have defined
\begin{equation}
\Delta X^{(2)} \equiv X^{(2)} - X^{(0)}, \quad \quad  \Delta X^{(i)}
\equiv X^{(i)} - X^{(i-1)} \quad
\mbox{for} \quad i \ge 3\,.
\end{equation}
Assuming that the chiral expansion of the nuclear force
translates into a similar expansion of the observable, one expects  
\begin{equation}
\label{PCAssumption}
\Delta X^{(i)} = \mathcal{O} (Q^i X^{(0)})\,. 
\end{equation}
In \cite{Epelbaum:2014efa}, the size of truncated contributions at a given order $Q^i$
was then 
estimated via 
\begin{equation}
\label{ErrorOrig}
\delta X^{(0)} = Q^2 | X^{(0)}|, \quad \quad \delta X^{(i)} =
\max_{2 \le j \le i} \Big( Q^{i+1} | X^{(0)} |, \; Q^{i+1-j} | \Delta
X^{(j)} | \Big) \; \;  \mbox{for} \; \; i \ge 2\,, 
\end{equation}
subject to the additional constraint 
\begin{equation}
\label{ErrorOrig2}
\delta X^{(i)} \ge \max \Big( | X^{(j \ge i)} - X^{(k \ge i)} | \Big)\,,
\end{equation}
where the expansion parameter $Q$ was chosen as 
\begin{equation}
\label{ExpPar}
Q = \max \bigg( \frac{p}{\Lambda_b}, \; \frac{M_\pi}{\Lambda_b} \bigg)\,.
\end{equation} 
For the breakdown scale of the chiral expansion $\Lambda_b$, the
values of $\Lambda_b = 600$~MeV for $R=0.8,~0.9$ and $1.0$~fm, $\Lambda_b
= 500$~MeV for $R=1.1$~fm and $\Lambda_b= 400$~MeV for $R=1.2$~fm  were
adopted based on an analysis of error plots~\cite{Epelbaum:2014efa}. Smaller
values of the breakdown scale for softer cutoffs reflect an 
increasing amount of regulator artifacts. 

The algorithm for uncertainty quantification specified above allows one to
circumvent some of the drawbacks of the previous 
approach based on cutoff variation~\cite{Epelbaum:2004fk}, such as the relatively
narrow available range of cutoffs  and the fact that
residual regulator dependence shows the impact of neglected contact
interactions, which contribute only at even orders $Q^{2n}$ of the
chiral expansion; see \cite{Epelbaum:2014efa} for a comprehensive
discussion.  In addition, it provides an independent estimation of the theoretical 
uncertainty for any given cutoff value.
This novel algorithm was already successfully applied in the
two-nucleon sector. In particular,  the actual size of the N$^4$LO 
corrections to NN phase shifts and scattering observables was shown in 
Ref.~\cite{Epelbaum:2014sza} to be in a good agreement with the
estimated uncertainty at N$^3$LO \cite{Epelbaum:2014efa}. 
A statistical interpretation of the theoretical error bars is discussed in Refs.~\cite{Furnstahl:2015rha,Melendez:2017phj}. 

The theoretical developments outlined above open the way for 
understanding and validating the details of the many-body forces and
exchange currents that constitute an important frontier in nuclear
physics. First steps along these lines were taken in
Ref.~\cite{Binder:2015mbz} by employing the SCS NN potentials of
Refs.~\cite{Epelbaum:2014efa,Epelbaum:2014sza} along with the novel algorithm for uncertainty
quantification to analyze elastic nucleon-deuteron scattering and
selected observables in $^3$H, $^4$He and $^6$Li. In order to allow for a
meaningful quantification of truncation errors in incomplete
calculations based on NN interactions only, a slightly modified
procedure for estimating the uncertainty at N$^2$LO and higher orders
was adopted, by using for $i \ge 3$ 
\begin{equation}
 \label{ErrorMod}
 \delta X^{(i)} = \max \Big( Q^{i+1} |X^{(0)}|, \;  Q^{i-1} |\Delta
  X^{(2)}|, \;  Q^{i-2} |\Delta X^{(3)}|, \; Q \delta X^{(i-1)} \Big)  
\end{equation}
instead of Eqs.~(\ref{ErrorOrig}), (\ref{ErrorOrig2}); see
Ref.~\cite{Binder:2015mbz} for more details. For many considered
observables, the results at N$^2$LO and higher orders were then found to 
differ from experiment well outside the range of quantified
uncertainties, thus providing a clear evidence for missing three-nucleon
forces.\footnote{We remind the reader that nuclear forces are scheme
  dependent and not directly measurable. Our conclusions regarding the
  expected size of three-body contributions refer to the framework we
  employ.} Furthermore, the magnitude of the
deviations was found to be in agreement with the expected size of the 
chiral three-nucleon force (3NF) whose first contributions appear at
N$^2$LO. 

The same modified approach to error analysis was used recently in
Ref.~\cite{Skibinski:2016dve} to analyze the cross section and selected
polarization observables for deuteron photodisintegration,
nucleon-deuteron radiative capture and three-body $^3$He
photodisintegration and to study the muon capture rates in $^2$H
and $^3$He.  While these calculations are also incomplete as the 3NF was
not included and the axial (electromagnetic) currents were only taken
into account at the single-nucleon level (up to the two-nucleon level
via the Siegert theorem), most of the considered observables were
found to be in a good agreement with experimental data. 
For recent applications of the approaches to error analysis outlined
above to nuclear matter properties and muonic deuterium see
Refs.~\cite{Hu:2016nkw} and  \cite{Hernandez:2017mof}, respectively. 
These
promising results provide important tests of the novel chiral
potentials.

In this paper we focus on the  novel SCS chiral
potentials of Refs.~\cite{Epelbaum:2014efa,Epelbaum:2014sza} and 
extend our earlier work \cite{Binder:2015mbz} in
various directions. First, in addition to elastic
nucleon-deuteron scattering, we also study some of the most
interesting breakup observables. We present the first applications of
the SCS chiral NN potentials to light- and medium-mass nuclei
beyond $^6$Li using a variety of \emph{ab initio} methods and discuss
in detail the corresponding convergence pattern with respect to
truncations of the model space. Last but not least, we address the
limitations and robustness of the 
approach for uncertainty quantifications and consider some possible
alternatives.   

Our paper is organized as follows. Section \ref{sec:Nd} is devoted to
the nucleon-deuteron 
elastic and breakup scattering reactions. Our results for light nuclei
calculated by solving the Faddeev-Yakubovsky equations and/or using
the No-Core Configuration Interaction (NCCI) method are presented in section
\ref{sec:LightNuclei}, while those for medium-mass nuclei obtained
within the coupled-cluster (CC) 
method and in-medium similarity
renormalization group (IM-SRG) method are given in section
\ref{sec:CoupledCluster}. Next, in section \ref{sec:Uncertainties}, we
explore some alternative approaches for uncertainty quantification. 
Finally, the results of our work are summarized in section \ref{sec:Summary}.

\section{Nucleon-deuteron scattering}
\def\theequation{\arabic{section}.\arabic{equation}}
\label{sec:Nd}

\subsection{Faddeev calculations}
\label{nucl_ham}

Neutron-deuteron (nd) scattering with neutrons and protons interacting
via pairwise-interactions  is
described in terms of an amplitude $T\vert \phi \rangle $ satisfying the
Faddeev-type integral equation~\cite{wit88,glo96}
\begin{eqnarray}
T\vert \phi \rangle  &=& t P \vert \phi \rangle 
 + t P G_0 T \vert \phi \rangle .
\label{eq1a}
\end{eqnarray}
Here $t$ represents the two-nucleon $t$-matrix, which is the solution of the
Lippmann-Schwinger equation with a given NN interaction. 
 The permutation operator $P=P_{12}P_{23} +
P_{13}P_{23}$ is given in terms of the transposition operators,
$P_{ij}$, which interchange nucleons $i$ and $j$.  The incoming state $
\vert \phi \rangle = \vert \vec q_0 \rangle \vert \phi_d \rangle $
describes the free nd motion with relative momentum
$\vec q_0$ and the deuteron state $\vert \phi_d \rangle$.
Finally, $G_0$ is the resolvent of the three-body center-of-mass kinetic
energy. 
The amplitude for elastic scattering leading to the corresponding
two-body final state $\vert \phi ' \rangle$ is then given by~\cite{glo96,hub97}
\begin{eqnarray}
\langle \phi' \vert U \vert \phi \rangle &=& \langle \phi' 
\vert PG_0^{-1} \vert 
\phi \rangle + 
\langle \phi' \vert PT \vert \phi \rangle ,
\label{eq3}
\end{eqnarray}
while for the breakup reaction one has
\begin{eqnarray}
\langle  \phi_0'\vert U_0 \vert \phi \rangle &=&\langle 
 \phi_0'\vert  (1 + P)T\vert
 \phi \rangle ,
\label{eq3_br}
\end{eqnarray}
where $\vert \phi_0' \rangle$ is the free three-body breakup channel state. 
We refer to  \cite{glo96,hub97,book} for a general overview of
3N scattering and for details on the practical implementation of
the Faddeev equations. 

When solving the 3N Faddeev equation (\ref{eq1a}),  we  include the  
NN force components with a total two-nucleon angular momenta $j \le 5$ 
 in 3N partial-wave states with the total 3N system angular momentum
 below $J \le 25/2$. This is sufficient to get converged results for incoming 
neutron energies of $E_{\rm lab, \, n} \le 200$ MeV.

\subsection{Elastic nd scattering scattering}
\label{nd_elas}

\begin{figure}[tb]
\includegraphics[width=0.6\textwidth,keepaspectratio,angle=0,clip]{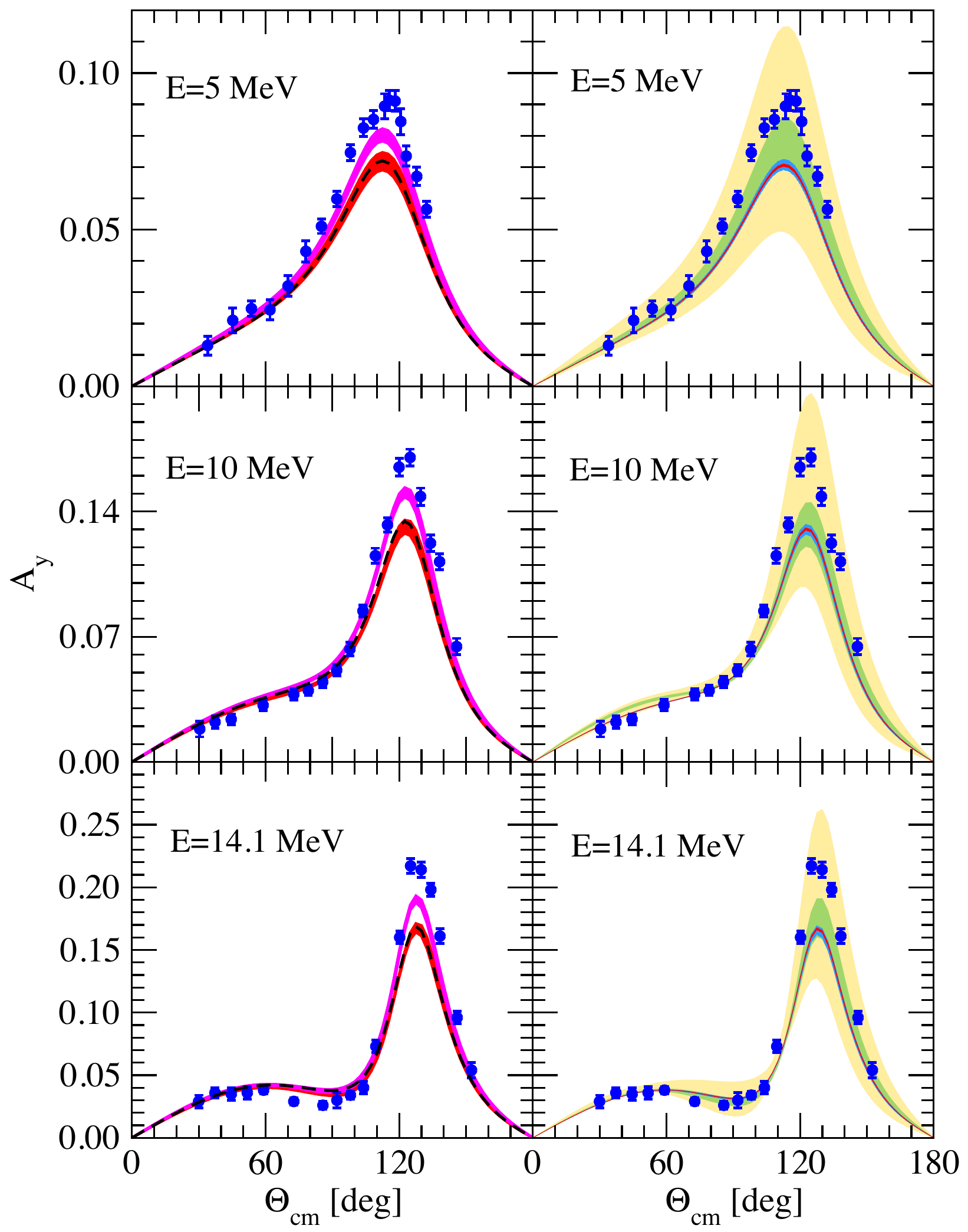}
\caption{(Color online) The nd elastic scattering analyzing power $A_y$  at
  $E_{\rm lab, \, n}=5$~MeV, $10$~MeV, and $14.1$~MeV.  
 In the left panels the bottom  (red) band covers 
  predictions of standard NN potentials: AV18, CD~Bonn, Nijm1 and
  Nijm2. The upper (magenta) band results when these potentials
  are combined with the TM99 3NF. The dashed (black) line shows prediction
  of the AV18+Urbana IX combination. 
 In the right panel, predictions based 
on the SCS chiral NN potentials of
Refs.~\cite{Epelbaum:2014efa,Epelbaum:2014sza} with the
coordinate-space cutoff parameter $R=0.9$~fm are shown. 
The bands of 
increasing width show estimated theoretical uncertainty at N$^4$LO (red), 
N$^3$LO (blue), N$^2$LO (green) and NLO (yellow).
 The filled circles are nd data from Ref.~\cite{tornow_ay} at
 $6.5$~MeV, from Ref.~\cite{tornow_ay_10} at $10$~MeV, and from 
Ref.~\cite{ay_14.1} at $14.1$~MeV.
}
\label{fig1}
\end{figure}

At low energies of the incoming neutron,
 the elastic nd scattering analyzing
 power $A_y$  with polarized neutrons has been a quantity of great interest
 because predictions using standard high-precision NN  
potentials (AV18~\cite{Wiringa:1994wb}, CDBonn~\cite{CDBOnucleon-nucleon}, 
 Nijm1 and Nijm2~\cite{NIJMI}) 
fail to explain the experimental data for $A_y$. The data are 
underestimated 
by $\sim 30 \%$ in the region of the $A_y$ maximum, which occurs 
 at c.m. angles $\Theta_{c.m.} \sim 125 ^o$. Combining standard NN 
potentials  
with commonly used models of a 3NF, such as the Tucson-Melbourne (TM99)~\cite{TM99} or 
Urbana IX~\cite{uIX} models, 
 removes approximately only half of the discrepancy (see left column in Fig.~\ref{fig1}).

\begin{figure}[tb]
\hskip -1.2 true cm
\includegraphics[scale=0.6]{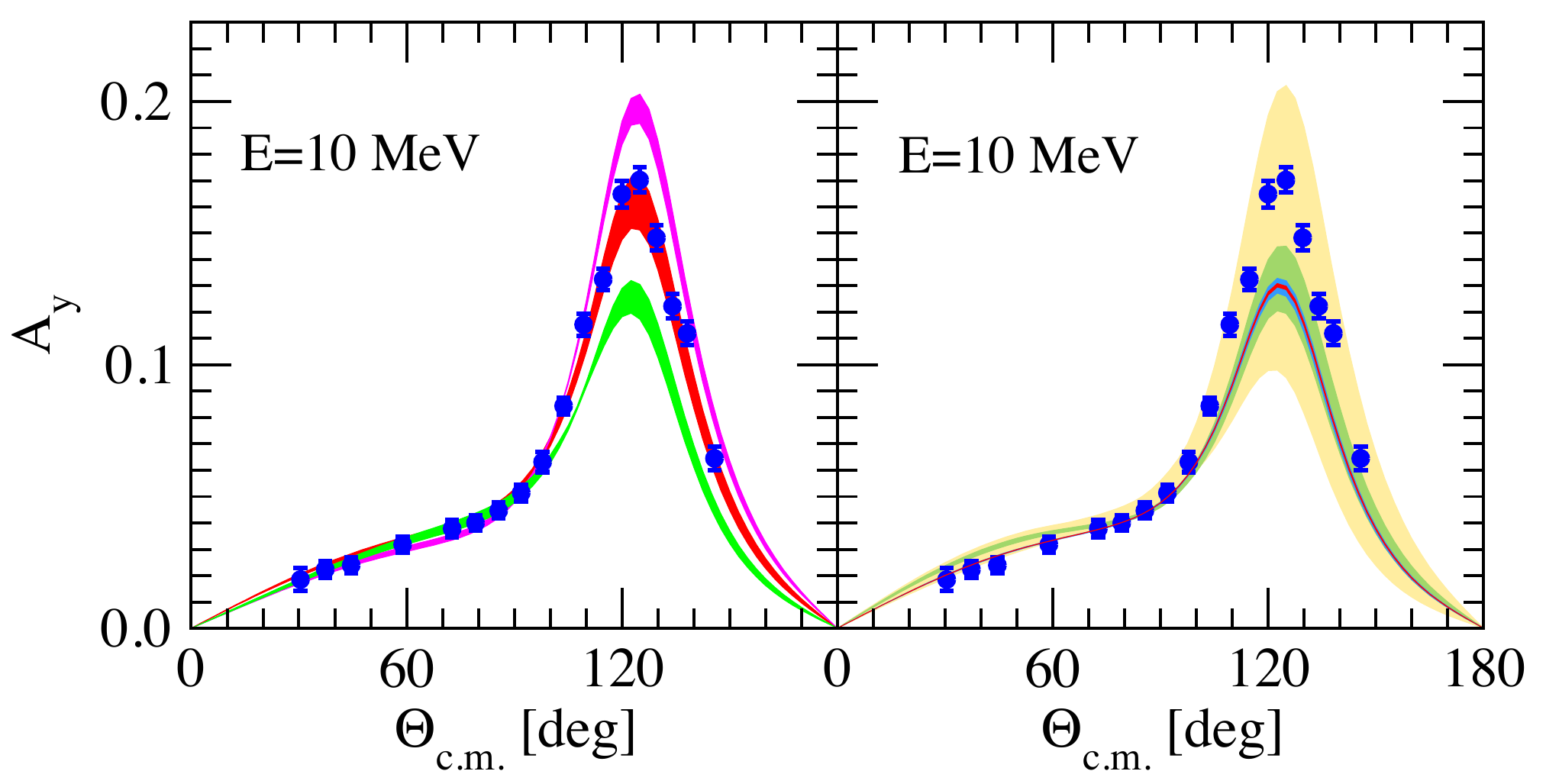}
\caption{
(Color online) The nd elastic scattering analyzing power $A_y$  at
  $E_{\rm lab, \, n}=10$~MeV. 
 In the left panel, bands of predictions for five versions of 
the old nonlocal chiral NN potentials of Refs.~\cite{Epelbaum:2004fk,epel_mod}
at different orders of the chiral expansion are
shown: NLO - the upper (magenta) band, 
 N$^2$LO - the middle (red) band, and N$^3$LO - the bottom 
 (green) band. These five versions correspond to different cutoff
 values used for the Lippmann-Schwinger equation 
 and the spectral function regularizations, namely $(450,500)$~MeV,
$(450,700)$~MeV, $(550,600)$~MeV, $(600,500)$~MeV, and
$(600,700)$~MeV.  
 In the right panel, predictions based 
on SCS chiral NN potentials of
Refs.~\cite{Epelbaum:2014efa,Epelbaum:2014sza} with the
coordinate-space cutoff parameter of $R=0.9$~fm are shown. 
The bands of 
increasing width show estimated theoretical uncertainty at N$^4$LO (red), 
N$^3$LO (blue), N$^2$LO (green) and NLO (yellow).
 The full circles are nd data from Ref.~\cite{tornow_ay_10} at $10$~MeV.
}
\label{fig2}
\end{figure}

\begin{figure}[tb]
\includegraphics[scale=0.7]{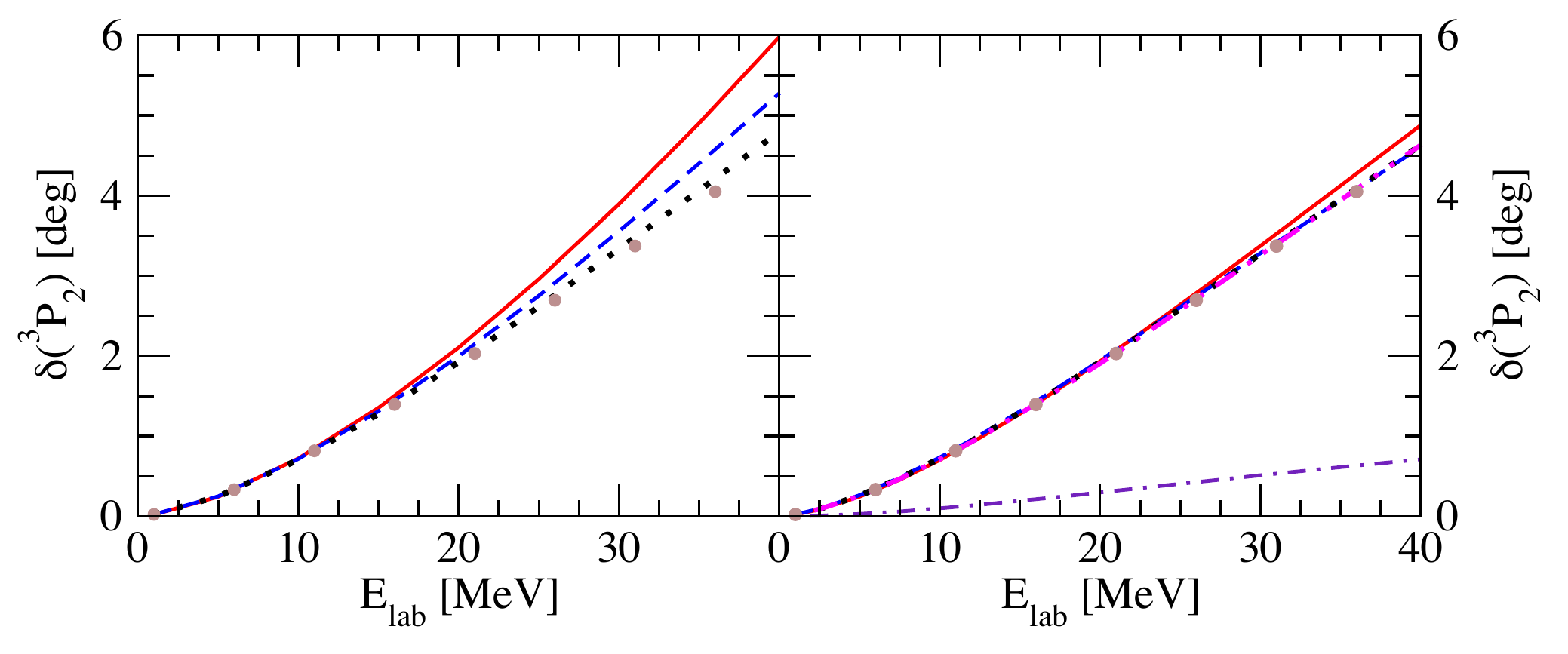}
\caption{(Color online)
The neutron-proton $^3P_j$ phase-shifts as 
 a function of laboratory energy $E_{\rm lab}$. 
In the left panel the solid (red), dashed (blue), and dotted (black) lines 
show predictions of 
the old chiral  Bochum NLO, N$^2$LO, and N$^3$LO NN potentials with
the cutoff parameters of $(600,500)$~MeV. 
In the right panel the dashed-dotted (indigo), solid (red), dashed (blue), 
dotted (black), and 
dashed-double-dotted (magenta) are predictions of SCS chiral 
potentials with local 
regulator and parameter $R=0.9$~fm at LO, NLO, N$^2$LO, N$^3$LO and N$^4$LO, 
respectively. 
The solid (brown) circles are experimental Nijmegen phase-shifts 
\cite{nijm_phase1,nijm_phase2}. 
}
\label{fig3}
\end{figure}

Using the old, nonlocally regularized chiral NN potentials of Refs.~\cite{Epelbaum:2004fk,epel_mod}, the 
predictions  for $A_y$ vary with the order of the chiral expansion. 
In particular, as reported in
Ref.~\cite{epel2002},  the NLO results overestimate the $A_y$ data 
while the N$^2$LO NN forces seem to be in quite a good 
agreement with experiment, see Fig.~\ref{fig2}.  
Only when the N$^3$LO NN chiral 
forces are used does 
a clear discrepancy between theory and data emerge in the region of $A_y$
maximum, which is similar to the one for standard forces. This is
visualized for $E_n=10$~MeV in the left panel of Fig.~\ref{fig2}, where 
bands of predictions correspond to five versions of 
the nonlocal NLO, N$^2$LO and N$^3$LO potentials of Refs.~\cite{Epelbaum:2004fk,epel_mod},
which differ from each other by the  
cutoff parameters used for the Lippmann-Schwinger equation 
and the spectral function
regularizations.  
Such a behavior of $A_y$ predictions at different orders in the chiral
expansion  can be traced back to a high sensitivity 
of $A_y$ to $^3P_j$ NN force components
\cite{nijm_phase1,nijm_phase2}, which are 
accurately 
reproduced for the old nonlocal chiral potentials 
only at N$^3$LO. This
is visualized in the left panel of Fig.~\ref{fig3}. 
Contrary to the observed behavior of old potentials, 
the predictions for  $A_y$ based on the SCS NN chiral forces
turn out to be similar to those of the high-precision  phenomenological
potentials already starting from NLO; see the right panel of
Fig.~\ref{fig2}. 
This reflects the considerably  improved convergence with the order of
the chiral 
expansion of the novel semilocal potentials, as visualized in the
right panel of Fig.~\ref{fig3} for the case of the $^3P_2$ phase shift. 
Only LO values are far away from the empirical  values while the NLO
results already turn out to be very close to those of the Nijmegen
partial wave analysis (NPWA)
at energies below $\approx 40$~MeV. The N$^2$LO, N$^3$LO, and  N$^4$LO
results for the phase shifts overlap with each other and  
with the NPWA values. The corresponding 
$A_y$ predictions at orders above LO are very close to each other 
as seen in the right panels of Figs.~\ref{fig1} and ~\ref{fig2}. 

It is instructive to look at the estimated theoretical uncertainty from the
truncation of the chiral expansion shown in the right panels of
Figs.~\ref{fig1} and \ref{fig2}. Notice that our calculations for
three- and more-nucleon observables are incomplete starting from
N$^2$LO due to the missing 3NFs. The width of the bands calculated
using Eqs.~(\ref{ErrorOrig}), (\ref{ErrorOrig2}) at LO and NLO and
using Eq.~(\ref{ErrorMod}) starting from N$^2$LO show our
estimations of the expected theoretical uncertainties \emph{after}
inclusion of the corresponding 3NF contributions. 
At the considered low energies, the theoretical uncertainty decreases
quite rapidly so that one expects precise predictions for $A_y$
starting from N$^3$LO.\footnote{We emphasize, however, that the usage
  of Eq.~(\ref{ErrorMod}) in the incomplete calculations presented
  here may lead to underestimation of the theoretical uncertainty at higher
  orders. A more reliable estimation of the truncation error is
  expected from
  performing complete calculations that include 3NFs and using
  Eqs.~(\ref{ErrorOrig}, \ref{ErrorOrig2}) at all orders. This work
  is in progress.}  Interestingly, our novel approach to uncertainty
quantification is capable of accounting for the already mentioned 
strongly fine-tuned nature of this observable which results in a large
theoretical uncertainty at NLO. Notice that the experimental data are correctly 
described at this order within the errors.
It remains to be seen upon the inclusion of the 3NF
and performing complete calculations whether the $A_y$-puzzle will
survive at higher orders of the chiral expansion. Notice further that
at N$^4$LO, the 3NF involves purely short-range contributions with two
derivatives, which affect nucleon-deuteron (Nd) P-waves \cite{Girlanda:2011fh}. It is conceivable
that the inclusion of such terms will lead to a proper description
of $A_y$ once the corresponding LECs are tuned to reproduce
Nd scattering observables.  

\begin{figure}[tb]
\includegraphics[width=\textwidth,keepaspectratio,angle=0,clip]{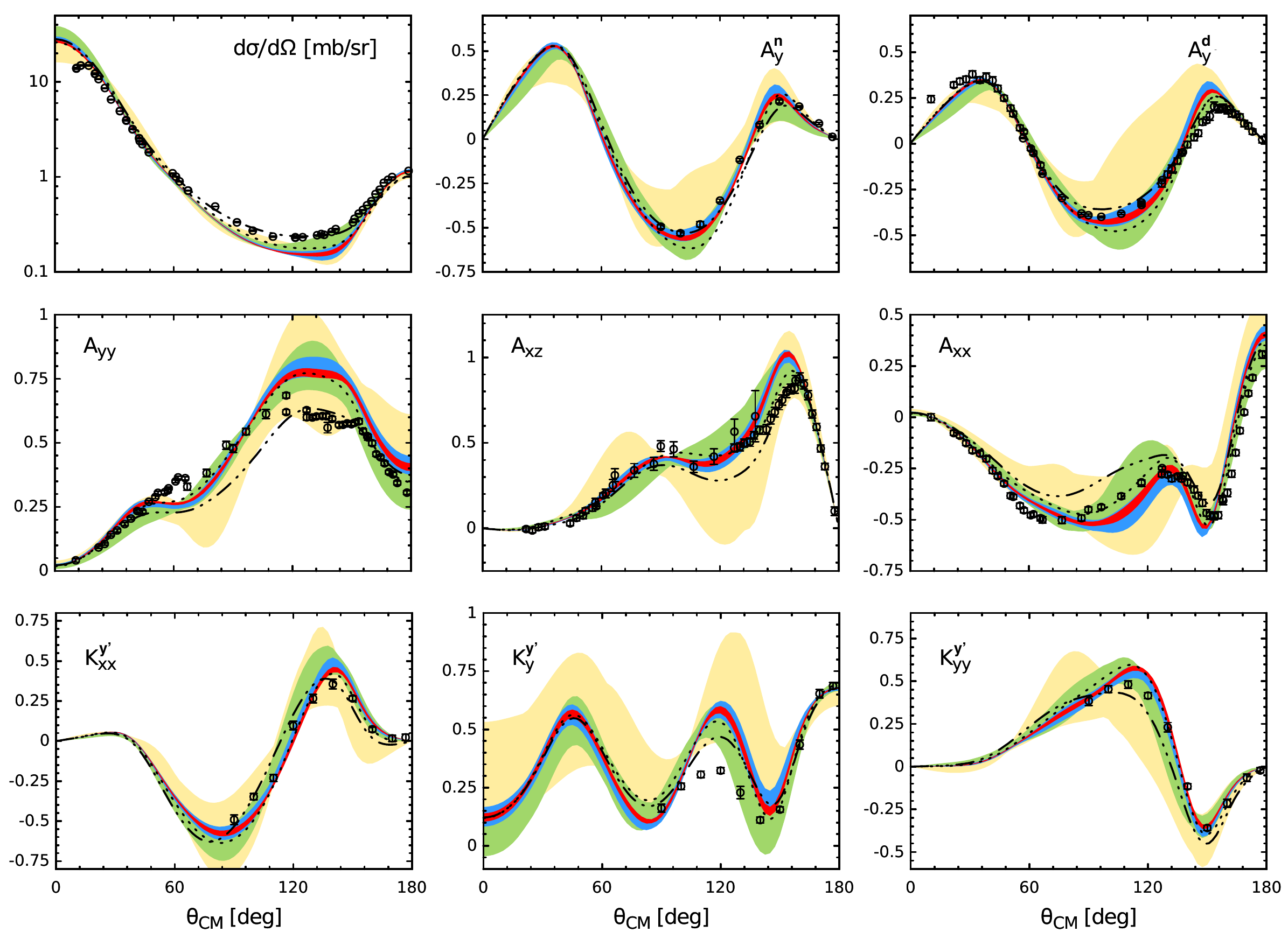}
\caption{(Color online) Predictions for the differential cross
  section, nucleon and deuteron vector analyzing powers $A_y^{n}$
  and  $A_y^{d}$, deuteron tensor analyzing powers $A_{yy}$,  $A_{xz}$ and
  $A_{xx}$ and polarization-transfer coefficients $K_{xx}^{y'}$,  $K_{y}^{y'}$, 
and  $K_{yy}^{y'}$ at the laboratory energy of $135$~MeV based on the
NN potentials of Refs.~\cite{Epelbaum:2014efa,Epelbaum:2014sza} for 
$R=0.9$~fm without including the 3NF.  
The bands of 
increasing width show estimated theoretical uncertainty at N$^4$LO (red), 
N$^3$LO (blue), N$^2$LO (green) and NLO (yellow). The dotted (dashed)
lines show the results based on the CD Bonn NN potential (CD Bonn NN
potential in combination with the Tucson-Melbourne 3NF).
Open circles are proton-deuteron data from
Refs.~\cite{Sakamoto:1996xdz,Sakai:2000mm,Sekiguchi:2002sf,Sekiguchi:2004yb}.  
}
\label{fig3a}
\end{figure}

Apart from $A_y$ and the deuteron tensor analyzing power $i
T_{11}$, which is known to show a similar behavior to $A_y$, 
there is not much room for three-nucleon force effects in
elastic Nd scattering at low energies; see Ref.~\cite{Binder:2015mbz}  for the
predictions of other observables at $10$~MeV.  On the other hand,
significant disagreements with the data start to appear at
intermediate energies of $\sim 50$~MeV and higher. 
As a representative example, we
show in Fig.~\ref{fig3a} our predictions for selected elastic scattering
observables at $135$~MeV. 
In addition to
the well-known underestimation of the differential cross section
minima, the spin-observables calculated using the NN interactions
only start to show deviations from the data.  These deviations tend to increase
with energy; see \cite{KalantarNayestanaki:2011wz} for a comprehensive
discussion. As shown in Ref.~\cite{Binder:2015mbz}, the 
theoretical uncertainty of the chiral EFT results at N$^3$LO and N$^4$LO
is considerably smaller than the observed disagreements between the
predictions based on the NN forces and the experimental data even at
energies of the order of $200$~MeV. Our results suggest that elastic Nd
scattering in the energy range of $\sim 50-200$~MeV is very well
suited to study the detailed structure  of the chiral 3NF.

\subsection{Nd breakup}
\label{nd_breakup}

Among numerous kinematically complete configurations of the Nd breakup 
reaction the so-called symmetric space star (SST) and quasi-free
scattering (QFS) configurations have attracted special
attention. 
 The cross sections for these geometries 
are very stable with respect to the underlying dynamics.
To be specific, different phenomenological potentials, alone or combined
with standard 3NFs, lead to very similar results for the 
cross sections \cite{din1} which deviate significantly from the available
 SST and neutron-neutron (nn) QFS data.  
 At low energies,  the cross sections in the SST and QFS configurations are
 dominated by the S-waves. For the SST configuration, the largest
contribution to the cross section comes from the $^3S_1$ partial 
 wave, while for the nn QFS
 the $^1S_0$ partial wave dominates.
Neglecting rescattering, the QFS configuration resembles free NN
scattering. 
For elastic low-energy neutron-proton (np) scattering one expects
contributions from the $^1S_0$ np and $^3S_1$ force components. For elastic nn
scattering, only the $^1S_0$ nn channel is allowed by the Pauli principle. This suggests that
the nn QFS
is a powerful tool to study the nn interaction.
The measurements of np QFS cross sections  have revealed a good agreement
 between the data and theory \cite{nnqfs1}, thus confirming the knowledge of
 the np force.
On the other hand, for the nn QFS, it was found that the theory underestimates the data by
$\sim 20\%$ \cite{nnqfs1,nnqfs2}. The 
stability of the QFS cross sections
 with respect to the underlying dynamics means that, assuming
 correctness of the nn QFS data,  the present day
 $^1S_0$ nn interaction is probably incorrect \cite{din1,din2,din3}. 

\begin{figure}[tb]
\includegraphics[scale=0.65]{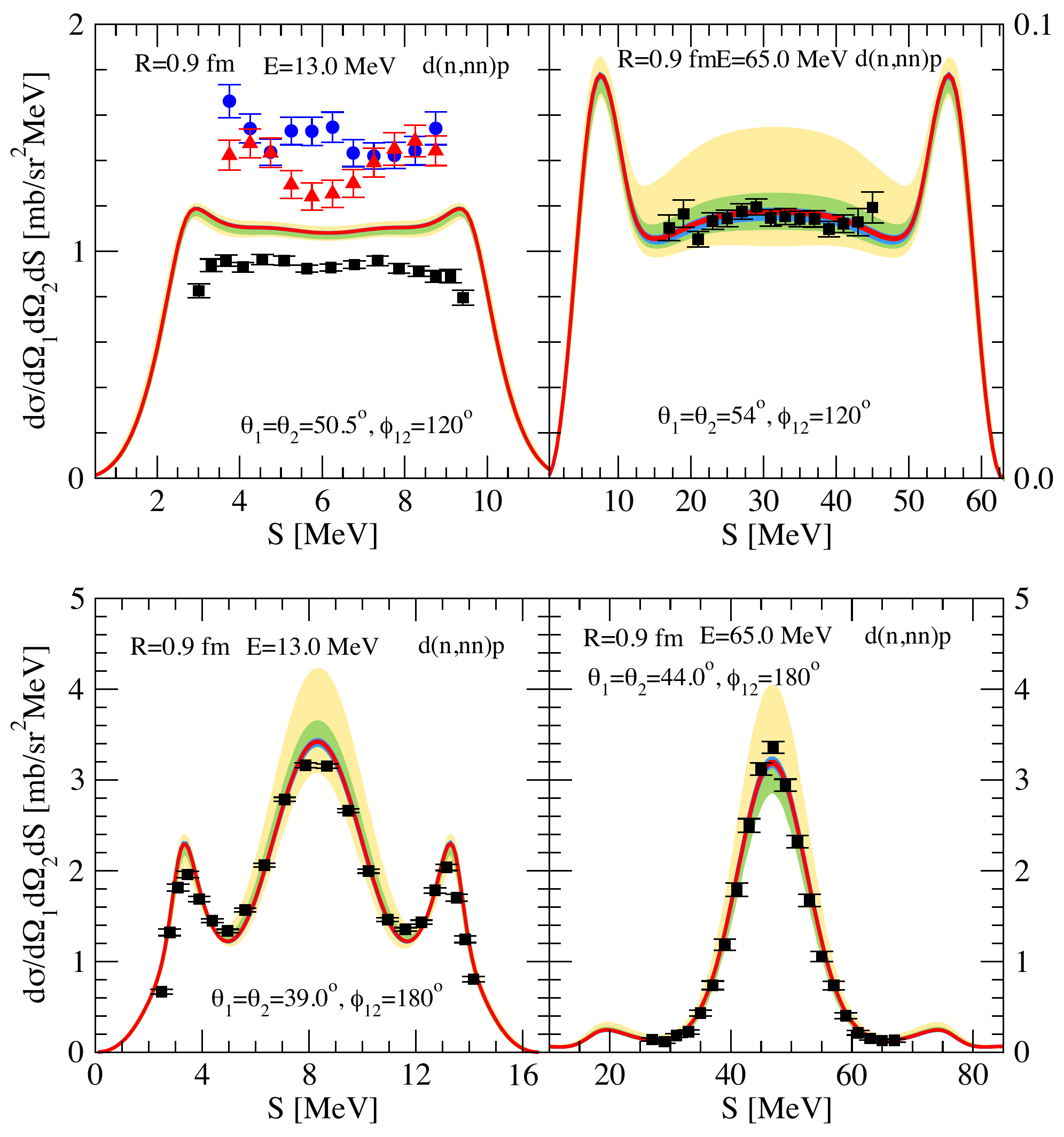}
\caption{(Color online) The SST (upper panel) and QFS (lower panel) nd breakup cross
  sections at incoming neutron laboratory energy $E_{\rm lab, \, n}=13$~MeV (left graphs) 
 and $65$~MeV (right graphs), as a
  function of the arc-length S along the kinematical locus in the
  $E_1-E_2$ plane.  
 The bands of 
increasing width show estimated theoretical uncertainty at N$^4$LO (red), 
N$^3$LO (blue), N$^2$LO (green) and NLO (yellow) based 
on SCS chiral NN potentials of Refs.~\cite{Epelbaum:2014efa,Epelbaum:2014sza} with  local 
regulator and parameter $R=0.9$~fm.
(Blue) circles and (red) triangles are nd data from
Ref.~\cite{sst} and \cite{erlangen13a,erlangen13b}, respectively. 
Proton-deuteron experimental data are shown as (black) squares. At $13$~MeV, 
the pd data are from Ref.~\cite{koln13}. 
At $65$~MeV, the pd data are from
Ref.~\cite{psi65sst} for the SST and from  Ref.~\cite{psi65qfs}
for the QFS configurations. 
}
\label{fig4}
\end{figure}

In the upper panel of Fig.~\ref{fig4}, we compare predictions of the SCS chiral potentials 
at different orders to the SST cross section data at two incoming nucleon 
energies $E=13$~MeV and $65$~MeV. 
At $65$~MeV the theoretical uncertainty is large at NLO but decreases
rapidly at higher orders of the chiral expansion. One expects accurate
predictions at N$^3$LO  and N$^4$LO. Given the good agreement with the
experimental data of Ref.~\cite{psi65sst} as visualized in the right panel of Fig.~\ref{fig4},
there is not much room for 3NF effects for this observable.  
At $13$~MeV, the uncertainty bands are rather narrow at all considered
orders, but the nd and proton-deuteron (pd) breakup data are far away from the theory. The two nd data 
sets are from different measurements and both show a significant
disagreement with our theoretical results, even though the data seem
to be inconsistent with each other 
for the values of the kinematical locus variable $S$ in the range
of $S=5 \ldots 7$ MeV.  

The pd data set shown in the upper-left panel of Fig.~\ref{fig4} 
is supported by 
other SST pd breakup measurements~\cite{sagarasst} in a similar energy range. The 
calculations of the pd breakup with inclusion of the pp Coulomb 
force \cite{deltuvabr}  
revealed only very small Coulomb force effects for this configuration. 
Since, at that energy, the SST configuration is practically dominated by
the S-wave NN force components, the big difference between pd and nd
data seems to indicate 
a large charge-symmetry breaking in the $^1S_0$ NN partial wave. We
anticipate it to be very difficult to explain the large difference
between the nd and pd data sets by the inclusion of a 3NF without
introducing large charge symmetry breaking interactions. 
Furthermore, the discrepancy between the theory and experimental pd
data is puzzling.  It remains to be seen whether the inclusion of the
chiral 3NF will affect the results for the pd SST configuration at
this energy.

\begin{figure}[tb]
\includegraphics[scale=0.7]{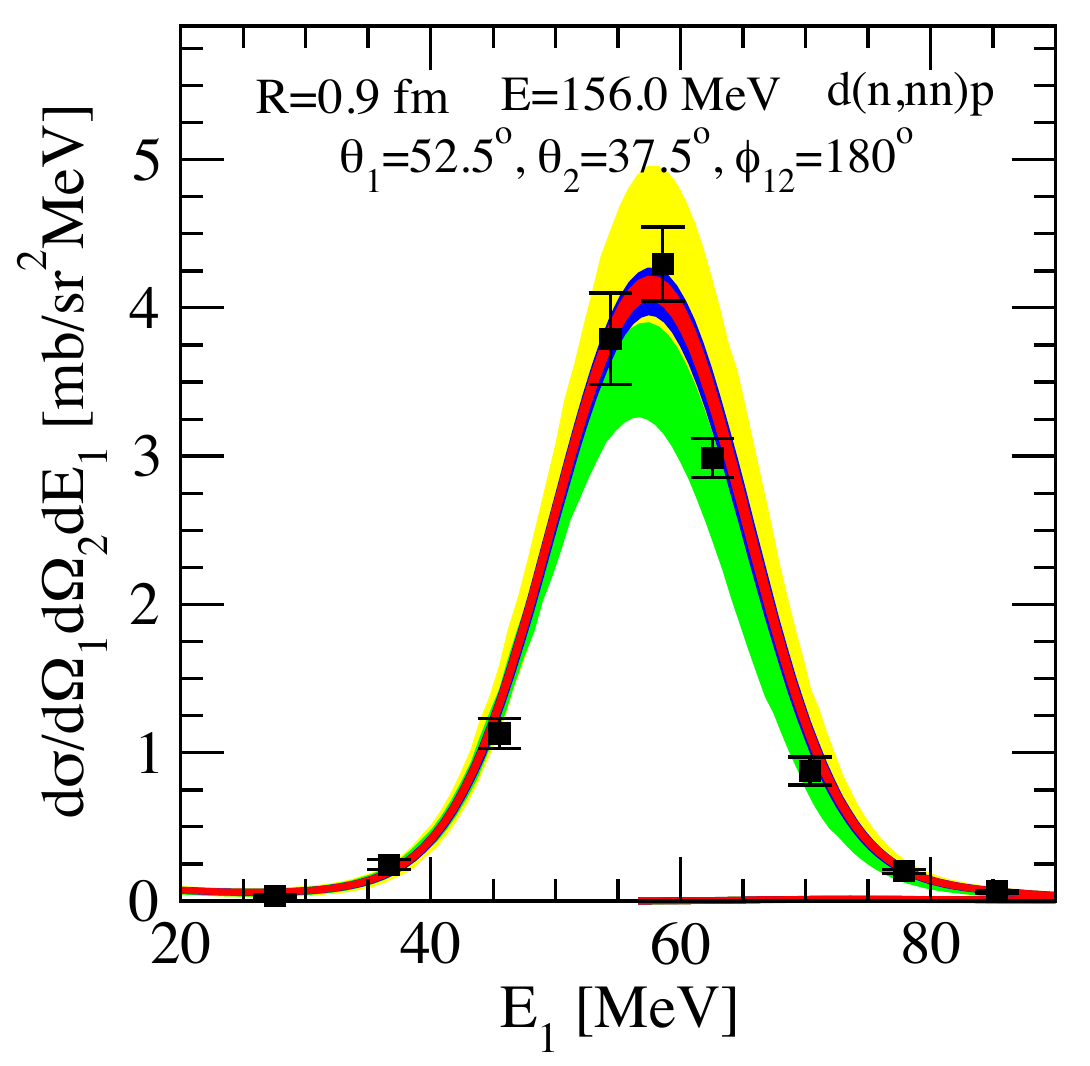}
\caption{(Color online) The pp QFS pd breakup cross
  section at incoming neutron laboratory energy $E_{\rm lab, \, n}=156$~MeV 
 as a  function of the energy $E_1$ in the
  $E_1-E_2$ plane.  
 The bands of 
increasing width show estimated theoretical uncertainty at N$^4$LO (red), 
N$^3$LO (blue), N$^2$LO (green) and NLO (yellow) based 
on the SCS chiral NN potentials of Refs.~\cite{Epelbaum:2014efa,Epelbaum:2014sza} with  local 
regulator and parameter $R=0.9$~fm.
 The (black) squares are pd data 
of Ref.~\cite{pd156}.
}
\label{fig5}
\end{figure}

For the pp QFS geometry, we show in the lower panel of Fig.~\ref{fig4}
and in Fig.~\ref{fig5} the predictions based on the 
SCS chiral potentials at $E=13$~MeV, $65$~MeV, and $156$~MeV,  
together with the available pd breakup data. Again the theoretical 
uncertainty rapidly decreases with an increasing order  
of the chiral  expansion, leading to very precise predictions 
at N$^3$LO and N$^4$LO, which, in addition, agree well 
with the pd breakup data. Assuming that the agreement will hold after
the inclusion of the corresponding 3NF, this provides,  together with 
the drastic $\approx 20 \%$ underestimation of nn QFS data 
found in \cite{nnqfs1,nnqfs2}, yet another indication of our poor 
knowledge of low energy $^1S_0$ neutron-neutron force. 

\begin{figure}[tb]
\includegraphics[scale=0.65]{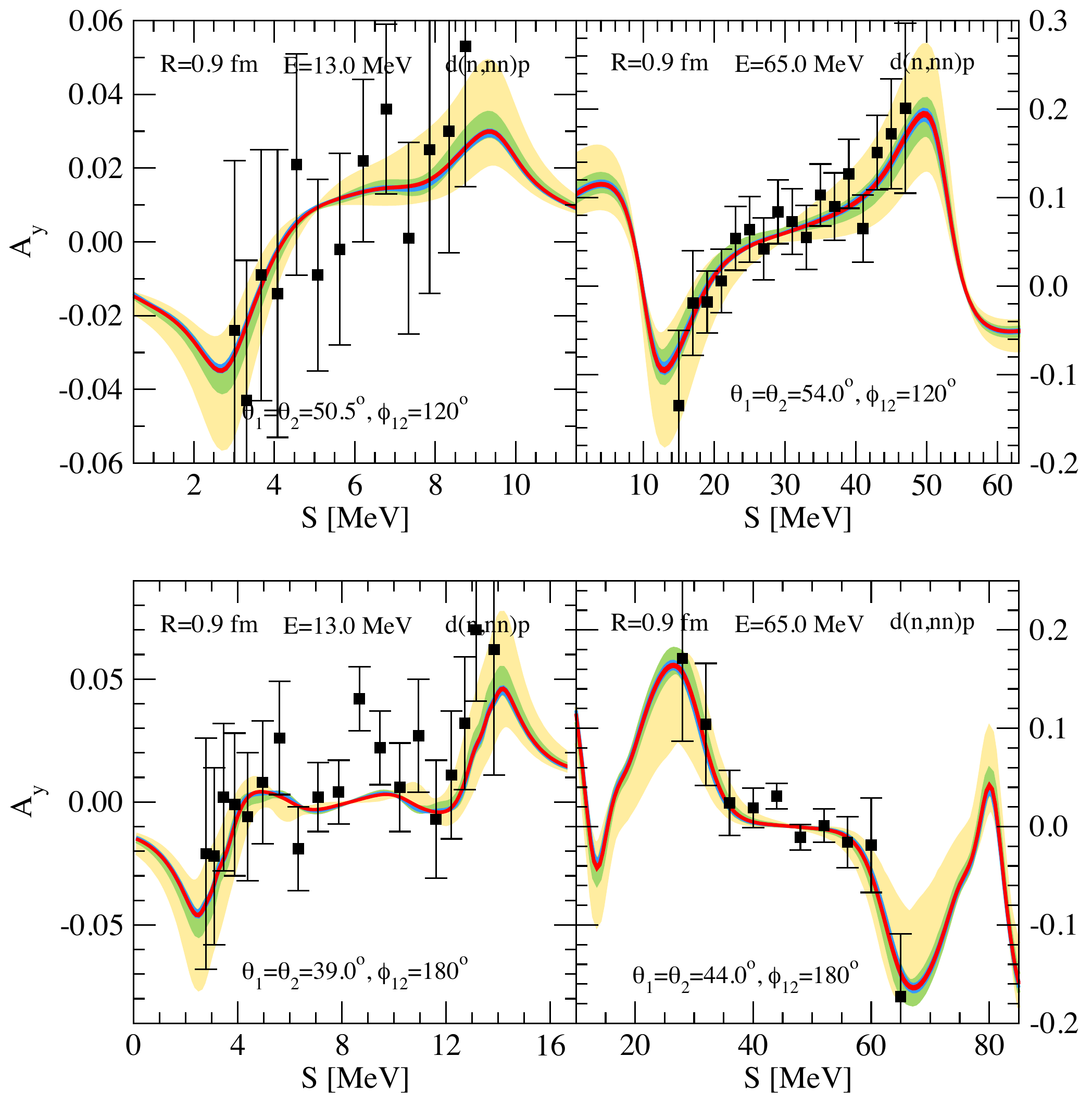}
\caption{(Color online) The SST pd analyzing power 
 at incoming neutron laboratory energy $E_{\rm lab, \, n}=13$~MeV (left panel) 
 and $65$~MeV (right panel), as a
  function of the arc-length S along the kinematical locus in the
  $E_1-E_2$ plane.  
 The bands of 
increasing width show estimated theoretical uncertainty at N$^4$LO (red), 
N$^3$LO (blue), N$^2$LO (green) and NLO (yellow) based 
on the SCS chiral NN potentials of Refs.~\cite{Epelbaum:2014efa,Epelbaum:2014sza} with  local 
regulator and parameter $R=0.9$~fm.
 In the left panel the (black) squares are pd data 
 of Ref.~\cite{koln13}. 
In the right panel the (black) squares are pd data of 
 Ref.~\cite{psi65sst}.
}
\label{fig6}
\end{figure}

Finally, in Fig.~\ref{fig6},  the results for the nucleon analyzing power 
$A_y$ in the SST and pp QFS geometries of the Nd 
 breakup reaction at $13$~MeV and $65$~MeV 
are presented. Again, the band widths 
  of theoretical uncertainties become quite narrow with an  
increasing order of chiral expansion. There 
appears to be reasonable agreement between experiment 
and theory without 3NF contributions given
the large error bars of the available data.

\section{Light nuclei}
\def\theequation{\arabic{section}.\arabic{equation}}
\label{sec:LightNuclei}

\subsection{Faddeev--Yakubovsky calculations}
For the $A=3$ and $A=4$ bound state calculations, we solve Faddeev and Yakubovsky equations, respectively. These 
calculations are performed in momentum space, which enables us to obtain high accuracies for binding energies and 
also for the properties of the wave function. 

For $A=3$, similar to the 3N scattering case, we rewrite the Schr\"odinger equation
into Faddeev equations 
\begin{equation}
| \psi \rangle = G_0 \, t \, P \, | \psi \rangle  \;. 
\end{equation}
Here, $|\psi \rangle$ denotes the Faddeev component. The 3N wave function is related to the Faddeev 
component by $| \Psi \rangle = (1+P)\,  | \psi \rangle$.  In contrast
to the 3N scattering problem, no singularities show up 
for bound states since the energy is negative and below the binding energy of the deuteron.  

We represent the equation using partial wave decomposed momentum eigenstates 
\begin{equation}
| \, p_{12} \, p_3 \, \alpha \, \rangle = | \, p_{12} \, p_3 \,   [
(l_{12} s_{12}) j_{12}  (l_3 \frac{1}{2}) I_3 ] J_3 \ (t_{12}
\frac{1}{2} ) T_3\rangle  
  \;,
\end{equation}
where $p_{12}$, $l_{12}$, $s_{12}$ , $t_{12}$ and $j_{12}$ are the
magnitude of the momentum, the orbital angular momentum, the spin, the
isospin   
and the total angular momentum of the subsystem of nucleons 1 and 2. 
$p_3$,  $l_3$ and $I_3$ denote the magnitude of the momentum, the
orbital angular momentum and the total angular momentum  
of the spectator nucleon relative to the (12) subsystem,
respectively. 
The angular momenta and isospin quantum numbers are coupled together
with the spin and isospin $\frac{1}{2}$  
of the third nucleon to the total angular momentum $J_3=\frac{1}{2}$
and isospin $T_3$. For the results shown below, we take angular
momenta  
up through $j_{12}=7$ and $T_3=\frac{1}{2}$ and $\frac{3}{2}$ states into account.  
We adopt $N_{12}=N_3=64$ mesh points 
for the discretization of the momenta between 0 and
$p_{\max}=15$~fm$^{-1}$. We note that the solution of the
Lippmann-Schwinger equation  
for $t$ requires a more extended momentum grid up to momenta of
$35$~fm$^{-1}$. We find that this choice of momenta guarantees that
our numerical  
accuracy is better than 1~keV for the binding energy and, for the 3N
systems, also the expectation values of the Hamiltonian $H$. The latter ones require
the calculation of  
wave functions and are therefore more difficult to obtain. 
We take isospin breaking of the nuclear interaction 
 into account.
For the SCS chiral interactions, we  add the point Coulomb 
 interaction in $pp$. The contribution 
of the neutron/proton mass difference is later treated perturbatively and given in Table~\ref{tab:3n-exp}.
For the calculation of the binding energies and wave functions, we use an 
averaged mass of $m_N=938.918$~MeV. More details on the computational aspects 
can be found in \cite{Nogga:2001cz}. Results for the binding energies
are summarized in Table~\ref{tab:3n-faddeev}. 

\begin{table}[t]
  \renewcommand{\arraystretch}{1.2}
  \begin{ruledtabular}
    \begin{tabular}{l|lllll}
 \multicolumn{1}{l}{$R$ [fm]} &
      \multicolumn{1}{c}{LO}  & 
      \multicolumn{1}{c}{NLO} &
      \multicolumn{1}{c}{N$^2$LO} &
      \multicolumn{1}{c}{N$^3$LO} &
      \multicolumn{1}{c}{N$^4$LO} 
      \\ \hline
      \multicolumn{6}{l}{$^3$H}
      \\ \hline
      0.8   &  $-$12.038  &   $-$8.044  &   $-$8.039 &   $-$7.569 &   $-$7.489 \\    
      0.9 &  $-$11.747 &   $-$8.216 &   $-$8.146 &   $-$7.575 &   $-$7.600 \\            
      1.0 &  $-$11.295  &   $-$8.380  &   $-$8.282  &   $-$7.534  &   $-$7.642 \\    
        1.1 &  $-$10.822  &   $-$8.554  &   $-$8.428  &   $-$7.514  &   $-$7.630 \\
      1.2 &  $-$10.394  &   $-$8.727 &   $-$8.579 &   $-$7.481  &   $-$7.580 \\   
      \hline
      \multicolumn{6}{l}{$^3$He}
      \\ \hline    
       0.8 &  $-$11.151  &   $-$7.312  &   $-$7.303  &   $-$6.867  &   $-$6.794 \\
       0.9 &  $-$10.862  &   $-$7.472  &   $-$7.402  &   $-$6.875  &   $-$6.897 \\
       1.0 &  $-$10.423  &   $-$7.624  &   $-$7.528  &   $-$6.837  &   $-$6.935 \\
       1.1 &  $-$9.968  &  $-$7.786  &  $-$7.664  &  $-$6.816  &  $-$6.923 \\
       1.2 &  $-$9.561  &  $-$7.948  &  $-$7.806  &  $-$6.783  &  $-$6.876 \\
    \end{tabular}  
  \end{ruledtabular} 
  \caption{\label{tab:3n-faddeev}
    Calculated $^3$H and $^3$He binding energies  
    using chiral NN interactions at different orders of the chiral
    expansion and at five different values of $R$. Energies are given
    in MeV.} 
\end{table}

In order to provide benchmark results, we also summarize expectation
values for the kinetic energy, the potential, the point proton and
neutron  
rms radii and probabilities for $S$-, $P$- and $D$-states in Table~\ref{tab:3n-exp}. 
Here, we restrict ourselves to N$^4$LO and $^3$H and compare to
results of two phenomenological interactions, AV18 and CDBonn, and  
to  ones based on the older series of chiral interactions of
Ref.~\cite{Epelbaum:2004fk} ($\Lambda,\tilde \Lambda=$(600, 700) MeV)
and the chiral interaction of Ref~\cite{Entem:2003ft}. 
Note that we have used, for the calculations with these forces, the EM
interaction of AV18 \cite{Wiringa:1994wb} acting in $pp$ and $nn$ in
order to be consistent  
with previous calculations.
The deviation of the binding energy $E$ and expectation value 
$\langle H \rangle$ of the Hamiltonian is due to the
contribution of the mass difference of the proton and neutron to the
kinetic  
energy $\langle T_{CSB} \rangle$, which we take into account for the
calculation  
of  $\langle H \rangle$ but which we do not consider for the solution
of the Faddeev equations.  
We checked that results for $^3$He are close to the results for $^3$H
except that the sign of the contribution of the  
proton/neutron mass difference is opposite and proton and neutron
radii are interchanged, as expected for mirror nuclei.  
Because the convergence with respect to partial waves of the Faddeev
component is faster, it is advantageous to project on  
Faddeev components whenever possible. Therefore, the wave function and
Faddeev component are normalized to $3 \langle \Psi | \psi \rangle=
1$.  
The results  for the norm $\langle \Psi | \Psi \rangle$ show that our
representation of the wave function includes 99.9\% of  
the wave function. Nevertheless, we evaluate the kinetic energy using
again the faster convergence  
for overlaps of Faddeev component and wave function by $\langle T
\rangle = 3 \langle \Psi | \, T \, | \psi \rangle$. 
A similar trick for the potential operator is not possible and not
necessary since the potential operator suppresses contributions of  
high angular momenta due to its finite range. Note that our choice of
normalization ensures that the relevant partial waves are properly  
normalized and, therefore, the calculation of the expectation values
does not require a division by the norm of the wave function.  
 
Comparing the new results to those from non-local chiral interactions of Ref.~\cite{Epelbaum:2004fk}, 
the kinetic energies tend to be larger now. 
But they only 
become comparable to a standard local phenomenological interaction
like AV18 for smaller configuration-space regulators $R$.   
For larger $R$, the expectation values are in better agreement with non-local interactions like CDBonn. 
Generally, the observed pattern indicates 
that the new interactions induce more NN short-range correlations than the chiral
interactions of Refs.~\cite{Epelbaum:2004fk,Entem:2003ft} 
but, at least for larger $R$, still less than 
phenomenological ones. 
Notice that the kinetic energies at
N$^3$LO, which are not shown explicitly, are found to take similar
values, while those at NLO and N$^2$LO appear to be
significantly smaller.  These findings are in line with the
nonperturbative nature of the SCS potentials at N$^3$LO and N$^4$LO
as found in the Weinberg eigenvalue analysis of
Ref.~\cite{Hoppe:2017lok}. As demonstrated in Ref.~\cite{Reinert:2017usi}, this feature can
be traced back to the large values of the LECs accompanying the
redundant N$^3$LO contact interactions in the $^1$S$_0$ and
$^3$S$_1$-$^3$D$_1$ channels.   

The contributions of the $D$-wave  component of the wave function is of the order of 6-7\%, which is comparable 
to the non-local and older chiral interactions but smaller than results for AV18. We note that the $D$-state 
probability increases with increasing $R$. This is a feature of the higher-order interactions  at N$^3$LO
and N$^4$LO. The lower order interactions show the opposite behavior. 
We found that 
the proton and neutron rms radii are not strongly affected by the
regulator $R$. This is somewhat surprising given that the  
kinetic energies are strongly dependent on $R$.  It is also
interesting that the radii do not appear to be strictly correlated to the binding energies. 

\begin{table}[t]
  \renewcommand{\arraystretch}{1.2}
  \begin{ruledtabular}
    \begin{tabular}{l|lllllllllll}
 \multicolumn{1}{c}{$R$ [fm]} &
      \multicolumn{1}{c}{$E$}  & 
      \multicolumn{1}{c}{$\langle H \rangle$} &
      \multicolumn{1}{c}{$\langle T \rangle$} &
      \multicolumn{1}{c}{$\langle V \rangle$} &
       \multicolumn{1}{c}{$\langle T_{CSB} \rangle$} & 
             \multicolumn{1}{c}{$\langle \Psi | \Psi \rangle $} &       
       \multicolumn{1}{c}{$P(S) $} &
       \multicolumn{1}{c}{$P(P) $} &
     \multicolumn{1}{c}{$P(D) $} & 
    \multicolumn{1}{c}{$r_p $} &  
      \multicolumn{1}{c}{$r_n$}  
      \\ \hline
0.8 &  $-$7.489 &    $-$7.499 &   53.59 &    $-$61.08 &     $-$9.48 &  0.9989 &    93.95 &   0.033 &     6.02 &    1.675 &    1.849 \\
0.9 &  $-$7.600 &    $-$7.608 &   48.45 &    $-$56.05 &     $-$8.45 &  0.9993 &    93.91 &   0.034 &     6.06 &    1.669 &    1.838 \\
1.0 &  $-$7.642 &    $-$7.649 &   44.30 &    $-$51.94 &     $-$7.70 &  0.9995 &    93.70 &   0.035 &     6.27 &    1.670 &    1.838 \\
1.1 &  $-$7.630 &    $-$7.637 &   40.74 &    $-$48.37 &     $-$7.08 &  0.9996 &    93.16 &   0.040 &     6.80 &    1.678 &    1.845 \\
1.2 & $-$7.580 &    $-$7.587 &   37.57 &    $-$45.15 &     $-$6.52 &  0.9998 &    92.58 &   0.046 &     7.37 &    1.689 &    1.858 \\
\hline
AV18    &  $-$7.620 & $-$7.626 & 46.71 &   $-$54.34 &    $-$6.75 &  0.9988 &   91.43 &    0.066 &    8.51 &     1.653 &     1.824 \\ 
CDBonn &  $-$7.981 & $-$7.987 &  37.59 &   $-$45.57 &    $-$5.85 &  0.9996 &   92.93 &    0.047 &    7.02 &     1.614 &     1.775 \\ 
N$^2$LO \cite{Epelbaum:2004fk}    & $-$7.867 &  $-$7.872 &  31.85 & $-$39.72 &    $-$5.11 &  0.9995 &   93.43 &    0.039 &    6.53 &     1.624 &     1.787 \\
Idaho N$^{3}$LO \cite{Entem:2003ft} & $-$7.840 &  $-$7.845 &  34.52 & $-$42.36 &   $-$5.52 &  0.9998 &   93.65 &    0.037 &   6.32  &     1.653 &    1.812 \\
  \end{tabular}  
  \end{ruledtabular} 
  \caption{\label{tab:3n-exp}
    Binding energy $E$, expectation values of the Hamiltonian
    $\langle H \rangle$, the kinetic energy $\langle T \rangle$, the
    potential energy 
    $\langle V \rangle$ and the contribution of the
    mass difference of proton and neutron to the kinetic energy
    $\langle T_{CSB} \rangle$ for $^3$H  at order N$^4$LO.   
    The calculated norm of the wave function  
    and probabilities for $S$-, $P$-, and $D$-states are also
    shown. Finally, we also list results for the point proton and neutron  
    rms radii $r_p$ and $r_n$.  Energies are given in MeV (except for
    the $\langle T_{CSB} \rangle$ which is given in keV), radii in fm
    and probabilities in \%. 
}
\end{table}

For $A=4$, we can rewrite the Schr\"odinger equation into Yakubovsky
equations for the two Yakubovsky components $|\psi_1\rangle $ and  
$|\psi_2\rangle $. Again, we can recover the wave function by applications of permutation operators 
$ | \Psi \rangle = [ 1- (1+P)P_{34}] (1+P) | \, \psi_1 \, \rangle +
(1+P)(1+ \tilde P) | \psi_2 \rangle $. In addition to the sum of
cyclic and anticyclic  
permutations used in the 3N system, we also need a transposition of
nucleons 3 and 4, $P_{34}$, and the interchange of the subsystems (12)
and (34)  
given by $\tilde P = P_{13} P_{24}$. The two coupled Yakubovsky equations then read 
\begin{eqnarray}
| \, \psi_1 \, \rangle  & = & G_0 \, t \, (1+P) \, [ \, (1-P_{34}) | \psi_1 \rangle + | \psi_2 \rangle \, ] \\
| \psi_2 \rangle        & = & G_0 \, t \, \tilde P \, \, [ \, (1-P_{34}) | \psi_1 \rangle + | \psi_2 \rangle \, ]  \ \ .
\end{eqnarray}
Here, $G_0$ and $t$ are, again, the free propagator and NN t-matrix
respectively. It is understood that they are embedded into the 4N
Hilbert space for this application.  

We again solve the equations in momentum space using a partial-wave
decomposed basis. The form of the equations  
guarantees a rather fast convergence with respect to partial waves if
the two Yakubovsky components are expressed   
in different basis sets. The first component is expanded in a set of
Jacobi momenta that separate the (12) subsystem ($p_{12}$), the motion  
of the third nucleon relative to (12) ($p_3$) and the fourth nucleon
relative to the (123) subsystem ($q_4$) 
\begin{equation}
| \, p_{12} \, p_3 \, q_4 \, \alpha \, \rangle = | \, p_{12} \, p_3 \,
q_4 \,  \{ [ (l_{12} s_{12}) j_{12}  (l_3 \frac{1}{2}) I_3 ] \, J_3
(l_4 \frac{1}{2}) \, I_4 \, \} J_4   
\ [ (t_{12} \frac{1}{2} ) T_3   \frac{1}{2} ] T_4 \rangle  \ \ .
\end{equation}
In addition to the quantities defined for the 3N system, 
we require  the orbital angular
momentum corresponding to the momentum of the fourth particle $l_4$,
its  
total angular momentum $I_4$ and the total angular momentum and
isospin of the 4N system, $J_4$ and $T_4$, respectively. We refer to
this set of 
basis states as 3+1 coordinates. 
$| \psi_2 \rangle$ is expanded in states introducing relative momenta
within the subsystems (12) and (34), $p_{12}$ and $p_{34}$,
respectively, and  
the relative momentum of these two subsystems $q$ 
\begin{equation}
| \, p_{12} \, p_{34} \, q \, \alpha \, \rangle = | \, p_{12} \,
p_{34} \,  q  \,  \{ [ (l_{12} s_{12}) j_{12}  \lambda ]  \, I
(l_{34} s_{34}) j_{34}  \,  \} J_4   
\ (t_{12}  t_{34} )  \frac{1}{2} ] T_4 \rangle  \ \ .
\end{equation}
The angular momenta, spin and isospin  of the (34) system are given by
$l_{34}$, $s_{34}$, $j_{34}$ and $t_{34}$.  
The angular dependence of the $q$ momentum is expanded in orbital
angular momenta $\lambda$.  The angular momenta  
are coupled as indicated by brackets to the total 4N angular momentum
$J_4=0$ and isospin $T_4=0$. Below, we refer to these basis  
states as 2+2 coordinates.  

We again use 64 mesh points for the discretization of the momenta up
to 15~fm$^{-1}$. The only exception is the $q$ momentum where 48  
mesh points were sufficient to get binding energies with a accuracy
better than 10~keV and the expectation value of the Hamiltonian  
with an accuracy of better than 50~keV. Again the two-body angular
momentum is restricted to $j_{12}^{\max} =7$. We also restrict all
orbital  
angular momenta to 8 and the sum of all angular momenta to 16.  For
the four-body calculations, we assume that the $^4$He system is a pure
$T_4=0$  
state.  The neutron/proton mass difference does not contribute in this
case. More details on the computational aspects can again be found in  
\cite{Nogga:2001cz}. Results for the binding energies of the ground
state are summarized in Table~\ref{tab:4n-yakubovsky}. Notice that we
predict  
a bound excited state for the leading order interactions. The binding
energies for these excited states vary between $-12.6$ and
$-10.4$\,MeV depending  
on the regulator parameter $R$. This second bound state disappears at higher orders. 

\begin{table}[t]
  \renewcommand{\arraystretch}{1.2}
  \begin{ruledtabular}
    \begin{tabular}{l|lllll}
      \multicolumn{1}{c}{$R$ [fm]} &
      \multicolumn{1}{c}{LO}  & 
      \multicolumn{1}{c}{NLO} &
      \multicolumn{1}{c}{N$^2$LO} &
      \multicolumn{1}{c}{N$^3$LO} &
      \multicolumn{1}{c}{N$^4$LO} 
      \\ \hline
      \multicolumn{6}{l}{$^4$He}
      \\ \hline
0.8  & $-$50.14 &   $-$26.50 &  $-$26.68 &  $-$23.93 &  $-$23.43 \\
0.9  & $-$48.39 &   $-$27.52 &	$-$27.28 &  $-$23.93 &  $-$24.02 \\
1.0  & $-$45.46 &   $-$28.55 &	$-$28.13 &  $-$23.77 &  $-$24.29 \\
1.1   &  $-$42.34 &   $-$29.72 &	$-$29.11 &  $-$23.73 &  $-$24.30 \\
1.2   &  $-$39.43 &   $-$30.92 &	$-$30.16 &  $-$23.64 &  $-$24.13 \\
    \end{tabular}  
  \end{ruledtabular} 
  \caption{\label{tab:4n-yakubovsky}
    Calculated $^4$He binding energies  
    using chiral NN interactions at different orders of the chiral expansion and at five different values of $R$. Energies are given in MeV. }
\end{table}

For N$^4$LO, we also show expectation values of the Hamiltonian, the kinetic energy, the potential and the point proton rms 
radius in Table~\ref{tab:4n-exp}. We again compare our results to AV18, CD-Bonn and the two older chiral interactions. 
The Yakubovsky components, as the Faddeev component for the 3N system, converge 
much faster with respect to partial waves than the wave function. Therefore, we normalize the wave function using the 
relation $12 \langle \psi_1 | \Psi \rangle + 6  \langle \psi_2 | \Psi \rangle = 1$ and calculate the kinetic energy using a 
corresponding overlap of the wave function and the Yakubovsky components in the coordinates natural for the 
Yakubovsky component involved. The wave function itself can be expanded in 3+1 or 2+2 coordinates. We therefore 
give two values for the expectation value of $H$ and $V$ in the table. The first ones are obtained using the wave function 
expressed in 3+1 coordinates. The second ones are based on 2+2 coordinates. Especially, for $H$, we observe small 
deviations of the results that indicate that higher partial wave contributions are not completely negligible when small 
cutoffs $R$ are used. The deviation of the binding energy and the expectation values is partly due to the missing 
angular momentum states but also due to the restriction to isospin $T=0$ states. Generally, 
the wave function seems to be better represented in 3+1 coordinates. Nevertheless, even in 2+2 coordinates, 
the agreement of expectation values and binding energies is excellent. This is a non-trivial confirmation of our 
results. We note that the N$^4$LO results are the numerically most demanding ones since they required denser
momentum grids and more partial waves for convergence. Finally, Table~\ref{tab:4n-exp} gives results for the 
point proton radii that, in our $T_4=0$ approximation,  exactly agree with the point neutron radii. Again we find 
that there is no strict correlation of the radii and binding energies. The radii are remarkably independent of the 
cutoff parameter $R$. In the following section, we extend these calculations towards more complex systems 
using the no-core configuration interaction (NCCI) approach.

\begin{table}[t]
  \renewcommand{\arraystretch}{1.2}
  \begin{ruledtabular}
    \begin{tabular}{l|lllllll}
 \multicolumn{1}{c}{$R$ [fm]} &
      \multicolumn{1}{c}{$E$}  & 
      \multicolumn{1}{c}{$\langle H \rangle_1$} &
      \multicolumn{1}{c}{$\langle H \rangle_2$} &      
      \multicolumn{1}{c}{$\langle T \rangle$} &
      \multicolumn{1}{c}{$\langle V \rangle_1$} &
      \multicolumn{1}{c}{$\langle V  \rangle_2$} &                    
    \multicolumn{1}{c}{$r_p $} 
      \\ \hline
0.8 & $-$23.43 &  $-$23.39 &  $-$23.37 &  112.9 &    $-$136.2 &    $-$136.2 &  1.557 \\
0.9 &  $-$24.02 &  $-$24.00 &  $-$23.99 &  101.4 &    $-$125.3 &    $-$125.3 &  1.545 \\
1.0 &  $-$24.29 &  $-$24.27 &  $-$24.27 &   91.9 &    $-$116.1 &    $-$116.2 &  1.546 \\
1.1  & $-$24.30 &  $-$24.28 &  $-$24.29 &   83.7 &    $-$108.0 &    $-$108.0 &  1.554 \\
1.2  & $-$24.13 &  $-$24.11 &  $-$24.12 &   76.5 &    $-$100.6 &    $-$100.6 &  1.568 \\
\hline 
AV18  & $-$24.25 &  $-$24.21 &  $-$24.16 &  97.7 &   $-$121.9  &   $-$121.9  &    1.515  \\
CDBonn & $-$26.16 & $-$26.08 &  $-$26.07 &  77.6 &   $-$103.6 &   $-$103.6 &   1.457 \\
N$^{2}$LO \cite{Epelbaum:2004fk} & $-$25.60 &  $-$25.58 &  $-$25.59 &  62.58 &  $-$88.16 &    $-$88.16 &   1.478  \\
Idaho N$^3$LO \cite{Entem:2003ft} & $-$25.38 & $-$25.37 &  $-$25.37 &  69.18 &   $-$94.55 &   $-$94.55 &  1.518 \\
  \end{tabular}  
  \end{ruledtabular} 
  \caption{\label{tab:4n-exp}
    Binding energy $E$, expectation values of the Hamiltonian in 3+1 (2+2) coordinates $\langle H \rangle_1$ ($\langle H \rangle_2)$,
    the kinetic energy $\langle T \rangle$, the potential energy in 3+1 (2+2) coordinates
    $\langle V \rangle_1$ ($\langle V \rangle_2$) for $^4$He  at order N$^4$LO.  We also give results for the point proton 
    rms radii $r_p$.  Energies are given in MeV and  radii in fm. }
\end{table}


\subsection{No-Core Configuration Interaction calculations}
\label{sec:ncci}

For larger nuclei, $A > 4$, we use 
NCCI methods to solve the many-body Schr\"odinger equation.  These
methods have advanced rapidly in recent years and one can now
accurately solve fundamental problems in nuclear structure and
reaction physics using realistic interactions, see
e.g., Ref.~\cite{Barrett:2013nh} and references therein.  In this
section we follow Refs.~\cite{Maris:2008ax,Maris:2013poa} where, for a
given interaction, we diagonalize the resulting many-body Hamiltonian
in a sequence of truncated harmonic-oscillator (HO) basis spaces.  The
basis spaces are characterized by two parameters: $N_{\max}$ specifies
the maximum number of total HO quanta beyond the HO Slater determinant
with all nucleons occupying their lowest-allowed orbitals and
$\hbar\omega$ specifies the HO energy.  The goal is to achieve
convergence as indicated by independence of these two basis
parameters, either directly or by extrapolation~\cite{Maris:2008ax,Coon:2012ab,Furnstahl:2012qg,More:2013rma,Wendt:2015nba}.

In order to improve the convergence behavior of the many-body
calculations we employ a consistent unitary transformation of the
chiral Hamiltonians.  Specifically, we use the Similarity
Renormalization Group
(SRG)~\cite{Glazek:1993rc,Wegner:1994,Bogner:2007rx,Bogner:2009bt}
approach that provides a straightforward and flexible framework for
consistently evolving (softening) the Hamiltonian and other operators,
including three-nucleon
interactions~\cite{Jurgenson:2009qs,Roth:2011ar,Jurgenson:2013yya,Roth:2013fqa}.
In particular, at N$^3$LO and N$^4$LO this additional ``softening'' of
the chiral NN potential is necessary in order to obtain sufficiently
converged results on current supercomputers.

In the SRG approach, the unitary transformation of an operator,
e.g., the Hamiltonian, is formulated in terms of a flow equation 
\begin{eqnarray}
  \frac{d}{d\alpha}H_{\alpha} &=& [\eta_{\alpha},H_{\alpha}] \,,
  \label{eq:flow}
\end{eqnarray}
with a continuous flow parameter $\alpha$.  The initial condition for
the solution of this flow equation is given by the `bare' chiral
Hamiltonian.  The physics of the SRG evolution is governed by the
anti-hermitian generator $\eta_{\alpha}$.  A specific form widely used
in nuclear physics~\cite{Bogner:2009bt} is given by
\begin{eqnarray}
  \eta_{\alpha} &=& m_N^2 [T_{\text{int}},H_{\alpha}]\,,
\end{eqnarray}
where $m_N$ is the nucleon mass and $T_{\text{int}}$ is the intrinsic
kinetic-energy operator.  This generator drives the Hamiltonian
towards a diagonal form in a basis of eigenstates of the intrinsic
kinetic energy, i.e., towards a diagonal in momentum space.

Along with the reduction in the coupling of low-momentum and
high-momentum components by the Hamiltonian, the SRG induces many-body
operators beyond the particle rank of the initial Hamiltonian.  In
principle, all induced terms up to the $A$-body level should be
retained to ensure that the transformation is unitary and the spectrum
of the Hamiltonian is independent of the flow parameter.  Here, we
truncate the evolution at the three-nucleon level, neglecting four-
and higher multi-nucleon induced interactions, which formally violates 
unitarity.  For consistency, we check that for $A=3$ we recover the
exact results (for a given input potential); and for $A \ge 4$ we
perform our calculations at two different values of $\alpha$ and
compare our results with calculations without SRG evolution.

The flow equation for the three-body system is solved using a HO
Jacobi-coordinate basis~\cite{Roth:2013fqa}.  The intermediate sums
in the three-body Jacobi basis are truncated at $N_{\max} = 40$ for
channels with $J \leq 7/2$, $N_{\max}=38$ for $J=9/2$, and $N_{\max} = 36$
for $J \geq 11/2$.  The SRG evolution and subsequent transformation to
single-particle coordinates were performed on a single node using an
efficient OpenMP parallelized code.

For the NCCI calculations we employ the code
MFDn~\cite{Maris:2010,Aktulga:2014}, which is highly optimized for
parallel computing on current high-performance computing platforms.  
The size of the largest
feasible basis space is constrained by the total number of three-body
matrix elements required as input, as well as by the number of
many-body matrix elements that are computed and stored for the
iterative Lanczos diagonalization procedure.
We can perform $^4$He calculations up to $N_{\max} = 14$ with 3N
interactions, but calculations of $A=6$ and $7$ nuclei are
restricted to $N_{\max} = 12$, and for $A>10$ we can only go up to
$N_{\max} = 8$ with (induced) 3N interactions.  Note that with bare NN
interactions, i.e., without the SRG evolution and the induced 3N terms,
we can go to significantly larger basis spaces, namely $N_{\max} = 20$
for $^4$He; $N_{\max} = 18$ for $A=6$; $N_{\max} = 16$ for $A=7$;
$N_{\max} = 14$ for $A=8$; $N_{\max} = 12$ for $A=9$ and $10$; and
$N_{\max} = 10$ for $A=16$.  The larger basis spaces achievable with
NN-only interactions arise due to the significantly smaller memory
footprint of the input Hamiltonian matrix element files and the
smaller memory footprint of the many-body Hamiltonian itself which is
stored completely in our calculations.
The latter issue has been reported as approximately a factor of 40
reduction in memory footprint with NN-only interactions compared to
NN+3N interactions~\cite{Vary:2009qp}.  The many-body calculations
were performed on the Cray XC30 Edison at NERSC and the IBM BG/Q Mira
at Argonne National Laboratory.

Finally, compared to the few-body bound state calculations presented
above, we use the following simplifications in our many-body
calculations: we employ the same (average) nucleon mass for the protons
and the neutrons, $m_N=938.92$~MeV.
Also, we do add the two-body Coulomb potential
between (point-like) protons, but we do not take any higher-order
electromagnetic effects into account.  
Furthermore, here and in what follows we
restrict ourselves to the intermediate values of the coordinate-space
regulator of $R=0.9$, $1.0$ and $1.1$~fm. The smallest available
cutoff choice of $R=0.8$~fm leads to highly nonperturbative NN
potentials \cite{Hoppe:2017lok,Reinert:2017usi}, which cannot be employed in many-body calculations without
SRG evolution or similar softening approaches. On the other hand, the softest regulator choice of $R=1.2$~fm is known to
lead to large finite-regulator  artifacts \cite{Epelbaum:2014efa,Furnstahl:2015rha,Melendez:2017phj}, and for this reason
we do not consider it in the following calculations.

\subsection{Results for ground state energies}

\begin{figure}
\includegraphics[width=0.99\columnwidth]{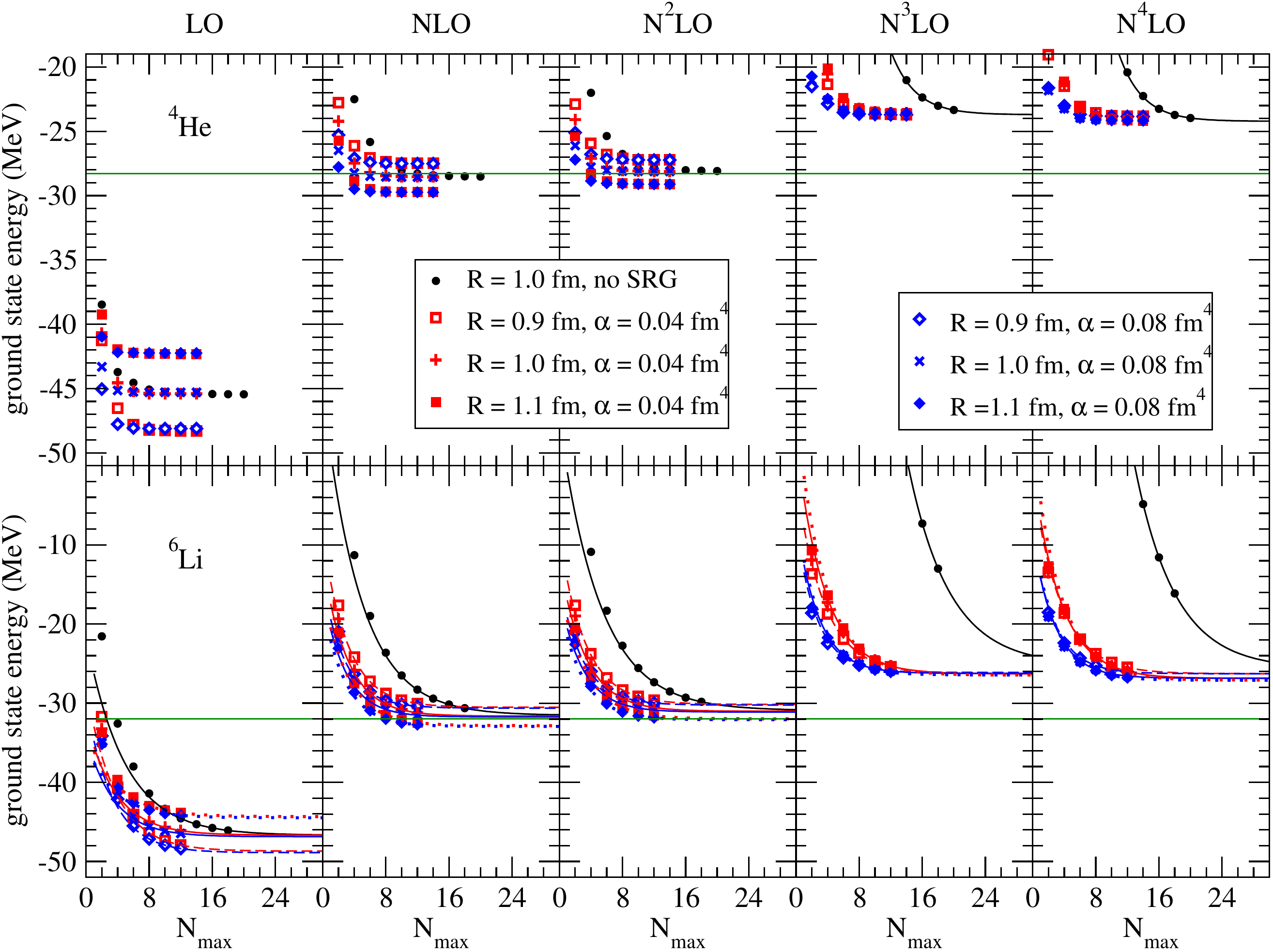}
\caption{\label{Fig:groundstate_extrapolated}
   (Color online) Ground state energies of $^4$He and $^6$Li at or just above the
  variational minimum in $\hbar\omega$ as a function of the basis truncation parameter
  $N_{\max}$ for LO to N$^4$LO chiral NN potentials:
  without SRG evolution (black dots) at $R=1.0$~fm;
  and for NN potentials at $R=0.9$~fm (open squares and open diamonds),
  $R=1.0$~fm (plusses and crosses),
  and $R=1.1$~fm (solid squares and solid diamonds), 
  SRG evolved to $0.04$~fm$^4$ (red) and $0.08$~fm$^4$ (blue),
  including induced 3NF.
  Dotted ($R=1.1$~fm), solid ($R=1.0$~fm), and dashed ($R=0.9$~fm)
  lines are exponential fits to the highest three $N_{\max}$ points for cases 
  where convergence is not well-established by direct calculation.}
\end{figure} 

In Fig.~\ref{Fig:groundstate_extrapolated} we show our results for the
ground state energies of $^4$He and $^6$Li at LO to N$^4$LO, both
without SRG evolution (for $R=1.0$~fm only) and with SRG evolution
(for $R=0.9$, $1.0$, and $1.1$~fm) including the induced 3N terms as
mentioned above.  (Note that, starting at N$^2$LO, there are also 3NFs
in the chiral expansion, which are not incorporated in the
calculations presented here.)  Before examining these results in
detail, we first make several qualitative observations:
(1) The overall trends are the same for the different chiral cutoffs:
significant overbinding at LO, close to the experimental values at NLO
and N$^2$LO, and underbinding at N$^3$LO and N$^4$LO.
(2) The dependence on the chiral cutoff $R$ decreases with increasing
chiral order, as expected.
(3) The convergence rate changes dramatically with the chiral order --
in particular when going from N$^2$LO to N$^3$LO, as anticipated by
the Weinberg eigenvalue analysis of Ref.~\cite{Hoppe:2017lok}.  However,
after applying the SRG evolution, convergence is reasonable, and the
dependence of the converged energies on the SRG parameter $\alpha$ is
negligible on the scale of these plots.

\begin{table}[t]
  \renewcommand{\arraystretch}{1.2}
  \begin{ruledtabular}
    \begin{tabular}{l|lllll}
      \multicolumn{1}{l}{$R$ [fm]} &
      \multicolumn{1}{c}{LO}  & 
      \multicolumn{1}{c}{NLO} &
      \multicolumn{1}{c}{N$^2$LO} &
      \multicolumn{1}{c}{N$^3$LO} &
      \multicolumn{1}{c}{N$^4$LO} 
      \\ \hline
      \multicolumn{6}{l}{$^4$He}
      \\ \hline
      $0.9$ & $-48.284\pm 0.002\pm 0.17$ & $-27.49 \pm 0.01 \pm 0.03$ & $-27.23 \pm 0.01 \pm 0.01$ & $-23.71 \pm 0.01 \pm 0.01$ & $-23.85\pm 0.01\pm 0.01$ 
      \\
      $1.0$ & $-45.407\pm 0.001\pm 0.12$ & $-28.542\pm 0.004\pm 0.03$ & $-28.113\pm 0.006\pm 0.01$ & $-23.59 \pm 0.01 \pm 0.01$ & $-24.14\pm 0.01\pm 0.01$
      \\
      $1.1$ & $-42.312\pm 0.001\pm 0.07$ & $-29.723\pm 0.002\pm 0.02$ & $-29.102\pm 0.003\pm 0.01$ & $-23.59 \pm 0.01 \pm 0.01$ & $-24.18\pm 0.01\pm 0.01$
      \\ \hline
      \multicolumn{6}{l}{$^6$Li}
      \\ \hline
      $0.9$ & $-48.7 \pm 0.4 \pm 0.2$ & $-30.5 \pm 0.1 \pm 0.1$ & $-30.2 \pm 0.1 \pm 0.1$ & $-26.2 \pm 0.2 \pm 0.1$ & $-26.3 \pm 0.2 \pm 0.1$
      \\
      $1.0$ & $-46.7 \pm 0.3 \pm 0.2$ & $-31.6 \pm 0.1 \pm 0.1$ & $-31.0 \pm 0.1 \pm 0.1$ & $-26.3 \pm 0.2 \pm 0.1$ & $-26.9 \pm 0.3 \pm 0.1$
      \\
      $1.1$ & $-44.4 \pm 0.3 \pm 0.1$ & $-32.8 \pm 0.1 \pm 0.1$ & $-32.0 \pm 0.1 \pm 0.1$ & $-26.4 \pm 0.2 \pm 0.1$ & $-27.1 \pm 0.3 \pm 0.1$
    \end{tabular}
  \end{ruledtabular}
  \caption{\label{Tab:groundstate_A46_R}
    Calculated $^4$He and $^6$Li ground state energies (in MeV) 
    using chiral NN interactions at three different values of $R$,
    SRG evolved to $\alpha=0.04$~fm$^4$ (including induced
    3NFs). The first theoretical error is the extrapolation
    uncertainty estimate following Ref.~\cite{Maris:2013poa},
    whereas the second is an estimate of the SRG error, based on
    the difference between results at $\alpha=0.04$~fm$^4$ and
    $\alpha=0.08$~fm$^4$, due to omitting the induced multi-nucleon
    interactions at and above 4NFs.}
\end{table}

Based on the results in finite basis spaces, we can use extrapolations
to the complete (infinite-dimensional) basis.  Here we use a three
parameter fit at fixed $\hbar\omega$ at or just above the variational
minimum
\begin{eqnarray}
 E(N_{\max}) &\approx&  E_{\infty} + a \exp{(-b N_{\max})} \,,
\end{eqnarray}
which seems to work well for a range of interactions and
nuclei~\cite{Maris:2008ax,Maris:2013poa}.  The lines in
Fig.~\ref{Fig:groundstate_extrapolated} correspond to the
extrapolating function fitted to the highest available $N_{\max}$
values.
Our estimate of the extrapolation uncertainty is based on the
difference with smaller $N_{\max}$ extrapolations, as well as the
basis $\hbar\omega$ dependence over an $8$ to $12$~MeV span in
$\hbar\omega$ values around the variational
minimum~\cite{Maris:2013poa}.
As a consistency check, we first performed calculations for $A=3$:
including induced 3N contributions the results with and without SRG
evolution are in agreement with each other, to within the estimated
convergence or extrapolation uncertainty.  Furthermore, they also
agree with the Faddeev binding energies of Table~\ref{tab:3n-faddeev}
to within the estimated accuracy.
Our results with SRG evolution to $\alpha=0.04$~fm$^4$ for the ground
state energies of $^4$He and $^6$Li are summarized in
Table~\ref{Tab:groundstate_A46_R}.  In addition to the extrapolation
uncertainty, we also give, as a second (systematic) contribution to
the uncertainties, the difference between the ground state energies at
$\alpha=0.04$~fm$^4$ and at $\alpha=0.08$~fm$^4$, which may serve as an
indication of the `error' made by neglecting induced many-body forces.

The $^4$He results of Table~\ref{Tab:groundstate_A46_R} agree within
the estimated uncertainties with the binding energies presented in
Table~\ref{tab:4n-yakubovsky}, at least at LO, NLO, and N$^2$LO.
However, at N$^3$LO and N$^4$LO there are systematic differences:
at N$^3$LO these differences are between $140$~keV and $220$~keV,
depending on $R$, and at N$^4$LO between $120$~keV and $170$~keV.
These differences are an order of magnitude larger than the estimated
numerical uncertainties, and are largest for $R=0.9$~fm and smallest
at $R=1.1$~fm.  That is, these difference are smallest for the softest
interactions.  A possible explanation for this discrepancy could be
induced 4N forces, which we have neglected in the SRG evolution.
This suggests that the difference between the ground state energies at
$\alpha=0.04$~fm$^4$ and at $\alpha=0.08$~fm$^4$, may indeed serve as an
indication of the `error' made by neglecting induced many-body forces
up to N$^2$LO, but is likely to underestimate the effect of neglected
many-body forces at N$^3$LO and N$^4$LO.  Note that without the SRG
evolution the many-body calculation of the binding energy does agree
with the Yakubovsky calculation, though the extrapolation uncertainty
is significantly larger, see Table~\ref{Tab:groundstate_A461016_alpha}
below.

As already mentioned, at LO we find considerable overbinding for all
three values of the chiral cutoff $R$.  This overbinding depends
significantly on $R$ and is strongest for $R=0.9$~fm.  At NLO (and
N$^2$LO), the $R$-dependence is reduced by a factor of about two to
three.  Furthermore, the pattern is reversed compared to the LO
results: At NLO and N$^2$LO, $R=0.9$~fm leads to a slight
underbinding, whereas $R=1.1$~fm gives slight overbinding for $^4$He
and $^6$Li.  At N$^3$LO the $R$-dependence is further reduced by about
an order of magnitude compared to NLO and N$^2$LO, and for $^6$Li
becomes of the same order as the many-body extrapolation uncertainty.
Interestingly, at LO in the chiral expansion, $^6$Li is not actually
bound with respect to the $\alpha + d$ breakup, whereas at NLO and
N$^2$LO it is bound by about $0.7$ to $0.9$~MeV (and it appears
to remain bound at higher orders as well).

\begin{table}[t]
  \renewcommand{\arraystretch}{1.2}
  \begin{ruledtabular}
    \begin{tabular}{l|lllll}
      \multicolumn{1}{l}{$\alpha$ [fm$^4$]} &
      \multicolumn{1}{c}{LO}  & 
      \multicolumn{1}{c}{NLO} &
      \multicolumn{1}{c}{N$^2$LO} &
      \multicolumn{1}{c}{N$^3$LO} &
      \multicolumn{1}{c}{N$^4$LO} 
      \\ \hline
      \multicolumn{6}{l}{$^4$He, $J^P=0^+$}
      \\ \hline
      $ 0 $ & $-45.453 \pm 0.006$ & $-28.533 \pm 0.004$ & $-28.11  \pm 0.01 $ & $-23.7  \pm 0.1 $ & $-24.2 \pm 0.1 $
      \\
      $0.04$& $-45.407 \pm 0.001$ & $-28.542 \pm 0.004$ & $-28.113 \pm 0.006$ & $-23.59 \pm 0.01$ & $-24.14 \pm 0.01$
      \\
      $0.08$& $-45.289 \pm 0.001$ & $-28.566 \pm 0.001$ & $-28.119 \pm 0.001$ & $-23.582\pm 0.002$& $-24.145\pm 0.002$
      \\ \hline
      \multicolumn{6}{l}{$^6$Li, $J^P=1^+$}
      \\ \hline
      $ 0 $  & $-46.7 \pm 0.1$ & $-31.6 \pm 0.2$ & $-31.0 \pm 0.2$ & $-24.4 \pm 2.3$ & $-25.7 \pm 1.9$
      \\
      $0.04$ & $-46.7 \pm 0.3$ & $-31.6 \pm 0.1$ & $-31.0 \pm 0.1$ & $-26.3 \pm 0.2$ & $-26.9 \pm 0.3$
      \\
      $0.08$ & $-46.9 \pm 0.3$ & $-31.7 \pm 0.1$ & $-31.1 \pm 0.1$ & $-26.3 \pm 0.2$ & $-26.9 \pm 0.3$
      \\ \hline
      \multicolumn{6}{l}{$^{10}$B, $J^P=1^+$}
      \\ \hline
      $ 0 $ & $ -93.9 \pm 0.8 $ & $ -64.9 \pm 1.8$ & $ -63.1 \pm 1.9$ & -- & --
      \\
      $0.04$& $ -94.0 \pm 1.5 $ & $ -64.5 \pm 0.8$ & $ -63.1 \pm 0.8$ & $-55. \pm 2. $ & $-55. \pm 2. $
      \\
      $0.08$& $ -94.9 \pm 0.9 $ & $ -64.3 \pm 0.8$ & $ -63.1 \pm 0.6$ & $-52.2\pm 0.8$ & $-53.3\pm 0.7$
      \\ \hline
      \multicolumn{6}{l}{$^{10}$B, $J^P=3^+$}
      \\ \hline
      $ 0 $ & $ -88.1 \pm 1.2 $ & $ -64.6 \pm 1.5$ & $ -62.3 \pm 1.7$ & -- & --
      \\
      $0.04$& $ -88.2 \pm 1.6 $ & $ -64.1 \pm 0.7$ & $ -62.1 \pm 0.8$ & $-51. \pm 4. $ & $-52. \pm 3. $
      \\
      $0.08$& $ -88.8 \pm 1.0 $ & $ -64.1 \pm 0.6$ & $ -61.9 \pm 0.6$ & $-50.1\pm 1.0$ & $-51.2\pm 0.9$
      \\ \hline
      \multicolumn{6}{l}{$^{16}$O, $J^P=0^+$}
      \\ \hline
      $ 0 $ & $-224.  \pm 2. $ & $-156.  \pm 5. $ & $-149.  \pm 5. $ & -- & --
      \\
      $0.04$& $-223.2 \pm 0.4$ & $-152.0 \pm 1.3$ & $-146.2 \pm 0.9$ & $-121.  \pm 4. $ & $-121. \pm 4.$
      \\
      $0.08$& $-220.9 \pm 0.2$ & $-150.1 \pm 0.8$ & $-144.8 \pm 0.6$ & $-113.  \pm 2. $ & $-114. \pm 2.$ 
    \end{tabular}
  \end{ruledtabular}
  \caption{\label{Tab:groundstate_A461016_alpha}
    Calculated $^4$He, $^6$Li, $^{10}$B, and $^{16}$O ground state
    energies (in MeV) using chiral NN interactions at $R=1.0$~fm
    without SRG evolution, and SRG evolved to $\alpha=0.04$~fm$^4$
    and $\alpha=0.08$~fm$^4$ (including induced 3NFs).
    The theoretical error is the extrapolation uncertainty estimate
    following Ref.~\cite{Maris:2013poa}, adjusted to be at least
    20\% of the difference with the variational minimum.}
\end{table}

In Table~\ref{Tab:groundstate_A461016_alpha} we summarize our results
with and without SRG evolution for several representative $p$-shell
nuclei at LO through N$^4$LO for $R=1.0$~fm.  The errors listed in
Table~\ref{Tab:groundstate_A461016_alpha} are our estimates of the
extrapolation uncertainties, adjusted to be at least 20\% of the
difference with the variational minimum.  Again, induced 3N contributions to
the SRG-evolved interaction are included, but induced 4N and higher
multi-nucleon induced interactions neglected.  The differences between
results without SRG evolution and at SRG values of
$\alpha=0.04$~fm$^4$ and at $\alpha=0.08$~fm$^4$ tend to be of the
same order as (or smaller than) the extrapolation uncertainties,
except at leading order.  When compared with the results at
$\alpha=0.04$~fm$^4$, the results at $\alpha=0.08$~fm$^4$ generally do
have smaller extrapolation uncertainties (i.e., are more converged in
the many-body basis expansion) as one would expect, but are slightly
further away from the results without SRG renormalization, where
available.

At N$^3$LO and N$^4$LO, we have to rely on SRG evolution (or other
renormalization schemes) for $p$-shell nuclei.  For $^6$Li we can do
an extrapolation of the bare interaction results, but the extrapolation
uncertainty is large, whereas the results at $\alpha=0.04$~fm$^4$ and
$\alpha=0.08$~fm$^4$ differ by less than $100$~keV.
For the upper half of the $p$-shell, SRG evolution also
becomes beneficial at NLO and N$^2$LO.

\begin{table}[t]
  \renewcommand{\arraystretch}{1.2}
  \begin{ruledtabular}
    \begin{tabular}{cc|lllll}
      \multicolumn{1}{c}{nucleus} &
      \multicolumn{1}{c|}{$J^P$}   &
      \multicolumn{1}{c}{LO}      & 
      \multicolumn{1}{c}{NLO}     &
      \multicolumn{1}{c}{N$^2$LO} &
      \multicolumn{1}{c}{expt} 
      \\ \hline
      $^3$H   &$\frac{1}{2}^+$& $-11.30 \pm 0.01 $  & $ -8.38 \pm 0.01 $  & $ -8.28 \pm 0.01 $ &  -8.482 \\
      $^4$He  & $0^+$         & $-45.453 \pm 0.006$ & $-28.533 \pm 0.004$ & $-28.11 \pm 0.01 $ & -28.296 \\
      $^6$He  & $0^+$         & $-43.2 \pm 0.2 $    & $-28.7 \pm 0.2 $    & $-27.9 \pm 0.2 $   & -29.27 \\
      $^6$Li  & $1^+$         & $-46.7 \pm 0.1 $    & $-31.6 \pm 0.2 $    & $-31.0 \pm 0.2 $   & -31.99 \\
      $^7$Li  &$\frac{3}{2}^-$& $-57.1 \pm 0.2$     & $-38.7 \pm 0.3 $    & $-38.0 \pm 0.4 $   & -39.24 \\ 
      $^8$He  & $0^+$         & $-39.8 \pm 0.6$     & $-29.7 \pm 0.5 $    & $-27.8 \pm 0.6 $   & -31.41 \\ 
      $^8$Li  & $2^+$         & $-55.7 \pm 0.5$     & $-40.3 \pm 0.7 $    & $-39.0 \pm 0.8 $   & -41.28 \\ 
      $^8$Be  & $0^+$         & $-87.7 \pm 0.4$     & $-56.0 \pm 0.7 $    & $-55.4 \pm 0.9 $   & -56.50 \\ 
      $^9$Li  &$\frac{3}{2}^-$& $-57.1 \pm 0.4$     & $-43.9 \pm 0.7 $    & $-41.7 \pm 0.8 $   & -45.34 \\ 
      $^9$Be  &$\frac{3}{2}^-$& $-84.7 \pm 0.7$     & $-58.0 \pm 1.4 $    & $-56.4 \pm 1.5 $   & -58.16 \\ 
      $^{10}$B & $1^+$         & $-93.9 \pm 0.8$     & $-64.9 \pm 1.8 $    & $-63.1 \pm 1.9 $   & -64.03 \\
      $^{10}$B & $3^+$         & $-88.1 \pm 1.2$     & $-64.6 \pm 1.5 $    & $-62.3 \pm 1.7 $   & -64.75 \\
      $^{16}$O & $0^+$         & $-224. \pm 2.$      & $-156. \pm 5.$    & $-149.\pm 5. $   & -127.62 \\ 
    \end{tabular}
  \end{ruledtabular}
  \caption{\label{Tab:groundstates}
    Calculated NCCI ground state energies, in MeV, using chiral NN
    interactions at $R=1.0$~fm (without SRG evolution).  Results are
    compared with experimental data in the last column.  The quoted
    theoretical errors are due to extrapolation uncertainties
    following Ref.~\cite{Maris:2013poa}, adjusted to be at least 20\%
    of the difference with the variational minimum.  }
\end{table}

\begin{figure}
\includegraphics[width=0.99\columnwidth]{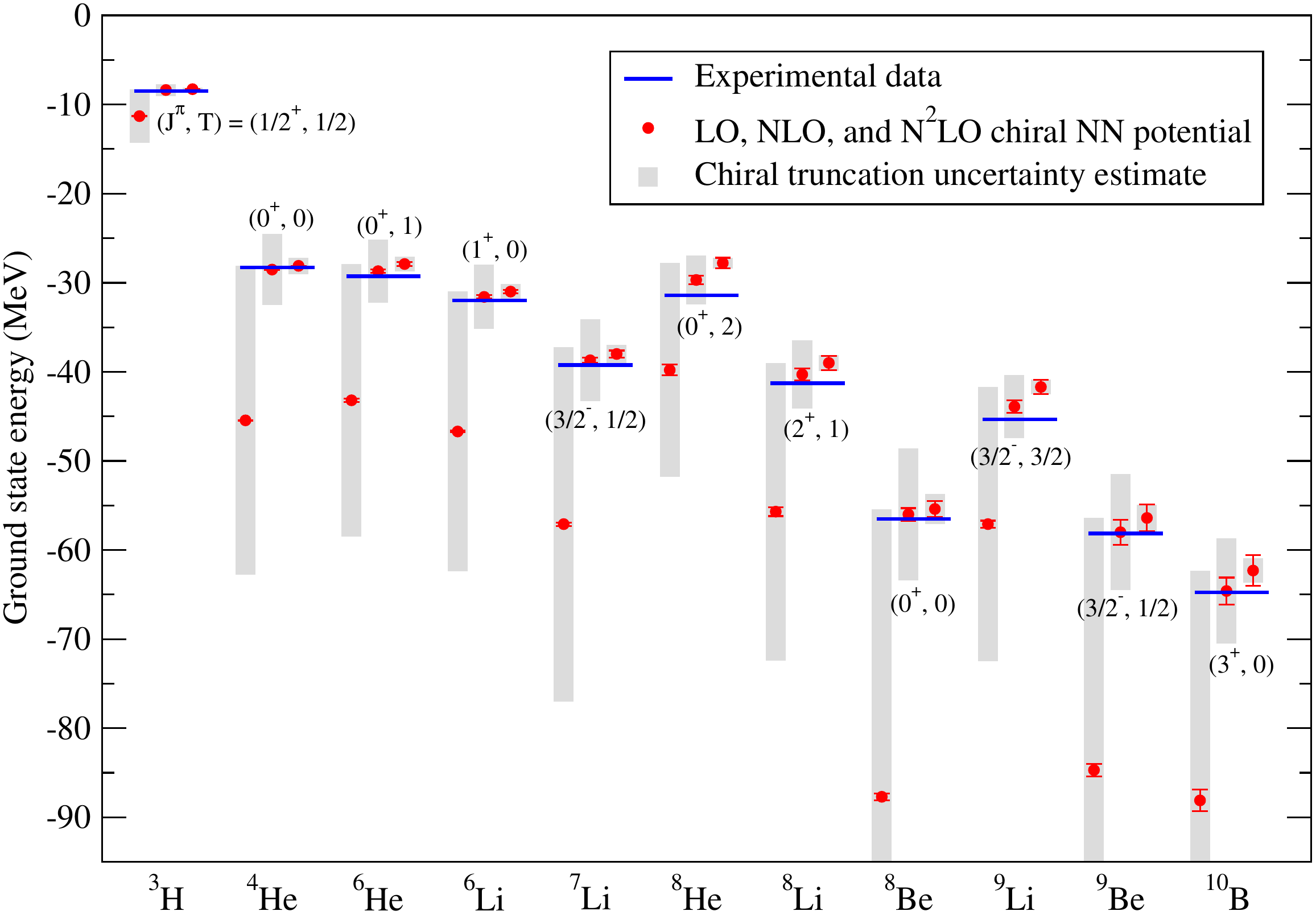}
\caption{\label{Fig:res_gs_Eb_A03A10}
  (Color online) Calculated (red dots) ground state energies in MeV
  using chiral LO, NLO, and N$^2$LO NN interactions at $R=1.0$~fm
  (without SRG evolution) based on the NN forces only in comparison
  with experimental values (blue levels).  Red error bars indicate
  NCCI extrapolation uncertainty and shaded bars indicate the
  estimated truncation error at each chiral order as defined in the
  Introduction.}
\end{figure}

In Table~\ref{Tab:groundstates} and 
Fig.~\ref{Fig:res_gs_Eb_A03A10} 
we
summarize our results for the ground state energies of $A=3$ to $A=10$
nuclei, as well as for $^{16}$O in Table~\ref{Tab:groundstates}, 
based on extrapolations of the chiral LO, NLO, and N$^2$LO
interactions without applying any further SRG renormalization.  
With the exception of $^7$Li at LO, and of $^{10}$B, the ground state
spins all agree with the experimental spin of the ground state.
The results with SRG evolution, through the limited range that we
investigate, (including induced 3NFs) are very similar, and fall
within the quoted uncertainty estimates for all cases.  Given this
similarity of results with and without SRG evolution we do not present
here the results with SRG evolution.

The ground state energies of all nuclei in
Table~\ref{Tab:groundstates} follow similar patterns:
significantly overbound at LO, closer to the experimental values
at NLO, and then shifted towards less binding at N$^2$LO.
E.g., the $J^\pi=\frac{3}{2}^-$ ground state of $^7$Li follows the same
overall pattern as that of $^4$He and $^3$H, and is actually bound
with respect to breakup into $^4$He plus $^3$H at LO, NLO and N$^2$LO.
However, at $A=8$ (and to a lesser extend also at $A=9$) we see that
the difference between LO and NLO results decreases significantly with
increasing isospin: it is much smaller for the $^8$He than it is for
$^8$Be.
Also note that the deviation from experiment at N$^2$LO is largest for
$^8$He, and smallest for $^8$Be.  (Similar effects can be seen for
$^9$Li and $^9$Be.)
Neither $^8$He nor $^8$Be are bound at LO ($^8$He is about
$5.5$~MeV above $^4$He, and $^8$Be is about $3.3$~MeV above two
$\alpha$-particles, and the applicability of the HO basis is rather
questionable for these states). On the other hand, at NLO $^8$He does become bound,
whereas $^8$Be remains unbound, both in qualitative agreement with
experiment.  Whether or not $^6$He (and $^8$He) are bound at
N$^2$LO (and higher orders) depends crucially on the chiral 3NFs --
without these, they are not bound.
Note that $^9$Be is also not bound at LO: despite the enormous
overbinding compared to experiment, it is not bound with respect to breakup
into two $\alpha$-particles plus a neutron, and its ground state
energy is even above that of $^8$Be.  Only at NLO does $^9$Be become
bound and it may remain bound at N$^2$LO but the uncertainties do not
allow us to make a definite statement.

Finally, the level ordering of the lowest states of $^{10}$B is known
to be sensitive to the details of the interaction~\cite{Navratil:2007we},
and typically one finds a $J^\pi=1^+$ ground state with NN-only
potentials, instead of a $3^+$ ground state.  With a 3NF one can
obtain the correct $J^\pi=3^+$ ground state spin for $^{10}$B, but the
convergence pattern of the lowest $1^+$ is different than that of the
lowest $3^+$ state; furthermore, the splitting between these two
states appears to be very sensitive both to the parameters of the
interaction and to the SRG evolution~\cite{Jurgenson:2013yya}.  In our
calculations, the $1^+$ is the ground state at LO, and about 6 MeV
below the $3^+$ state, but at NLO and at N$^2$LO the level splitting
between these two states is less than our estimated extrapolation
uncertainties.

We show the chiral truncation error estimate for the ground state energies of
light nuclei up to $A=10$ using the methods reviewed in
Sec.~\ref{sec:intro} but limited to N$^2$LO in
Fig.~\ref{Fig:res_gs_Eb_A03A10}.  We remind the reader that the shown
results at N$^2$LO are incomplete as the corresponding 3NF are not
included.  Accordingly, at leading order, the chiral error estimate
appears to be given by $\delta E^{(0)} = | E^{(3)} - E^{(0)} |$, and at NLO and
N$^2$LO by $Q \delta E^{(0)}$ and $Q^2 \delta E^{(0)}$ respectively,
for all 10 nuclei.  
As in Ref.~\cite{Binder:2015mbz}, the expansion parameter for these
light nuclei is estimated here
as $Q=M_\pi / \Lambda_b$ (see Section~\ref{sec:chiraltruncationerror}).
Note that if we were to include results up to
N$^4$LO without including 3NFs (and possibly 4NFs), all chiral error
estimates following this prescription would increase noticeably,
because the N$^3$LO and N$^4$LO results without consistent 3NFs leads
to a larger $\max( | E^{(i)} - E^{(0)} | )$ that appears in
Eq.~(\ref{ErrorOrig2}).  Alternative chiral truncation error estimates for these
results are discussed in Section~\ref{sec:Uncertainties} below.

Looking further into the results in Fig.~\ref{Fig:res_gs_Eb_A03A10},
one notices that at N$^2$LO, where omitted 3NFs may have an impact,
we see significant differences between the current results and
experiment that go beyond the estimated chiral truncation uncertainty.
These differences are easily visible for $^6$He, $^8$He, $^8$Li and
$^9$Li.  Future work that includes the 3NFs is needed to discern their
role and to understand if they resolve these differences while not
creating significant differences in the cases where little difference
is currently found.

\subsection{Magnetic moments}

\begin{figure}
  \includegraphics[width=0.99\columnwidth]{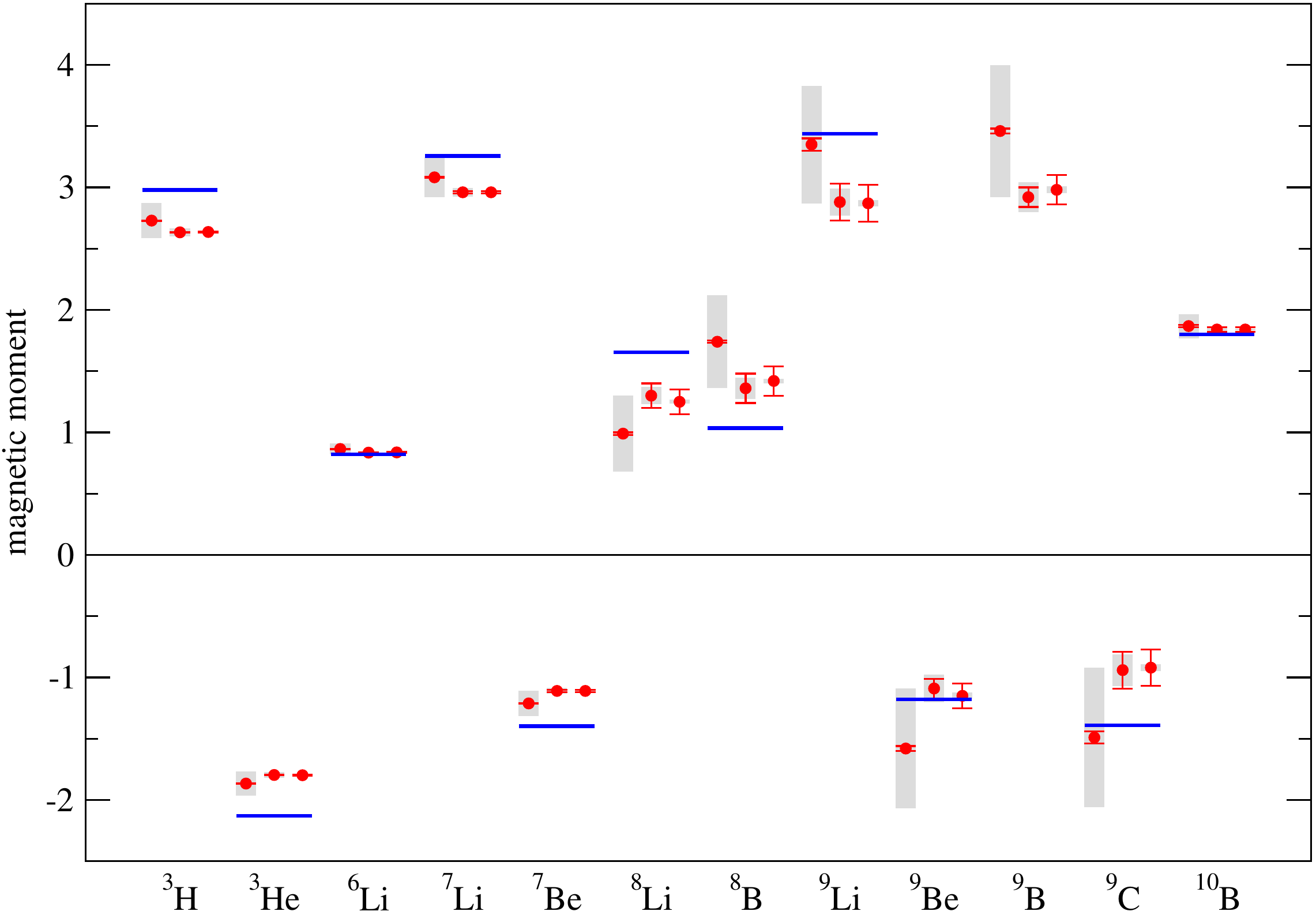}
  \caption{\label{Fig:res_gs_mu_A03A10} (Color online) Calculated (red
    dots) ground state magnetic moments of light nuclei up to $A=10$
    at LO, NLO, and N$^2$LO with $R=1.0$~fm in comparison with
    experimental values (blue horizontal lines).  Red error bars
    indicate NCCI extrapolation uncertainty and shaded bars indicate
    the estimated truncation error at each chiral order as defined in
    the Introduction.}
\end{figure}

In addition to binding energies we also calculated the magnetic
moments for the ground states of $p$-shell nuclei up to $A=10$.
In contrast to long-range observables such as radii, 
magnetic moments tend to converge rapidly in a HO basis.
Indeed, the magnetic moments for the ground states of $^6$Li, $^7$Li,
and $^7$Be are very well converged.  Furthermore, the dependence on
the chiral order is very weak, and the results are remarkably close to
the experimental values.  For $A=8$ and $9$, the convergence is not as
good, and there is a stronger dependence on the chiral order, but the
magnetic moment of the ground state of $^{10}$B is again very well
converged, and only very weakly dependent on the chiral order.

Note that here we only used the canonical one-body current operator
and we defer to a future effort the development and application of
consistent chiral current operators at each order.  We expect that
with such improved current operators, including meson-exchange
currents~\cite{Kolling:2009iq,Kolling:2011mt,Krebs:2016rqz,Pastore:2008ui,Pastore:2012rp,Baroni:2015uza},
the calculated magnetic moments (as well as magnetic transition matrix
elements) will be in good agreement with experimental values -- the
deviation with experimental magnetic moments that we find here are of the same sign and
magnitude as suggested by phenomenological meson-exchange
contributions~\cite{Pastore:2012rp}.

\section{Medium-mass nuclei}
\label{sec:CoupledCluster}

Over the past few years, several \emph{ab initio} methods have been
developed to address ground states of nuclei in the medium-mass
regime, beyond the reach of standard NCCI calculations. Already the
simplest observables, like ground-state energies and radii for
medium-mass nuclei, e.g., the doubly magic calcium isotopes, provide a
valuable testing ground for chiral interactions, 
far away from the few-body domain that
was used to constrain the Hamiltonians.  

For a first characterization of the new generation of chiral NN
interactions in the medium-mass regime, we employ the most advanced
coupled-cluster (CC) formulations and state-of-the-art in-medium
similarity renormalization group (IM-SRG) calculations for
ground-state observables of $^{16,24}$O and $^{40,48}$Ca. We mirror
the discussion of the previous section and analyze the order-by-order
behavior and the theoretical uncertainties. In addition we compare 
to results with other, widely used chiral forces.

\subsection{Coupled-Cluster Theory}

Single-reference CC theory expresses the exact
many-body state as $| \Psi \rangle = e^T| \Phi\rangle$, where
$|\Phi\rangle$ is a single-Slater-determinant reference state based on
a Hartree-Fock calculation
\cite{Hagen:2007ew,Hagen:2010gd,Binder:2012mk,Jansen:2012ey,Binder:2013oea,Hagen:2012rq,Baardsen:2013vwa,Hagen:2012fb,Hagen:2012sh,Hagen:2013nca}.  
Correlations are introduced by the action of the exponential $e^T$ of
the particle-hole excitation operator $T=T_1 + T_2 + \dots + T_A$ on
the reference state. In practical calculations, the cluster operator
$T$ is truncated at some low $n$-particle-$n$-hole ($n$p$n$h)
excitation level, such as the 2p2h excitations, 
$T\approx T_1+T_2$. This constitutes the very popular CC with singles
and doubles excitations (CCSD) approach. 
Due to the exponential ansatz, all powers of $T_1$, $T_2$ and
mixed products of these are present in the description of the wave
function, resulting in the facility to describe many-body correlations 
of considerable complexity that may be difficult to achieve 
in alternative many-body methods.  

The essential ingredient in CC theory is the similarity-transformed
Hamiltonian $\bar{H} = e^{-T}H e^T$. In terms of $\bar{H}$,
one can solve for the $T$ amplitudes by projecting from the left with
particle-hole excited reference states $| \Phi^{ab\dots}_{ij\dots}\rangle$ in
order to obtain the set of equations $0=\langle
\Phi^{ab\dots}_{ij\dots} | \bar{H} | \Phi \rangle$ which
determines the cluster amplitudes. The energy is obtained from
calculating the closed diagrams of $\bar{H}$ according to
$E=\langle \Phi| \bar{H} | \Phi\rangle$~\cite{ShBa09}.  

Going beyond the singles and doubles approximation in CC calculations
leads to an increased complexity of the equations to be solved
and to increased computational cost. Therefore, the current approach in
nuclear structure theory to incorporate higher-than-doubles
excitations in ground-state CC calculations is by a non-iterative
inclusion of triples excitation effects to the ground-state energy
(but not the wave function) via the CCSD(T)~\cite{RaTr89},
$\Lambda$CCSD(T)~\cite{TaBa08,TaBa08-2}, or the
CR-CC(2,3)~\cite{PiGo09} method.  

Three-body interactions can be included in CC calculations using the
normal-ordering approximation at the 
two-body level (NO2B)~\cite{Hagen:2007ew,Roth:2011vt}. Alternatively, the CC
method can straightforwardly be  
extended to the full treatment of three-body Hamiltonians, however,
often at prohibitively large computational 
cost \cite{Binder:2012mk,Binder:2013oea}. In this work, we will work with the CCSD approach combined with the
CR-CC(2,3) energy correction including 3N interactions in the NO2B approximation.

\subsection{In-Medium Similarity Renormalization Group}

The IM-SRG aims at decoupling an $A$-body reference state $\ket{\Phi}$ from all particle-hole
excitations or, equivalently, at suppressing a specific
``off-diagonal'' part of the Hamiltonian
\cite{Tsukiyama:2010rj,Hergert:2012nb,Hergert:2015awm,Hergert:2016etg}. This
decoupling at the $A$-body level can be implemented using the concepts
of the similarity renormalization group, that we already exploited in
few-body spaces (cf. Section \ref{sec:ncci}). 
We formulate a continuous unitary transformation of the Hamiltonian 
$H(s) = U^{\dagger}(s) H(0) U(s)$ in $A$-body space, where $s$ denotes the flow parameter of the IM-SRG.
This transformation is rewritten into the following operator differential equation
\begin{equation}
	\totdiff{H}{s}(s) = \comm{\eta(s)}{H(s)}
	\text{   ,}
	\label{eq:operator_flow_equation_hamiltonian}
\end{equation}
where $\eta(s)$ refers to the so-called generator of the transformation.
The Hamiltonian $H(s)$ and the generator $\eta(s)$ are normal-ordered
with respect to the reference state
$\ket{\Phi}$ and truncated at 
the normal-ordered two-body level, e.g.,
\begin{equation}
	H(s) 
	= E(s) 
	+ \sum_{pq} f^p_q(s) \{a_p^{\dagger} a_q\} 
	+ \frac{1}{4} \sum_{pqrs} \Gamma^{pq}_{rs}(s) \{a_p^{\dagger} a_q^{\dagger} a_r a_s\}
	\text{  ,}
\end{equation}
where normal-ordered products of single-particle creation and annihilation operators appear. 
Evaluating the right-hand side of equation
(\ref{eq:operator_flow_equation_hamiltonian}) via Wick's theorem, one
can derive the flow equations for the 
matrix elements of the normal-ordered zero-, one-, and two-body part,
i.e., $E(s)$, $f^p_q(s)$ and $\Gamma^{pq}_{rs}(s)$, respectively, of
the Hamiltonian. 
As an example, the flow equation for zero-body part, which represents
energy expectation value in the reference state, reads 
\begin{equation}
	\totdiff{E(s)}{s}=
	\sum_{pq}\
	\left(n_p-n_q\right) \
	\opme{\eta}{p}{q}(s) \ 
	\opme{f}{q}{p}(s)
	+
	\frac{1}{4} \ 
	\sum_{pqrs} \
	\left(
	\tpme{\eta}{p}{q}{r}{s}(s) \ 
	\tpme{\Gamma}{r}{s}{p}{q}(s) \  
	n_{p}n_{q} (1-{n}_{r})(1-n_{s})
	- \ichange{\eta}{\Gamma}
	\right)
	\text{  ,}	
\end{equation}
where $n_{p}$ is the occupation number w.r.t.\ the reference state $\ket{\Phi}$.
Formally, the flow equations of the IM-SRG are a coupled system of
first-order ordinary differential equations 
which can be solved numerically as an initial value problem until decoupling is reached.

A great advantage of the IM-SRG is the simplicity and flexibility of its basic concept.
Through different choices for the generator $\eta(s)$, 
we obtain different decoupling patterns, numerical characteristics and efficiencies.
As a consequence, we can tailor the IM-SRG for specific applications,
e.g., the derivation of valence-space shell model interactions
\cite{Tsukiyama:2012sm,Bogner:2014baa,Stroberg:2016ung}. 
Furthermore, it is straightforward to use the formalism of the IM-SRG for a consistent evolution of observables
since the flow equation for an observable is similar to the one given
in equation (\ref{eq:operator_flow_equation_hamiltonian}). 
The IM-SRG was first applied for the study of ground-state energies of closed-shell nuclei 
but can be easily extended to open-shell nuclei via multi-reference generalizations of normal 
ordering and Wick's theorem \cite{Hergert:2013vag,Hergert:2014iaa,Gebrerufael:2016xih}.

\subsection{Chiral Truncation Error}
\label{sec:chiraltruncationerror}

In order to quantify the truncation errors in nuclear ground-state energies at various chiral orders, 
we recall the approach introduced in 
Ref.~\cite{Binder:2015mbz}, see the discussion in section \ref{sec:intro}, and  employ Eqs.~(\ref{ErrorOrig}) and 
(\ref{ErrorOrig2}) at LO and NLO and Eq.~(\ref{ErrorMod}) at N$^2$LO and higher chiral orders. 
In Ref.~\cite{Binder:2015mbz}, the expansion parameter $Q$ of the chiral 
expansion defined in Eq.~(\ref{ExpPar}), which enters Eqs.~(\ref{ErrorOrig}), (\ref{ErrorOrig2}) and (\ref{ErrorMod}), 
was estimated for $^3$H, $^4$He and $^6$Li as $Q=M_\pi / \Lambda_b$.  While
this is reasonable for very light nuclei as seen in the discussion of chiral truncation errors 
in light nuclei in Sec. \ref{sec:LightNuclei}, one may expect the typical momentum 
to increase in heavier systems due to the increased role of Pauli blocking. 

In order to estimate these effects, we employ two different methods to evaluate a
nucleus-dependent characteristic momentum scale: the Hartree-Fock (HF) approximation
and the NCCI method.  We use the resulting ground-state wave function, in each case, 
to evaluate the expectation value of the relative
kinetic energy operator $\langle T_{\hbox{\scriptsize rel}} \rangle$ given by
\begin{eqnarray}
  T_{\hbox{\scriptsize rel}} & \equiv & \sum_{i<j}\frac{(\vec{p}_i-\vec{p}_j)^2}{2 \, A \, m} 
  \; = \; \frac{2}{A} \sum_{i<j}\frac{(\vec{p}_{ij})^2}{2 \, \mu}
\end{eqnarray}
in terms of the relative momenta $\vec{p}_{ij} = (\vec{p}_i-\vec{p}_j)/2$
and the reduced two-nucleon mass $\mu = m/2$.  
Based on this expectation value, we define the average relative momentum scale as follows:
\begin{eqnarray}
\label{pavg}
  p_{\hbox{\scriptsize avg}} & = & \sqrt{ \frac{2\mu}{(A-1)} \langle T_{\hbox{\scriptsize rel}} \rangle } 
  = \sqrt{ \frac{2}{A(A-1)} \bigg\langle \sum_{i<j} (\vec{p}_{ij})^2 \bigg\rangle } \;.
\end{eqnarray}
As the last expression shows, this simply corresponds to the
root-mean-square relative momentum of all nucleon pairs, i.e., the
square root of the expectation value of the squared relative momenta
summed over all particle pairs and divided by the number of
pairs. Thus, this quantity reflects a characteristic scale for
relative two-body momenta in the nucleus, which will depend on the
nucleus under consideration and on the underlying interaction.  

The results for $p_{\hbox{\scriptsize avg}}$ obtained in HF and NCCI are summarized in Tables \ref{tab:MomScale} and \ref{tab:MomScale2} of Appendix \ref{ExpVal}. For HF, we employ the SRG-evolved Hamiltonian with the SRG parameter $\alpha = 0.08~$fm$^4$
and evaluate the expectation value of the SRG-transformed relative kinetic energy operator for
input to the calculation of $p_{\hbox{\scriptsize avg}}$.  We also employ the spherical HF approximation
and the filling fraction approximation for open shell nuclei. For NCCI, we extrapolate the expectation
value of the relative kinetic energy to the infinite basis limit using NCCI results from 
currently attainable $N_{\rm max}$ values.  

The HF results are available for all chiral orders and show a systematic decrease in $p_{\hbox{\scriptsize avg}}$ with increasing order. Pronounced changes appear from LO to NLO and from N${}^2$LO to N${}^3$LO. This general trend can be explained in a simple mean-field type picture, keeping the  behaviour of the ground-state  charge radii in mind. With increasing chiral order the radius of a given nucleus shows a systematic increase (including the more pronounced changes, cf. Figs. \ref{fig:ccimsrg_a1} and \ref{fig:ccimsrg_a2}), which translates into a decrease of the Fermi energy and the associated momentum scale in a mean-field picture. The $p_{\hbox{\scriptsize avg}}$ scale evaluated at the HF level captures exactly this mean-field or low-momentum physics.

It is interesting to compare this to the NCCI calculations, which converge to the exact solution of the many-body problem, including all correlations beyond the mean-field level. These results are available up to N${}^2$LO for the p-shell nuclei and up to N${}^4$LO for s-shell nuclei (both, from Faddeev-Yakubovsky and NCCI calculations). Up to N${}^2$LO the $p_{\hbox{\scriptsize avg}}$ scales extracted from the NCCI kinetic energies for the bare Hamiltonian agree surprisingly well with scales extracted from HF expectation values based on SRG-evolved operators. This indicates the the SRG transformation does capture the main beyond-HF correlations such that the kinetic energy expectation values are very similar to the full NCCI values. Still, even with the SRG transformation, not all correlations are covered and the HF ground-state energies differ significantly from the converged NCCI result.

This difference becomes apparent at N${}^3$LO and N${}^4$LO, where the
SCS NN interactions are significantly harder and much more difficult
to converge in the NCCI than at lower orders (see e.g.,
Fig. \ref{Fig:groundstate_extrapolated}). This is the reason why no
NCCI scales can be extracted for p-shell nuclei beyond N${}^2$LO. For
s-shell nuclei the $p_{\hbox{\scriptsize avg}}$ scales obtained from
NCCI at N${}^3$LO and N${}^4$LO are significantly larger than for the
lower orders, in contrast to the mean-field trend shown by the
HF-based scale estimates. At this point, short-range or high-momentum
physics explicitly affects the momentum scales extracted from NCCI
wave functions, which is absent in the HF treatment. Such short-range
correlation effects are regulator scale and scheme dependent and represent specific
high-momentum aspects of the wave function and not a gross momentum
scale corresponding to the Fermi-momentum in a homogeneous
system. 
We do not have a strong physics reason for preferring one or another
approach to estimating the nucleus-dependent momentum scale
$p_{\hbox{\scriptsize avg}}$. In the following, we adopt the HF-based
scale estimate as input for the uncertainty quantification out of
convenience. 

Given that the $p_{\hbox{\scriptsize avg}}$ values
show significant variations at different chiral orders, we average over the 
available results from LO to N$^4$LO to arrive at a single nucleus-dependent and $R$-dependent 
value for $p_{\hbox{\scriptsize avg}}$ quoted in the last column in Tables \ref{tab:MomScale} and \ref{tab:MomScale2}.  
Then, for a given nucleus, the expansion parameter $Q$ is estimated as   
\begin{equation}
\label{newQ}
Q = \frac{\max( p_{\hbox{\scriptsize avg}}, \, M_\pi)}{\Lambda_b},
\end{equation}  
where $p_{\hbox{\scriptsize avg}}$ is the result in the last column of 
Tables \ref{tab:MomScale} and \ref{tab:MomScale2}.

Another feature of the results in Tables \ref{tab:MomScale} and \ref{tab:MomScale2} is the increase in  
$p_{\hbox{\scriptsize avg}}$ with increasing $A$. For very heavy nuclei, 
the relevant momentum scale should be closer to 
the Fermi momentum $p_F \sim 260$~MeV corresponding to the 
saturation density of nuclear matter.   The trend in the results of the last
column of Tables \ref{tab:MomScale} and \ref{tab:MomScale2} appears consistent with that 
expectation. However, for light nuclei $p_{\hbox{\scriptsize avg}}$ 
is within a few percent of of $M_\pi$, at least up to $A = 9$ for $R=1.0~$fm.
Since the chiral uncertainty estimates shown in
Figs.~\ref{Fig:res_gs_Eb_A03A10} and \ref{Fig:res_gs_mu_A03A10} 
would change only
minimally by adopting $p_{\hbox{\scriptsize avg}}$ for the definition of Q,
we do not show them for light nuclei. However, for the following
discussion of ground-state observables of medium-mass nuclei, we will
adopt the nucleus-dependent momentum scales for the order-by-order
uncertainty quantification.


\subsection{Results}

\begin{figure}[h]
	\begin{center}
		\includegraphics[width=0.45\textwidth]{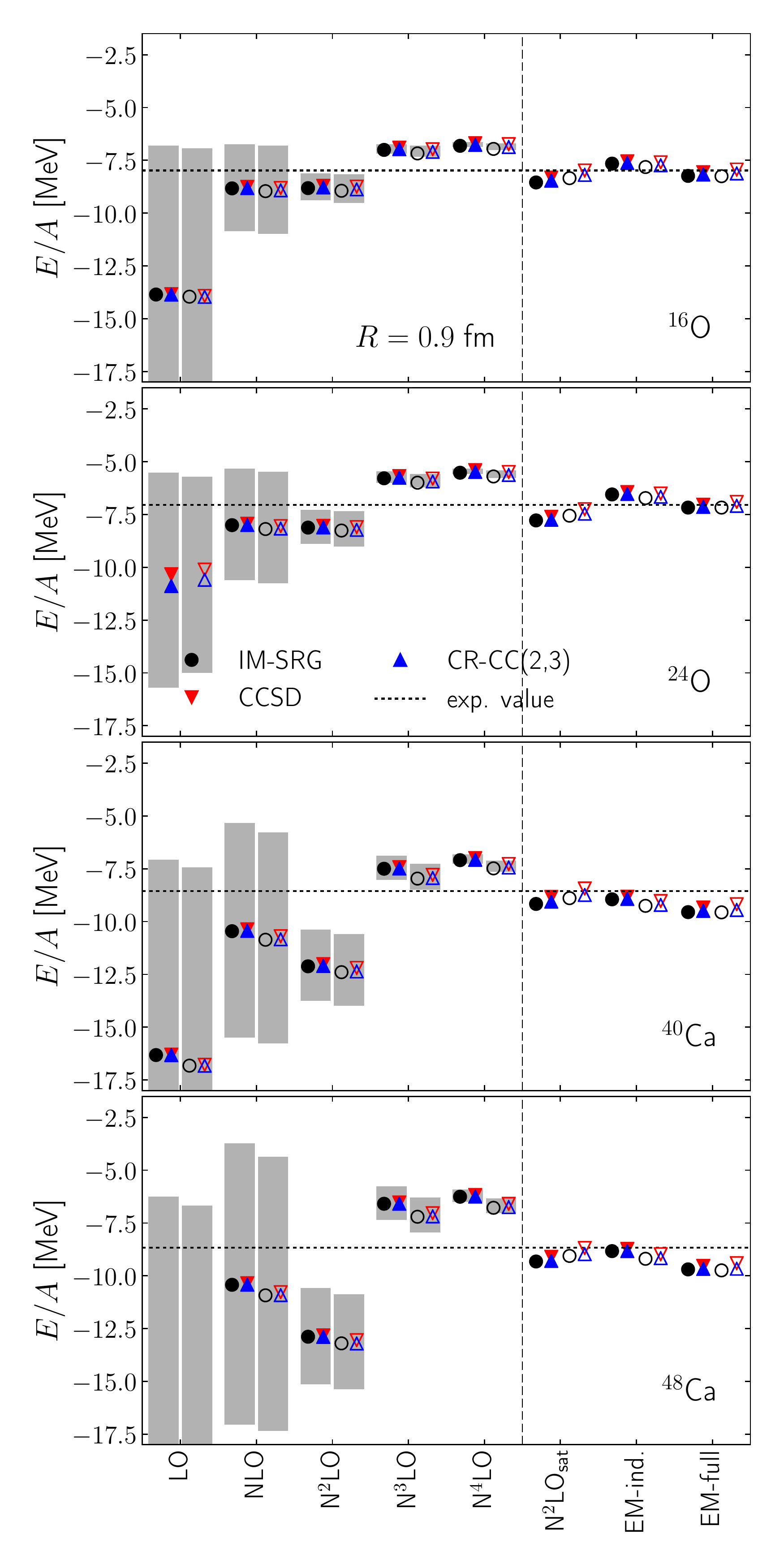}
		\includegraphics[width=0.45\textwidth]{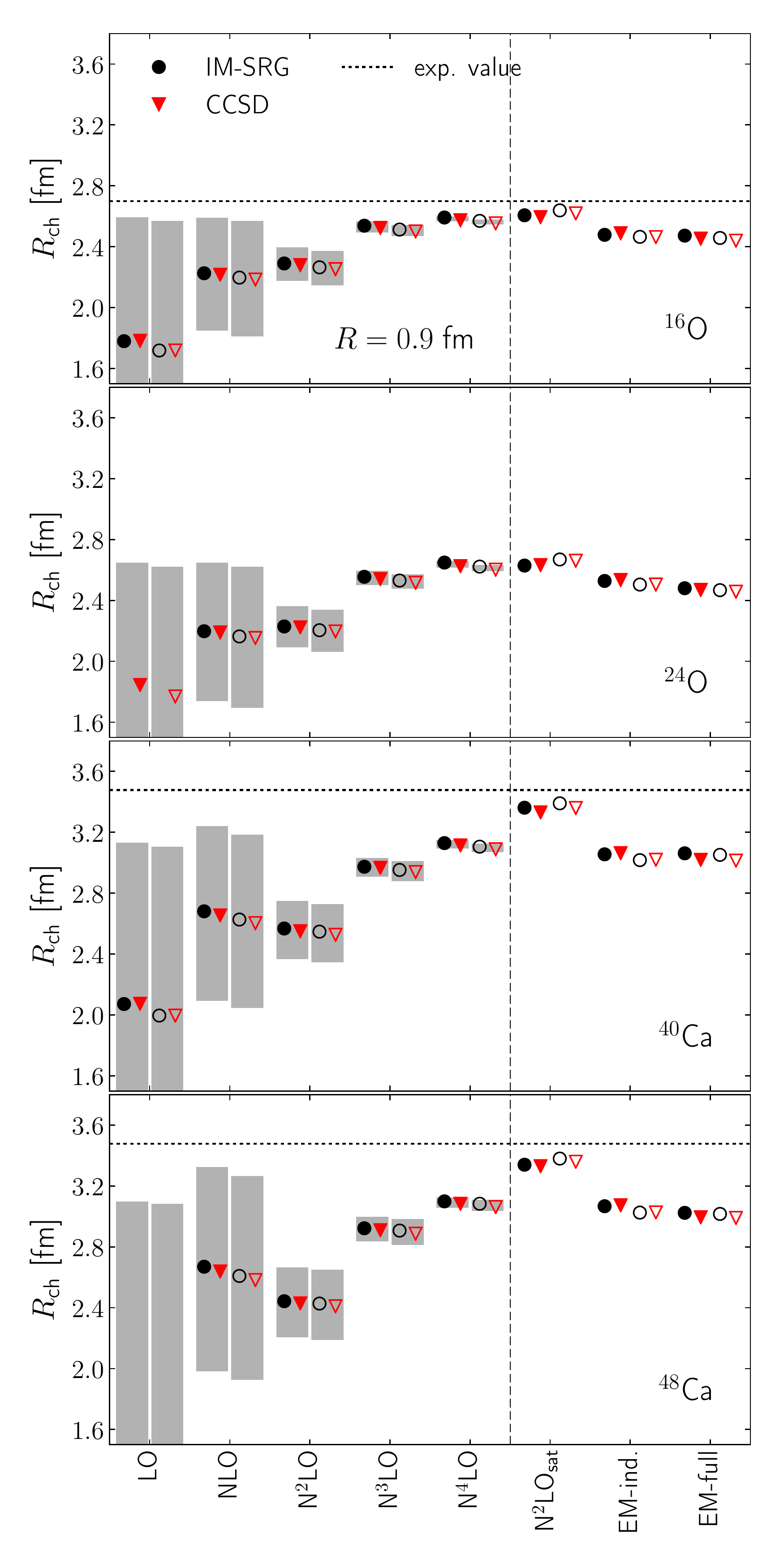}
	\end{center}
	\caption{(Color online) 
		Ground-state energies and charge radii for the
                ground-state of $^{16,24}$O and $^{40,48}$Ca obtained
                from CC and IM-SRG based on HF reference states. The different columns correspond
                to different initial interactions, starting with the
                SCS  chiral NN interaction from LO to N$^4$LO with
                the cutoff $R=0.9\,\text{fm}$, followed by the
                N$^2$LO-SAT NN+3N interaction \cite{Ekstrom:2015rta}, chiral NN interaction at
                N$^3$LO by Entem and Machleidt \cite{Entem:2003ft} without (EM-ind) and with (EM-full) an
                additional local chiral 3N interaction at N$^2$LO \cite{Navratil:2007zn} with
                reduced cutoff $\Lambda_{3N}=400\;\text{MeV}$ \cite{Roth:2011vt}. Solid symbols refer to a free-space SRG
                parameter $\alpha=0.08\;\text{fm}^{4}$ whereas open
                symbols refer to $\alpha=0.04\;\text{fm}^{4}$. The
                grey bars indicate the estimated theoretical
                uncertainties at various chiral orders. 
               } 
		\label{fig:ccimsrg_a1}
\end{figure}

\begin{figure}
	\begin{center}
		\includegraphics[width=0.45\textwidth]{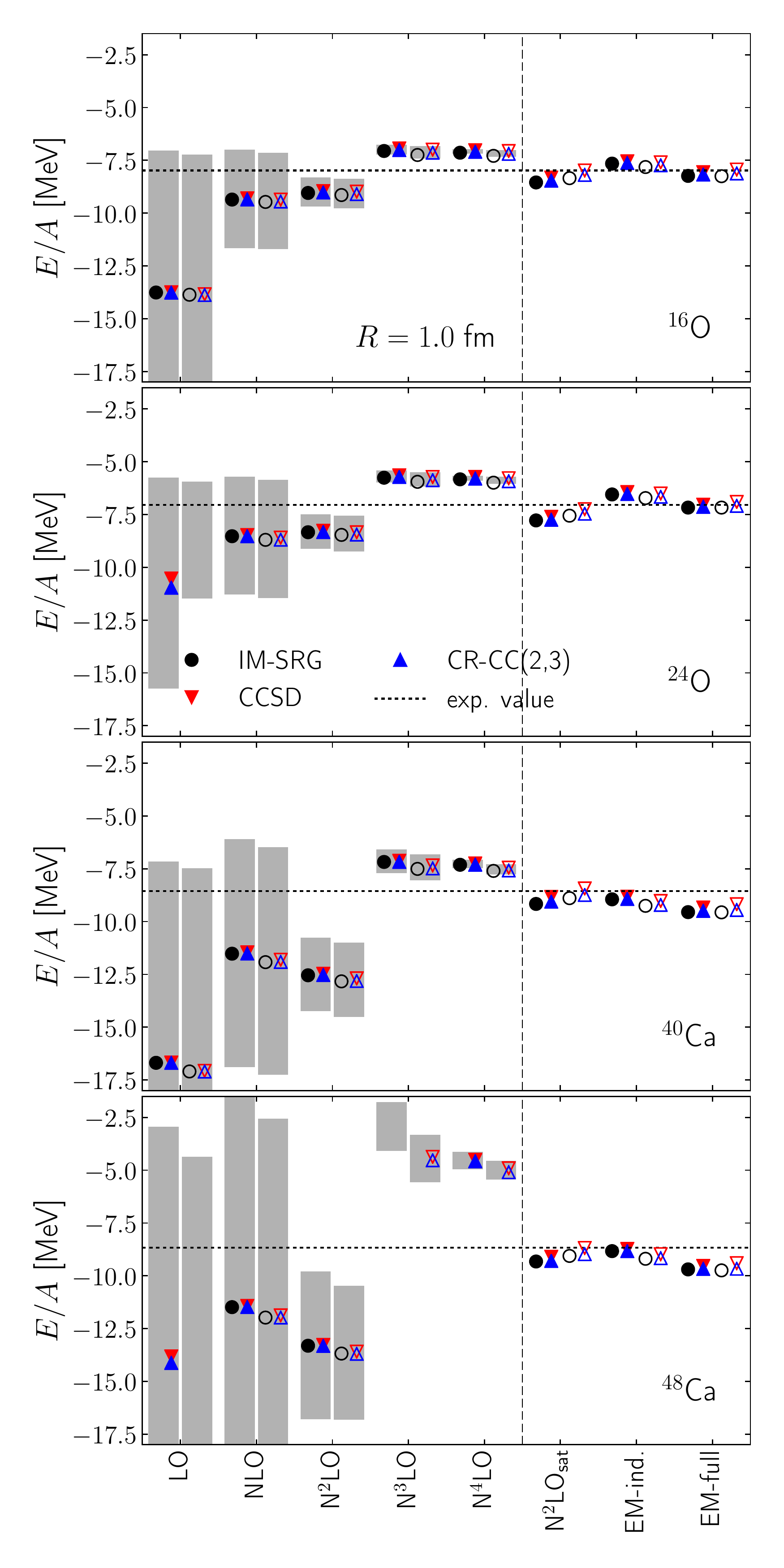}
		\includegraphics[width=0.45\textwidth]{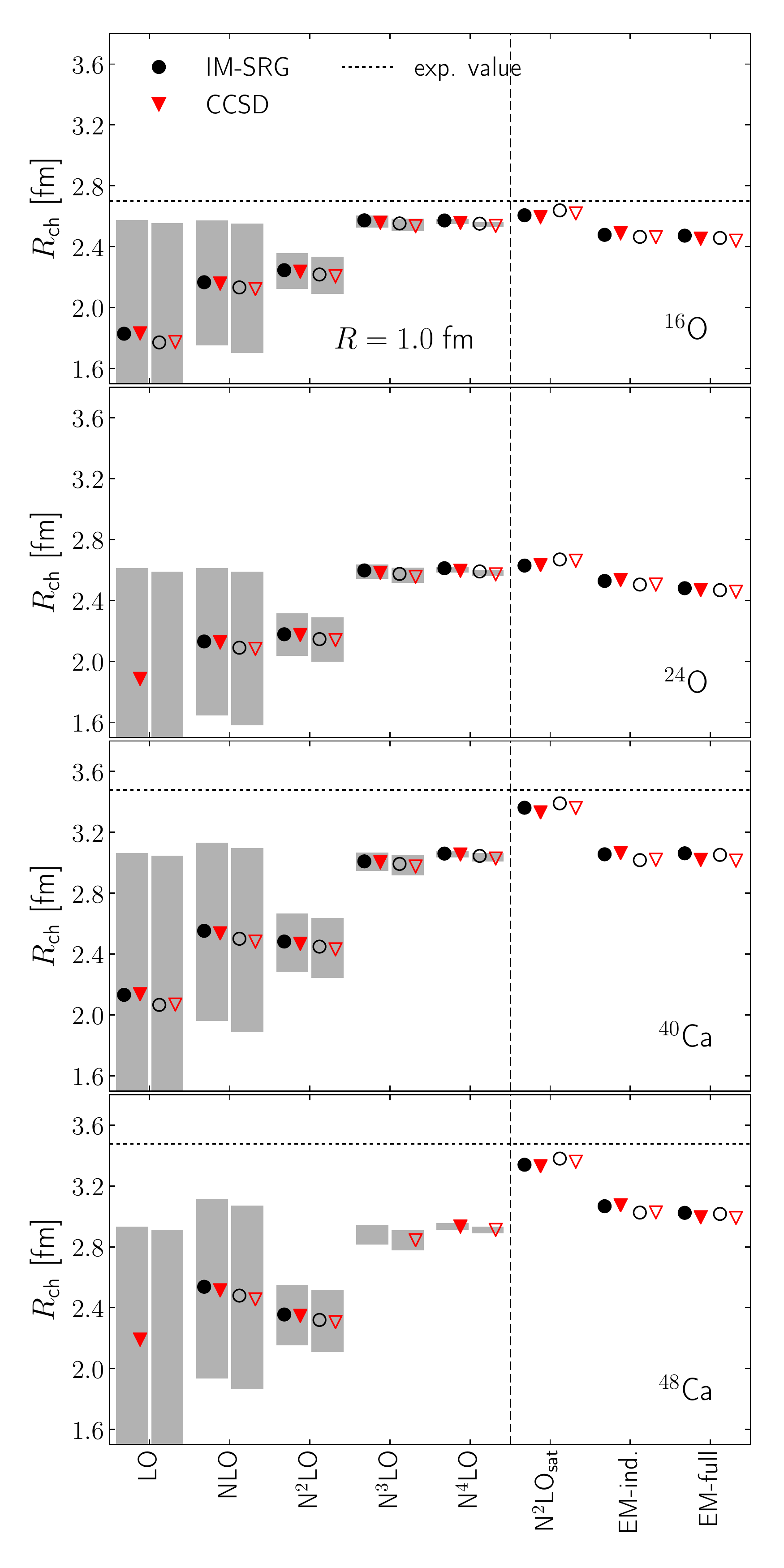}
	\end{center}
	\caption{ Same as Figure \ref{fig:ccimsrg_a1} but for the cutoff value of $R=1.0~$fm.} 
		\label{fig:ccimsrg_a2}
\end{figure}

Using CC and IM-SRG we explore the ground-state energies and charge radii of the doubly magic 
nuclei $^{16,24}$O and $^{40,48}$Ca with the SCS NN
interactions from LO to N$^4$LO. The focus of these calculations is
the investigation of the order-by-order behavior of the chiral
expansion in the medium-mass regime and the theory uncertainties
derived from it.  

For all calculations presented in this section we use SRG-evolved
interactions including the induced three-nucleon contributions. We use
two different SRG flow parameters, $\alpha=0.04\,\text{fm}^4$ and
$0.08\,\text{fm}^4$, to probe the contributions of higher-order
induced forces that are not explicitly included. For the specific
interaction and nucleus under consideration, we first perform a
Hartree-Fock calculation for the full Hamiltonian in a HO basis
truncated with respect to the maximum single-particle principal
quantum number $e_{\max}=(2n+l)_{\max}$. The HF solution defines the
reference state and an optimized single-particle basis, which
eliminates the dependence of the subsequent many-body solutions on the
oscillator frequency. The full Hamiltonian is normal-ordered with
respect to the reference Slater determinant and residual
normal-ordered three-body terms are discarded. We have explored the
accuracy of the normal-ordered two-body approximation in the
medium-mass regime, e.g., through direct comparisons of CC
calculations with and without the residual three-body terms and found
agreement at the $1\%$ level or better
\cite{Binder:2012mk,Roth:2011vt}.  

With these inputs, we perform CC calculations at the level of CCSD and
CR-CC(2,3), which provide a direct way to quantify the residual
uncertainty due to the cluster truncation. In addition, we perform
single-reference IM-SRG(2) calculations. The results for the
ground-states energies and the charge radii of  $^{16,24}$O and
$^{40,48}$Ca are summarized in  Fig.~\ref{fig:ccimsrg_a1} for the
sequence of SCS NN interaction at cutoff $R=0.9\,\text{fm}$ and
in Fig.~\ref{fig:ccimsrg_a2} for $R=1.0\,\text{fm}$. For comparison
each panel also shows the corresponding results with established
chiral interactions, i.e., the N$^2$LO-sat NN+3N interaction by
Ekstr\"om et al. \cite{Ekstrom:2015rta}, the N$^3$LO NN interaction by
Entem and Machleidt \cite{Entem:2003ft} without (EM-ind) and with
(EM-full) a supplementary local 3N interaction at N$^2$LO with cutoff
400 MeV \cite{Navratil:2007zn,Roth:2011vt}.  
The numerical values for the ground-state energies and charge radii obtained 
with the SCS NN interactions at cutoff values of $R=0.9\,\text{fm}$
and $R=1.0\,\text{fm}$ can be found 
in Tables \ref{tab:ccimsrg_energy} and \ref{tab:ccimsrg_radii}, respectively.

The different symbol shapes and colors distinguish the three many-body
methods while solid and open symbols indicate the two SRG flow-parameters
we use. The variation within the set of six calculations for any given
chiral interaction and nucleus provides an estimate for the
uncertainties in the solution of the many-body problem, including the
free-space SRG evolution and the many-body truncations.  

These many-body uncertainties can be compared to the uncertainties
inherent to the chiral interaction at any given order, which are
quantified using the protocol discussed in
Sec. \ref{sec:chiraltruncationerror}. We use the intrinsic kinetic
energy expectation value obtained in HF calculations with SRG
transformed operators to define a momentum scale. The uncertainties
for the ground-state energies and the charge radii are then determined
from Eqs.~(\ref{ErrorOrig}) and (\ref{ErrorOrig2}) for LO and NLO and
Eq.~(\ref{ErrorMod}) from N$^2$LO on. The gray bands in
Figs.~\ref{fig:ccimsrg_a1} and \ref{fig:ccimsrg_a2} indicate these
uncertainties extracted from the IM-SRG results as representatives for
the three different many-body approaches. For the neutron-rich
isotopes $^{24}$O and $^{48}$Ca, the LO interaction does not reproduce
the correct shell closures at the Hartree-Fock level and, thus, the
closed-shell formulations of CC and IM-SRG typically fail to
converge. In these cases we simply use the HF ground-state energy for
the uncertainty quantification. 
  
Generally we find a systematic decrease of the uncertainties with
increasing chiral order, as expected. For the lower orders up to
N$^2$LO, the interaction uncertainties are significantly larger than
the many-body uncertainties. Only at N$^3$LO and N$^4$LO the
interaction and many-body uncertainties are of comparable size. We
conclude from this observation that the many-body methods and their
truncation uncertainties are sufficiently well controlled in order to
address nuclei in the medium-mass regime with chiral
interactions. Even at the highest available order of the chiral
expansion, the different sources of uncertainties are comparable in
size, so that a significant improvement on the total uncertainty would
require improvements on all aspects of the calculation.  

The sequence of ground-state energies from LO to N$^4$LO for these
medium-mass nuclei shows the same systematic pattern observed in light
nuclei: The LO interactions for both cutoffs produce drastic
overbinding and unrealistic ground states. Going to NLO the energy
jumps and the overbinding is reduced significantly. The step to
N$^2$LO does not affect the ground-state energies for the oxygen
isotopes, but lowers the ground-state energies for the calcium
isotopes again. Going to N$^3$LO the ground-state energies exhibit another
jump leading to a moderate underbinding compared to experiment. From
N$^3$LO to N$^4$LO the energies remain stable for all nuclei.  
 
As repeatedly emphasized, 
one has to keep in mind that the 3N interactions, which appear from
N$^2$LO on, are not included in these calculations. Therefore, we
cannot draw rigorous conclusions about the convergence of the chiral
expansion at this stage. It will be very interesting to explore how
the inclusion of a consistent 3N interaction fitted in the few-body
sector for N$^2$LO and beyond will change the observed trends in
ground-state energies of medium-mass nuclei. This is the prime goal of
our ongoing research program. 

The charge radii mirror the pattern observed for the ground-state
energies. As the ground-state energy increases and the binding
decreases, the charge radii increase as expected from a naive
mean-field picture. For N$^3$LO and N$^4$LO the charge radii for
$^{16}$O are close to the experimental value, however, for
$^{40,48}$Ca the radii are underestimated by about $0.4\,\text{fm}$
although the nuclei are underbound. It remains to be seen, how the 3N
contributions affect the radii, but it is unlikely that the inclusion
of the consistent 3N interactions alone will resolve this discrepancy.  
 
\begin{table}[t]
  \renewcommand{\arraystretch}{1.2}
  \begin{ruledtabular}
  \begin{tabular}{ c | c | c | c | c | c | c | c }
Nucleus & Method & LO & NLO & N${}^2$LO & N${}^3$LO & N${}^4$LO &
                                                                  Exp. \\ \hline
\multicolumn{8}{l}{$R = 0.9\mbox{ fm}$}
 \\ \hline
$^{16}$O & CCSD       & $-$13.92 ; $-$13.84 &  $-$8.81 ;  $-$8.76 &  $-$8.74 ;  $-$8.71 &  $-$6.97 ;  $-$6.90 &  $-$6.74 ;  $-$6.69 &  $-$7.98 \\ 
 & CR-CC(2,3) & $-$13.97 ; $-$13.86 &  $-$8.93 ;  $-$8.82 &  $-$8.88 ;  $-$8.79 &  $-$7.10 ;  $-$6.97 &  $-$6.87 ;  $-$6.77 &  \\ 
 & IM-SRG     & $-$13.96 ; $-$13.86 &  $-$8.96 ;  $-$8.83 &  $-$8.94 ;  $-$8.82 &  $-$7.17 ;  $-$7.01 &  $-$6.96 ;  $-$6.81 &\\ \hline 

$^{24}$O & CCSD       & $-$10.11 ; $-$10.33 &  $-$8.05 ;  $-$7.93 &  $-$8.09 ;  $-$8.03 &  $-$5.80 ;  $-$5.68 &  $-$5.48 ;  $-$5.40 & $-$7.04 \\ 
 & CR-CC(2,3) & $-$10.59 ; $-$10.89 &  $-$8.17 ;  $-$8.00 &  $-$8.23 ;  $-$8.11 &  $-$5.94 ;  $-$5.76 &  $-$5.63 ;  $-$5.48 &  \\ 
 & IM-SRG     & ---    ; ---    &  $-$8.19 ;  $-$8.00 &  $-$8.26 ;  $-$8.12 &  $-$5.99 ;  $-$5.78 &  $-$5.70 ;  $-$5.52 & \\ \hline 

$^{40}$Ca & CCSD       & $-$16.78 ; $-$16.30 & $-$10.69 ; $-$10.37 & $-$12.19 ; $-$12.01 &  $-$7.79 ;  $-$7.42 &  $-$7.27 ;  $-$6.99 & $-$8.55 \\ 
 & CR-CC(2,3) & $-$16.83 ; $-$16.33 & $-$10.84 ; $-$10.44 & $-$12.37 ; $-$12.10 &  $-$7.94 ;  $-$7.49 &  $-$7.44 ;  $-$7.07 &  \\ 
 & IM-SRG     & $-$16.82 ; $-$16.32 & $-$10.86 ; $-$10.46 & $-$12.40 ; $-$12.12 &  $-$7.96 ;  $-$7.50 &  $-$7.48 ;  $-$7.09 & \\ \hline 

$^{48}$Ca & CCSD       & ---    ; ---    & $-$10.77 ; $-$10.35 & $-$13.05 ; $-$12.82 &  $-$7.03 ;  $-$6.52 &  $-$6.59 ;  $-$6.17 & $-$8.67 \\ 
 & CR-CC(2,3) & ---    ; ---    & $-$10.92 ; $-$10.42 & $-$13.21 ; $-$12.89 &  $-$7.19 ;  $-$6.59 &  $-$6.75 ;  $-$6.25 &  \\ 
 & IM-SRG     & ---    ; ---    & $-$10.93 ; $-$10.43 & $-$13.20 ; $-$12.89 &  $-$7.20 ;  $-$6.59 &  $-$6.78 ;  $-$6.26 & \\ \hline 
\multicolumn{8}{l}{$R = 1.0\mbox{ fm}$}
 \\ \hline
$^{16}$O & CCSD       & $-$13.84 ; $-$13.75 &  $-$9.36 ;  $-$9.30 &  $-$8.98 ;  $-$8.95 &  $-$7.00 ;  $-$6.93 &  $-$7.06 ;  $-$7.02 & $-$7.98\\ 
         & CR-CC(2,3) & $-$13.88 ; $-$13.77 &  $-$9.45 ;  $-$9.36 &  $-$9.10 ;  $-$9.02 &  $-$7.14 ;  $-$7.01 &  $-$7.20 ;  $-$7.10 &  \\ 
         & IM-SRG     & $-$13.87 ; $-$13.76 &  $-$9.47 ;  $-$9.36 &  $-$9.15 ;  $-$9.05 &  $-$7.25 ;  $-$7.06 &  $-$7.29 ;  $-$7.14 &  \\ \hline 

$^{24}$O & CCSD       & ---    ; $-$10.53 &  $-$8.59 ;  $-$8.47 &  $-$8.34 ;  $-$8.27 &  $-$5.72 ;  $-$5.64 &  $-$5.78 ;  $-$5.72 & $-$7.04\\ 
         & CR-CC(2,3) & ---    ; $-$10.97 &  $-$8.69 ;  $-$8.53 &  $-$8.45 ;  $-$8.33 &  $-$5.87 ;  $-$5.72 &  $-$5.92 ;  $-$5.80 &  \\ 
         & IM-SRG     & ---    ; ---    &  $-$8.70 ;  $-$8.53 &  $-$8.46 ;  $-$8.34 &  $-$5.95 ;  $-$5.76 &  $-$5.99 ;  $-$5.83 &  \\ \hline 

$^{40}$Ca & CCSD       & $-$17.07 ; $-$16.68 & $-$11.81 ; $-$11.46 & $-$12.69 ; $-$12.47 &  $-$7.34 ;  $-$7.12 &  $-$7.45 ;  $-$7.25 & $-$8.55 \\ 
          & CR-CC(2,3) & $-$17.10 ; $-$16.69 & $-$11.91 ; $-$11.52 & $-$12.81 ; $-$12.53 &  $-$7.49 ;  $-$7.18 &  $-$7.58 ;  $-$7.31 &  \\ 
          & IM-SRG     & $-$17.10 ; $-$16.69 & $-$11.92 ; $-$11.52 & $-$12.83 ; $-$12.54 &  $-$7.51 ;  $-$7.18 &  $-$7.60 ;  $-$7.31 &  \\ \hline 

$^{48}$Ca & CCSD       & ---    ; $-$13.82 & $-$11.87 ; $-$11.42 & $-$13.59 ; $-$13.27 &  $-$4.37 ; ---    &  $-$4.91 ;  $-$4.51 & $-$8.67\\ 
          & CR-CC(2,3) & ---    ; $-$14.13 & $-$11.98 ; $-$11.48 & $-$13.70 ; $-$13.32 &  $-$4.53 ; ---    &  $-$5.10 ;  $-$4.58 &  \\ 
          & IM-SRG     & ---    ; ---    & $-$11.98 ; $-$11.48 & $-$13.68 ; $-$13.31 & ---    ; ---    & ---    ; ---    &  \\ \hline 

  \end{tabular}
  \end{ruledtabular}
  \caption{\label{tab:ccimsrg_energy}
    Ground-state energies per nucleon, in MeV, using SRG-evolved SCS NN
    interactions from LO to N${}^{4}$LO at $R=0.9$~fm and $R=1.0$~fm obtained form CCSD, CR-CC(2,3) and IM-SRG calculations. For each isotope, method and chiral truncation two numbers are given, where the first corresponds to an SRG-flow parameter $\alpha = 0.04 \text{fm}^4$ and the second to  $\alpha = 0.08 \text{fm}^4$. If no result is quoted then the CC or IM-SRG equations did not provide a stable solution, because the initial HF single-particle spectrum does not exhibit the correct shell closures.}
\end{table}
\begin{table}[t]
  \renewcommand{\arraystretch}{1.2}
  \begin{ruledtabular}
  \begin{tabular}{ c | c | c | c | c | c | c | c }
Nucleus & Method & LO & NLO & N${}^2$LO & N${}^3$LO & N${}^4$LO & Exp. \\ \hline
\multicolumn{8}{l}{$R = 0.9\mbox{ fm}$}
 \\ \hline
$^{16}$O & CCSD       &   1.72 ;   1.78 &   2.18 ;   2.22 &   2.25 ;   2.28 &   2.50 ;   2.52 &   2.55 ;   2.57 & 2.70 \\ 
 & IM-SRG     &   1.72 ;   1.78 &   2.20 ;   2.23 &   2.26 ;   2.29 &   2.51 ;   2.54 &   2.57 ;   2.59 & \\ \hline 

$^{24}$O & CCSD       &   1.77 ;   1.84 &   2.15 ;   2.19 &   2.20 ;   2.22 &   2.52 ;   2.54 &   2.60 ;   2.63 & --- \\ 
 & IM-SRG     & ---    ; ---    &   2.16 ;   2.20 &   2.20 ;   2.23 &   2.53 ;   2.55 &   2.62 ;   2.65 & \\ \hline 

$^{40}$Ca & CCSD       &   2.00 ;   2.07 &   2.60 ;   2.65 &   2.53 ;   2.55 &   2.94 ;   2.97 &   3.09 ;   3.11 & 3.48 \\ 
 & IM-SRG     &   2.00 ;   2.07 &   2.63 ;   2.68 &   2.55 ;   2.57 &   2.95 ;   2.97 &   3.11 ;   3.13 & \\ \hline 

$^{48}$Ca & CCSD       & ---    ; ---    &   2.58 ;   2.64 &   2.41 ;   2.43 &   2.89 ;   2.91 &   3.06 ;   3.08 & 3.48 \\ 
 & IM-SRG     & ---    ; ---    &   2.61 ;   2.67 &   2.43 ;   2.44 &   2.91 ;   2.92 &   3.08 ;   3.10 & \\ \hline
\multicolumn{8}{l}{$R = 1.0\mbox{ fm}$}
 \\ \hline
$^{16}$O & CCSD       &   1.77 ;   1.83 &   2.12 ;   2.16 &   2.21 ;   2.24 &   2.54 ;   2.56 &   2.54 ;   2.56 &  2.70\\ 
         & IM-SRG     &   1.77 ;   1.83 &   2.13 ;   2.17 &   2.22 ;   2.25 &   2.55 ;   2.57 &   2.55 ;   2.57 & \\ \hline 

$^{24}$O & CCSD       & ---    ;   1.88 &   2.08 ;   2.12 &   2.14 ;   2.17 &   2.56 ;   2.58 &   2.57 ;   2.59 &  ---\\ 
         & IM-SRG     & ---    ; ---    &   2.09 ;   2.13 &   2.15 ;   2.18 &   2.57 ;   2.60 &   2.59 ;   2.61 & \\ \hline 

$^{40}$Ca & CCSD       &   2.07 ;   2.14 &   2.48 ;   2.54 &   2.43 ;   2.47 &   2.98 ;   3.00 &   3.03 ;   3.05 &  3.48\\ 
          & IM-SRG     &   2.07 ;   2.13 &   2.50 ;   2.55 &   2.45 ;   2.48 &   2.99 ;   3.01 &   3.04 ;   3.06 & \\ \hline 

$^{48}$Ca & CCSD       & ---    ;   2.19 &   2.46 ;   2.51 &   2.31 ;   2.35 &   2.84 ; ---    &   2.91 ;   2.93 &  3.48\\ 
          & IM-SRG     & ---    ; ---    &   2.48 ;   2.54 &   2.32 ;   2.36 & ---    ; ---    & ---    ; ---    & \\ \hline 

  \end{tabular}
  \end{ruledtabular}
  \caption{\label{tab:ccimsrg_radii}
    Charge radii, in fm, using SRG-evolved SCS NN
    interactions from LO to N${}^{4}$LO at $R=0.9$~fm and $R=1.0$~fm obtained form CCSD and IM-SRG calculations. For each isotope, method and chiral truncation two numbers are given, where the first corresponds to an SRG-flow parameter $\alpha = 0.04 \text{fm}^4$ and the second to  $\alpha = 0.08 \text{fm}^4$. If no result is quoted then the CC or IM-SRG equations did not provide a stable solution, because the initial HF single-particle spectrum does not exhibit the correct shell closures.}
\end{table}

\section{Alternative approaches for uncertainty quantification}
\def\theequation{\arabic{section}.\arabic{equation}}
\label{sec:Uncertainties}

As explained in the introduction, our simple and universal approach to estimating 
truncation errors assumes that the chiral expansion of the nuclear forces translates into 
a similar expansion for the calculated observables, see Eq.~(\ref{PCAssumption}). 
While this assumption holds true for the scattering amplitude in a
perturbative regime, it is violated in the near-threshold kinematics
if the corresponding scattering lengths take large values
\cite{Epelbaum:2017byx}, as is the case for the NN $^1$S$_0$ and $^3$S$_1$
partial
waves. The large S-wave NN scattering lengths also result in the strong
cancellations between the kinetic and potential energies 
when calculating the spectra of light nuclei \cite{Weinberg:1991um}. Instead of  
trying to account for all relevant dynamically generated fine-tuned scales in
all partial waves and for all kinematical conditions, 
we use a simplistic, universal approach to uncertainty quantification by
incorporating the
information about the actual pattern of the chiral expansion for a
given observable in order to account for the above-mentioned departures from
naive dimensional analysis. In the following, we address the reliability
of the resulting error estimations, discuss the robustness of our
approach and consider two alternative formulations. 
\begin{itemize}
\item
{\bf Alternative approach 1}\\
We first explore the possibility of relaxing the constraints in Eq.~(\ref{ErrorOrig2}). To 
retain a realistic estimation of the truncation error especially at
low orders of the chiral expansion, we still make use of the information about
the explicit size of the order-$Q^i$ contributions to an observable of
interest for all available chiral orders. Specifically, we replace 
Eqs.~(\ref{ErrorOrig}) and (\ref{ErrorOrig2})  by 
\begin{equation}
\label{ErrorDickComplete}
\delta X^{(0)} = \max_{i \geq 2} \Big( Q^2 | X^{(0)} |, \; Q^{2 - i} | \Delta
X^{(i)} | \Big), \quad \quad \delta X^{(j)} = Q^{j-1} \delta X^{(0)},
\quad \text{for} \; j \geq 2
\end{equation}
for the case of complete calculations. Such an approach
may be expected to provide a more realistic estimation of
uncertainties at lower orders in the chiral expansion as compared to
the standard method. 
For incomplete calculations
based on two-nucleon forces only, we rather estimate $\delta X^{(0)}$
via 
\begin{equation}
\label{ErrorDickIncomplete}
\delta X^{(0)} = \max_{i \geq 3} \Big( Q^2 |X^{(0)}|,  |\Delta X^{(2)}|, Q^{-1} |\Delta X^{(i)}| \Big)
, \quad \quad \delta X^{(j)} = Q^{j-1} \delta X^{(0)},
\quad \text{for} \; j \geq 2 \,.
\end{equation}
In practice, the above modifications are found to lead to very small changes in the
estimated theoretical uncertainties. For example, using
Eq.~(\ref{ErrorDickComplete}), we obtain for the neutron-proton total cross section
at $E_{\rm lab} = 143~$MeV for the cutoff of $R=0.9~$fm  
\begin{equation}
52.5 \pm 11.8_{[Q^0]} \; \rightarrow \; 49.1 \pm 5.1_{[Q^2]} \; \rightarrow
\; 54.2 \pm 2.2_{[Q^3]} \; \rightarrow \; 53.7 \pm 1.0_{[Q^4]}
\; \rightarrow \; 53.9 \pm 0.4_{[Q^5]}  \,,
\end{equation}
which has to be compared with the estimation based on the original
approach using Eqs.~(\ref{ErrorOrig}) and (\ref{ErrorOrig2}):
\begin{equation}
52.5 \pm 9.8_{[Q^0]} \; \rightarrow \; 49.1 \pm 5.1_{[Q^2]} \; \rightarrow
\; 54.2 \pm 2.2_{[Q^3]} \; \rightarrow \; 53.7 \pm 1.0_{[Q^4]}
\; \rightarrow \; 53.9 \pm 0.4_{[Q^5]}  \,.
\end{equation}
Thus, in this particular case, the modification only amounts to a
slight increase of the theoretical uncertainty at LO. Similarly, we
find very minor changes when using Eq.~(\ref{ErrorDickIncomplete})
instead of Eq.~(\ref{ErrorMod}) to estimate truncation errors in incomplete few-body
calculations based on two-nucleon interactions only, see Fig.~\ref{fig:Nd_Errors}
for representative examples. 
\begin{figure}[tb]
\includegraphics[width=\textwidth,keepaspectratio,angle=0,clip]{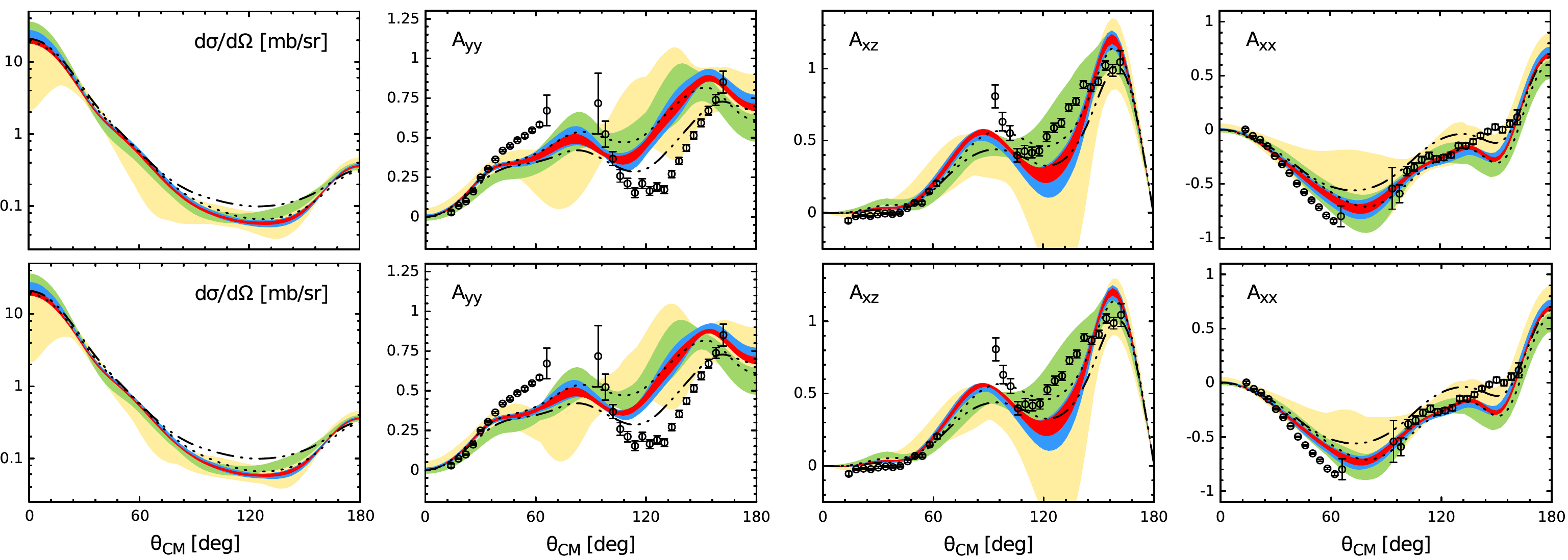}
\caption{(Color online) Predictions for the differential cross
  section, and deuteron tensor analyzing powers $A_{yy}$,  $A_{xz}$ and
  $A_{xx}$ at the laboratory energy of $200$~MeV based on the
NN potentials of Refs.~\cite{Epelbaum:2014efa,Epelbaum:2014sza} for 
$R=0.9$~fm without including the 3NF.  
The bands of 
increasing width show estimated theoretical uncertainty at N$^4$LO (red), 
N$^3$LO (blue), N$^2$LO (green) and NLO (yellow). 
The theoretical uncertainties in the upper and lower rows are estimated
using Eq.~(\ref{ErrorMod}) and (\ref{ErrorDickIncomplete}),
respectively. 
The dotted (dashed)
lines show the results based on the CD Bonn NN potential (CD Bonn NN
potential in combination with the Tucson-Melbourne 3NF).
Open circles are proton-deuteron data from
Refs.~\cite{vonPrzewoski:2003ig}.  
}
\label{fig:Nd_Errors}
\end{figure}

\item
{\bf Alternative approach 2}\\
Furthermore, we consider a minimalistic approach for uncertainty
quantification of calculated ground state energies that does not rely on
the knowledge of contributions beyond the leading order by assigning
the uncertainties as
\begin{equation}
\label{approach2}
\delta E^{(0)} =  Q^2 | \langle V \rangle^{(0)} |, \quad \quad \delta E^{(i
 \ge 2)} =  Q^{i+1} | \langle V \rangle^{(0)} |\,.
\end{equation}
without any further constraints.  Also the momentum scale $Q$ is based
on the calculated $p_{\hbox{\scriptsize avg}}$ at leading order given in
Table~\ref{tab:MomScale}, whereas in the original approach and
Alternative 1 we used the average of $p_{\hbox{\scriptsize avg}}$ over
all available chiral orders.  Thus, the uncertainties are estimated
entirely based on the leading order information.

Notice that using the expectation value of the potential energy rather
than the binding energy as done in the original approach and
Alternative 1 is crucial in order to account for the fine-tuning
associated with the NN interaction being close to the unitary limit
(large S-wave scattering lengths).  While the ignorance of the
fine-tuned nature of the binding energies in the other two approaches
is, to a large extent, effectively corrected by employing the
available information about the actual pattern of the chiral
expansion, an attempt to use the binding energy instead of
$ \langle V \rangle^{(0)} $ in Eq.~(\ref{approach2}) will
yield drastically underestimated truncation errors.

This simple minimalistic approach has an appealing feature that the
estimated uncertainties for the energies beyond the leading order do
not involve any information on the specific behavior at higher orders
as it only builds upon the expected suppression of higher-order
contributions of the chiral EFT expansion.  
On the other hand,
this method is less universal than the other two approaches 
since it is defined specifically for the bound state energy. 
\end{itemize}

\begin{figure}
  \includegraphics[width=0.99\columnwidth]{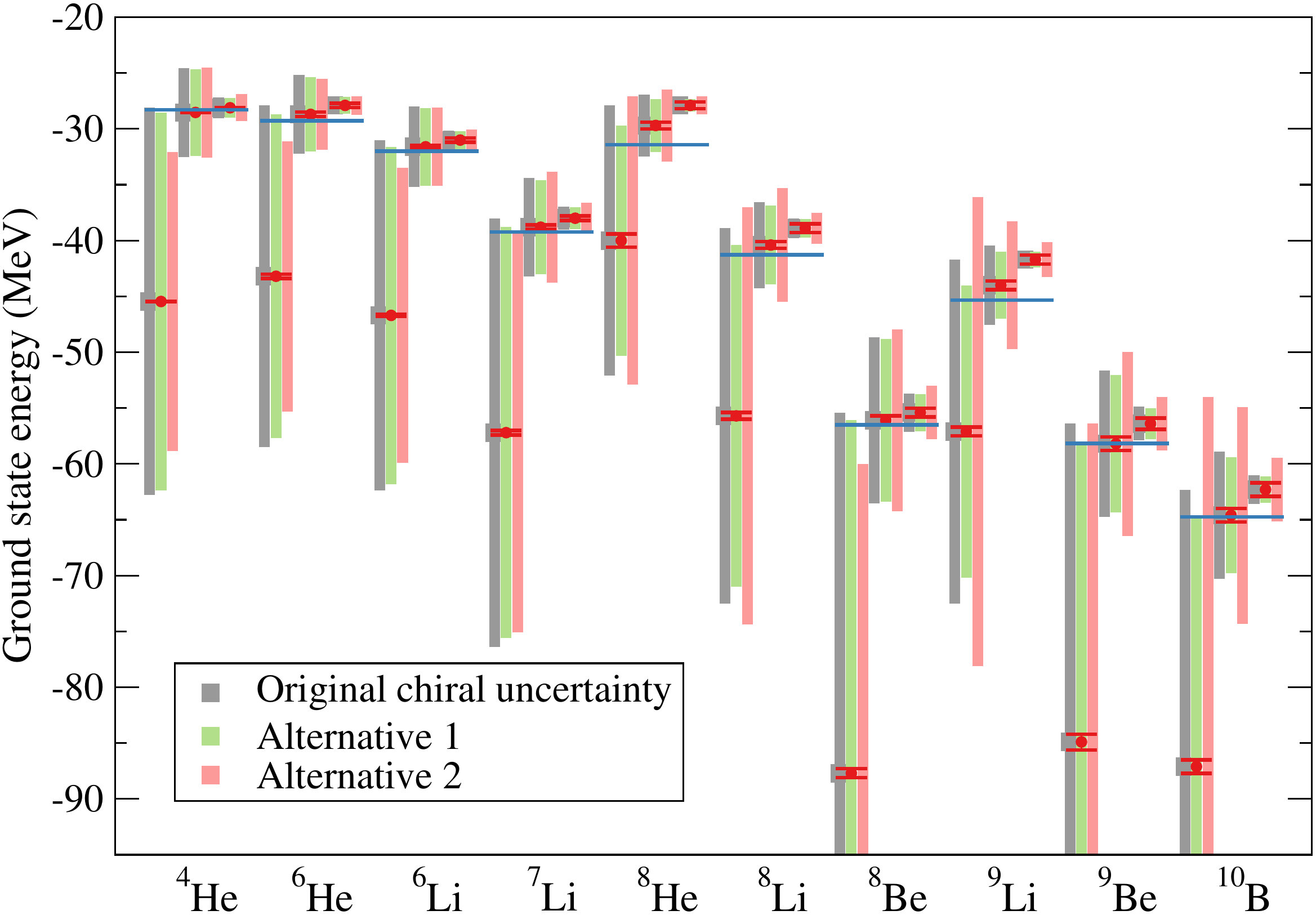}
  \caption{(Color online)
    Results from Fig.~\ref{Fig:res_gs_Eb_A03A10} showing chiral
    uncertainties as presented in the Introduction (grey bars)
    compared with the two alternative uncertainty estimates (green and
    pale red bars), discussed in the text.  The red error bars indicate
    the many-body uncertainties.  For comparison, the experimental
    ground state energies are also shown as the blue bars.
\label{Fig:res_gs_Eb_light_R10_originalA1A2}}
\end{figure} 
\begin{figure}
  \includegraphics[width=0.99\columnwidth]{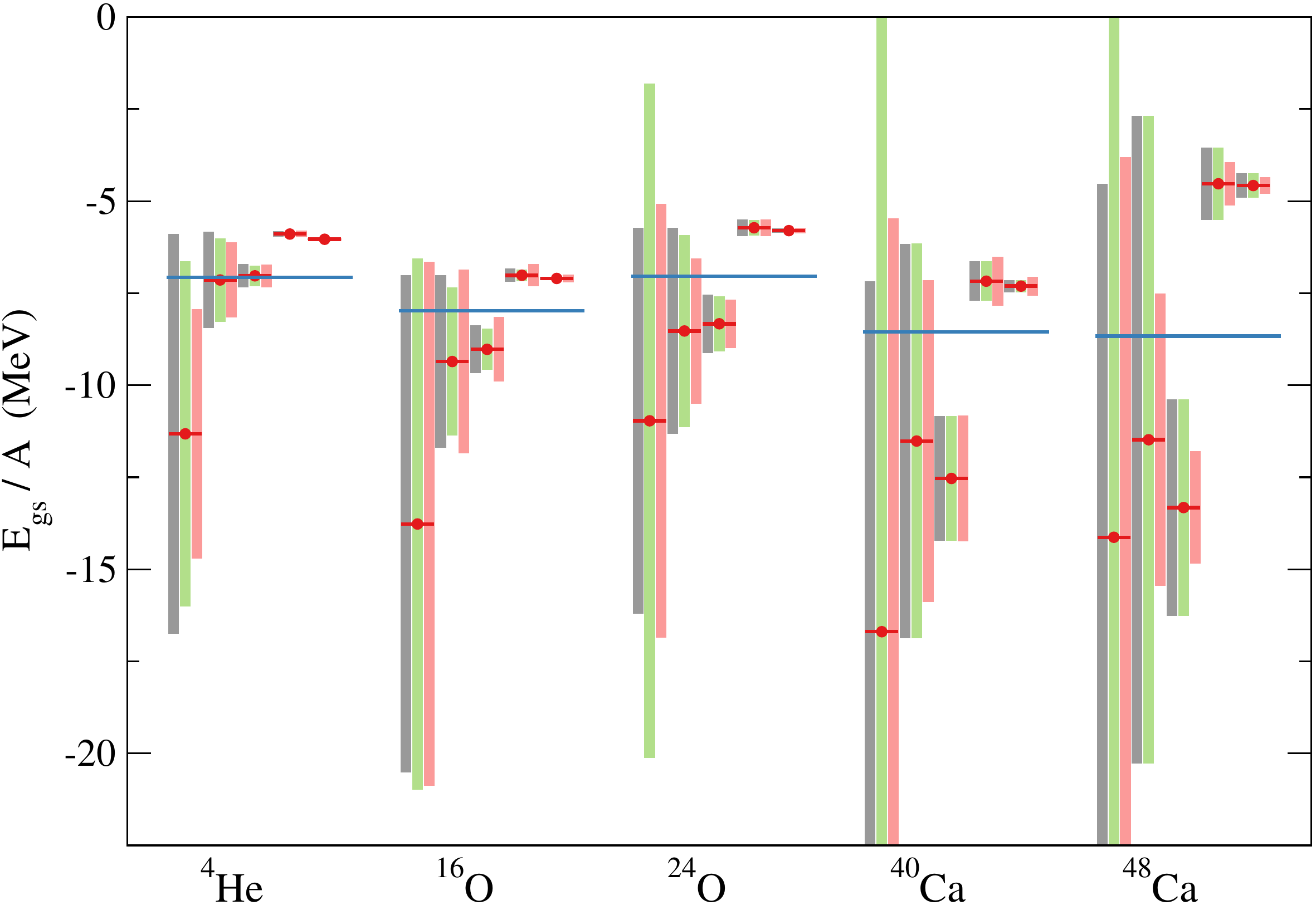}
  \caption{(Color online)
    Results for ground state energies per nucleon of closed (sub)shell
    nuclei, showing chiral 
    uncertainties as presented in the Introduction (gray bars)
    compared with the two alternative uncertainty estimates (green
    and pale red bars), discussed in the text.  No numerical
    many-body uncertainties are shown. 
All results correspond to $R=1.0$~fm and SRG $\alpha=0.08$~fm$^4$ except for
    $^{48}$Ca at N$^3$LO, where the
    results for $\alpha=0.04$~fm$^4$ were taken due to the
    unavailability of the ones for
    $\alpha=0.08$~fm$^4$.   
For comparison, the
    experimental values are also shown as the blue bars.
\label{Fig:res_gs_EA_medium_R10a08_originalA1A2}}
\end{figure} 

In Fig.~\ref{Fig:res_gs_Eb_light_R10_originalA1A2} we show the results
for light nuclei along with the uncertainty estimates presented in
Fig.~\ref{Fig:res_gs_Eb_A03A10} and the uncertainty estimates from these
alternative approaches.  Overall, Alternative 1 produces
very similar uncertainty estimates as the original approach for these
calculations which are truncated at N$^2$LO, but there are some
significant differences in the error estimates with Alternative 2.
One of the most notable differences is for $^{10}$B, where Alternative
2 produced the largest chiral uncertainty, and in general, Alternative
2 suggests larger chiral uncertainties than the original approach or
Alternative 1 as $A$ increases.  Another significant difference is for
$A = 8$ ($^8$He, $^8$Li, and $^8$Be), where the original error
estimates increase significantly as one proceeds towards $N=Z$ at
fixed $A$, but this does not happen as strongly with Alternative 2.

In Fig.~\ref{Fig:res_gs_EA_medium_R10a08_originalA1A2} we show the
results for ground state energies per nucleon of light and medium
nuclei with closed (sub)shells up to N$^4$LO with the different chiral
error estimates.  Overall, Alternative 2 produces very similar
uncertainty estimates as the original approach for these calculations,
but there are significant differences in the error estimates with
Alternative 1.  In particular, for the medium-mass nuclei $^{24}$O,
$^{40}$Ca, and $^{48}$Ca, Alternative 1 produces very large
uncertainties at LO.  This can be traced back to the large differences
between the N$^2$LO results and N$^3$LO and N$^4$LO results for these
nuclei.  
We emphasize, that the original error estimates and Alternative 1
are significantly influenced by the missing 3N (and possibly 4N)
forces at N$^3$LO and N$^4$LO.
Clearly the role of 3N (and possibly 4N) forces becomes more
important for these medium-mass nuclei, not only for the actual ground
state energies, but also for the chiral truncation uncertainty
estimates.

We interpret the differences  in the estimated truncation errors,
emerging from using the considered schemes, as an intrinsic
uncertainty of our approach to error analysis. It would be interesting to see if
it can be reduced by performing a more refined analysis using Bayesian
methods, which would also provide a statistical interpretation of the
theoretical error bars~\cite{Furnstahl:2015rha,Melendez:2017phj}.

\section{Summary and conclusions}
\def\theequation{\arabic{section}.\arabic{equation}}
\label{sec:Summary}

In this paper, we performed a comprehensive study of few- and
many-nucleon observables based on the novel SCS chiral NN potentials
of Refs.~\cite{Epelbaum:2014efa,Epelbaum:2014sza}. 
The pertinent results of our calculations can be summarized as follows:

\begin{itemize}
\item
We have analyzed various
nd elastic scattering and breakup observables and estimated 
truncation errors at different orders of the chiral
expansion. Similarly to other calculations, we
observe a considerable underprediction of the nd elastic scattering
analyzing power $A_y$ at low energy starting from N$^2$LO, the feature
commonly referred to as the $A_y$-puzzle. At intermediate energies, 
the discrepancies between the calculated elastic scattering
observables based on NN forces only and experimental data are, in many
cases, significantly larger than the theoretical uncertainty
at N$^3$LO and N$^4$LO and agree well
with the expected size of the 3NF contributions. 
This makes elastic nucleon-deuteron scattering
in this energy range a particularly promising testing ground for the
chiral 3NF. On the other hand, the considered breakup observables 
are well reproduced, leaving little room for possible 3NF effects
except for the symmetric space star configuration at low energy, 
where large deviations are observed. For these observables,
known to represent another low-energy puzzle, our
calculations agree with the ones based on other phenomenological and
chiral EFT nuclear forces, and the truncation errors turn out to be very
small. 
\item
We have calculated various properties of $A=3,4$ nuclei in the
framework of the Faddeev-Yakubovsky equations and studied light
p-shell nuclei using the NCCI method. In the latter case, we were able
to perform calculations at all chiral orders without SRG
transformations for $A \le 6$
using the cutoffs $R=0.9$, $1.0$ and $1.1$~fm. For heavier nuclei, we
had to rely on SRG evolution starting from N$^3$LO in order to
achieve converged results. We found a qualitatively similar
convergence pattern in all considered cases, namely a significant
overbinding at LO, results close to the experimental values at NLO and
N$^2$LO and underbinding at N$^3$LO and N$^4$LO. We have also
calculated ground state magnetic moments of light nuclei based 
on the single-nucleon current operator and estimated the corresponding
NCCI extrapolation and truncation errors. 
\item
To obtain results for medium-mass nuclei, we have performed 
state-of-the-art calculations within the coupled-cluster and in-medium
similarity renormalization group frameworks. The obtained results for
the ground state energies of $^{16,24}$O and $^{40,48}$Ca show a
similar pattern to that for light nuclei with the amount of
overbinding (underbinding) at LO, NLO and N$^2$LO  (N$^3$LO and
N$^4$LO) tending to increase with the number of nucleons $A$. The slower
convergence of the chiral expansion for heavier nuclei is to be
expected and reflects the increasing sensitivity to higher-momentum
components of the interaction. The calculated charge radii of the considered
medium-mass nuclei show a systematic improvement with the chiral
order, but remain underestimated at N$^4$LO.    
\item
Finally, we have addressed the reliability of our error analysis by
exploring alternative approaches for uncertainty
quantifications. We found, in general, a satisfactory agreement
between all considered methods. 
\end{itemize}

Our results demonstrate that the SCS chiral NN potentials are well
suited for \emph{ab initio} few- and many-body calculations and provide a
natural reference point for systematic studies of 3NF effects and 
specific details of the NN interactions such as the choice of the
basis of contact interactions and regularization schemes. It would be
 interesting to perform similar calculations using the new
SMS chiral NN potentials of Ref.~\cite{Reinert:2017usi}, which
provide an outstanding description of neutron-proton and
proton-proton scattering data below
$E_{\rm lab} = 300$~MeV and are considerably
softer than the SCS potentials starting from N$^3$LO. Such a study would, in
particular, bring insights into the role of the redundant contact
interactions at N$^3$LO.  Notice that the new SMS interactions
also provide the flexibility to propagate statistical uncertainties of
the NN LECs and to quantify the error from the uncertainty in the $\pi N$ LECs and the
choice of the energy range used in the determination of the NN contact
interactions. Finally and most importantly, the calculations should
be extended by the inclusion of the consistent 3NFs
\cite{vanKolck:1994yi,Epelbaum:2002vt,Bernard:2007sp,Bernard:2011zr,Krebs:2012yv,Krebs:2013kha,Epelbaum:2014sea}. Work
along these 
lines is in progress by the LENPIC Collaboration \cite{Hebeler:2015wxa}.

\section*{Acknowledgments}

This work was supported by BMBF (contracts No.~05P2015 - NUSTAR R\&D and No. 05P15RDFN1 - NUSTAR.DA),
by the European
Community-Research Infrastructure Integrating Activity ``Study of
Strongly Interacting Matter'' (acronym HadronPhysics3,
Grant Agreement n. 283286) under the Seventh Framework Programme of EU,
 the ERC project 259218 NUCLEAREFT, by the DFG (SFB 1245), by DFG and
 NSFC (CRC 110), by the Polish National Science
Center under Grants No. 2016/22/M/ST2/00173
and 2016/21/D/ST2/01120
and by the Chinese Academy of Sciences (CAS) President’s
International Fellowship Initiative (PIFI) (Grant No. 2018DM0034)
In addition, this
research was supported in part by the National Science Foundation
under Grants No.~NSF PHY11-25915 and NSF PHY16-14460, by the US Department of Energy under
Grants DE-FG02-87ER40371, DESC0008485, DE-SC0008533, DESC0018223 and DESC0015376.   
This research used resources of the National 
Energy Research Scientific Computing Center (NERSC) and
the Argonne Leadership Computing Facility (ALCF), which
are US Department of Energy Office of Science user facilities,
supported under Contracts No. DE-AC02-05CH11231 and
No. DE-AC02-06CH11357, and computing resources provided 
under the INCITE award `Nuclear Structure and Nuclear Reactions' from
the US Department of Energy, Office of Advanced Scientific Computing
Research. Further computing resources were provided by the TU Darmstadt (lichtenberg) and on JUQUEEN and JURECA 
of the J\"ulich Supercomputing Center, J\"ulich, Germany.

\appendix

\section{Expectation values of the kinetic energy in light- and
  medium-mass nuclei}
\def\theequation{\Alph{section}.\arabic{equation}}
\setcounter{equation}{0}
\label{ExpVal}

In this appendix we provide some details on the estimation of the expansion 
parameter used to quantify the theoretical uncertainties for the ground state 
properties of light- and medium-mass nuclei. As explained in section \ref{sec:LightNuclei}, 
this is achieved by inferring the relevant momentum scale from the expectation 
values of the SRG-transformed kinetic energy operator $T$. In Tables
\ref{tab:MomScale} and \ref{tab:MomScale2}, 
\begin{table}[bt]
  \renewcommand{\arraystretch}{1.2}
  \begin{ruledtabular}
    \begin{tabular}{l|rrrrr|r}
 \multicolumn{1}{l}{$R$ [fm]} &
      \multicolumn{1}{l}{$p_{\hbox{\scriptsize avg}}^{(0)}$}  & 
      \multicolumn{1}{l}{$p_{\hbox{\scriptsize avg}}^{(2)}$} &
      \multicolumn{1}{l}{$p_{\hbox{\scriptsize avg}}^{(3)}$} &
      \multicolumn{1}{l}{$p_{\hbox{\scriptsize avg}}^{(4)}$} &
      \multicolumn{1}{l}{$p_{\hbox{\scriptsize avg}}^{(5)}$} &
      \multicolumn{1}{l}{$p_{\hbox{\scriptsize avg}}$} 
      \\ \hline
      \multicolumn{7}{l}{$^3$H}
      \\ \hline
      0.9   & 158.3  & 125.2 & 123.0 & 116.7 & 116.8 & 128.0  \\
      1.0   & 149.8  & 123.8 & 121.5 & 111.6 & 112.9 & 123.9  \\
      1.0* & 144.0  &  126.0 & 124.0 &          &           &            \\
      \hline
      \multicolumn{7}{l}{$^3$He}
      \\ \hline    
      0.9   & 157.3  & 123.9 & 121.6 & 115.4 & 115.6 & 126.8  \\
      1.0   & 148.7  & 122.5 & 120.1 & 110.4 & 111.7 & 122.7  \\
      \hline
      \multicolumn{7}{l}{$^4$He}
      \\ \hline    
      0.9   & 192.9  & 147.9 & 145.8 & 132.8 & 133.1 & 150.5  \\
      1.0   & 180.6  & 146.3 & 144.0 & 126.7 & 128.4 & 145.2  \\
      1.0* & 180.0  &  147.0 & 145.0 & 176.8 & 169.6 & 163.6  \\
      \hline
      \multicolumn{7}{l}{$^6$He}
      \\ \hline    
      0.9   & 163.8  & 141.4 & 135.1 & 124.0 & 123.7 & 137.6  \\
      1.0   & 156.3  & 139.6 & 133.9 & 119.0 & 120.4 & 133.8  \\
      1.0* & 157.0  &  137.0 & 132.0 &          &           &            \\
      \hline
      \multicolumn{7}{l}{$^8$He}
      \\ \hline    
      0.9   & 153.3  & 146.8 & 136.0 & 124.0 & 123.1 & 136.6  \\
      1.0   & 148.1  & 144.4 & 135.1 & 118.9 & 120.1 & 133.3  \\
      1.0* & 150.0  &  141.0 & 131.0 &          &           &            \\
      \hline
      \multicolumn{7}{l}{$^6$Li}
      \\ \hline    
      0.9   & 166.6  & 143.5 & 137.1 & 125.6 & 125.3 & 139.6  \\
      1.0   & 159.4  & 141.8 & 135.9 & 120.4 & 121.9 & 135.9  \\
      1.0* & 160.0  &  140.0 & 136.0 &          &           &
      \\
      \hline
      \multicolumn{7}{l}{$^8$Li}
      \\ \hline    
      0.9   & 165.9  & 151.7 & 141.9 & 128.3 & 127.3 & 143.0  \\
      1.0   & 159.5  & 149.5 & 140.9 & 122.9 & 124.3 & 139.4  \\
      1.0* & 163.0  &  148.0 & 140.0 &          &           &            \\
      \hline
      \multicolumn{7}{l}{$^9$Li}
      \\ \hline    
      0.9   & 166.5  & 155.8 & 144.4 & 129.8 & 128.4 & 145.0  \\
      1.0   & 160.2  & 153.5 & 143.4 & 124.1 & 125.4 & 141.3  \\
      1.0* & 163.0  &  152.0 & 142.0 &          &           &            \\
    \end{tabular}  
  \end{ruledtabular} 
  \caption{\label{tab:MomScale}
Hartree-Fock results (rows without asterisk) and NCCI results
(rows with asterisk) for the average relative momentum (in MeV/c)
between a pair of nucleons in $^3$H, $^{3,4,6,8}$He and $^{6,8,9}$Li 
 as defined in Eq.~(\ref{pavg}). 
The first 5 columns of results correspond to the chiral orders from LO
to N$^4$LO 
while the last column is the average over the 5 columns of
results.  The rows are labeled by the value of the regularization
parameter $R$. 
Hartree-Fock results are from the expectation value of the
SRG-transformed kinetic energy operator with the SRG parameter $\alpha
= 0.08~$fm$^4$. 
The first two rows for each nucleus are obtained 
with spherical Hartree-Fock using the filling fraction approximation
appropriate to the specified nucleus.  Rows with results from an NCCI
calculation (labeled with an asterisk) 
use the bare NN interaction and the bare relative kinetic energy operator, 
extrapolated to the infinite matrix limit.  We quote the Yakubovsky results for 
$^4$He at N$^3$LO and N$^4$LO for their higher precision.
}
\end{table}
we list our Hartree-Fock results for all nuclei considered in this paper 
for the cutoff values of $R=0.9~$fm and $R=1.0~$fm.  The resulting values of the momentum 
scale given in the last column of this table are obtained by taking the average over 
all chiral orders $i$ of the quantity $\sqrt{2 m_N \langle T \rangle^{(i)}/A}$.

\begin{table}[tb]
  \renewcommand{\arraystretch}{1.2}
  \begin{ruledtabular}
    \begin{tabular}{l|rrrrr|r}
 \multicolumn{1}{l}{$R$ [fm]} &
      \multicolumn{1}{l}{$p_{\hbox{\scriptsize avg}}^{(0)}$}  & 
      \multicolumn{1}{l}{$p_{\hbox{\scriptsize avg}}^{(2)}$} &
      \multicolumn{1}{l}{$p_{\hbox{\scriptsize avg}}^{(3)}$} &
      \multicolumn{1}{l}{$p_{\hbox{\scriptsize avg}}^{(4)}$} &
      \multicolumn{1}{l}{$p_{\hbox{\scriptsize avg}}^{(5)}$} &
      \multicolumn{1}{l}{$p_{\hbox{\scriptsize avg}}$} 
      \\ 
      \hline
      \multicolumn{7}{l}{$^8$Be}
      \\ \hline    
      0.9   & 174.8  & 150.3 & 142.2 & 128.7 & 127.8 & 144.8  \\
      1.0   & 166.7  & 148.7 & 141.2 & 123.2 & 124.6 & 140.9  \\
      1.0* & 176.0  &  149.0 & 144.0 &          &           &            \\

      \hline
      \multicolumn{7}{l}{$^9$Be}
      \\ \hline    
      0.9   & 175.5  & 156.7 & 146.7 & 131.5 & 130.2 & 148.1  \\
      1.0   & 167.8  & 154.7 & 145.7 & 125.6 & 127.0 & 144.2  \\
      1.0* & 173.0  &  153.0 & 146.0 &          &           &            \\
\hline
      \multicolumn{7}{l}{$^{16}$O}
      \\ \hline    
      0.9   & 223.1  & 177.2 & 169.2 & 148.3 & 145.0 & 172.5  \\
      1.0   & 210.3  & 176.5 & 167.6 & 140.6 & 141.0 & 167.2  \\
      1.0* & 209.0  &  173.0 & 164.0 &          &           &            \\
      \hline
      \multicolumn{7}{l}{$^{24}$O}
      \\ \hline    
      0.9   & 211.7  & 186.1 & 179.8 & 153.6 & 147.9 & 175.8  \\
      1.0   & 201.3  & 185.4 & 178.7 & 145.2 & 144.4 & 171.0  \\
      \hline
      \multicolumn{7}{l}{$^{40}$Ca}
      \\ \hline    
      0.9   & 249.5  & 196.1 & 203.5 & 168.0 & 159.0 & 195.2  \\
      1.0   & 234.4  & 198.2 & 203.2 & 158.5 & 155.7 & 190.0  \\
      \hline
      \multicolumn{7}{l}{$^{48}$Ca}
      \\ \hline    
      0.9   & 244.8  & 203.8 & 222.4 & 178.1 & 167.1 & 203.2  \\
      1.0   & 230.9  & 206.2 & 221.9 & 174.2 & 169.8 & 200.6  \\
    \end{tabular}  
  \end{ruledtabular} 
  \caption{\label{tab:MomScale2}
Same as table \ref{tab:MomScale} but for $^{8,9}$Be, $^{16,24}$O and
$^{40,48}$Ca. 
}
\end{table}

\end{document}